\input amstex
\input amsppt.sty
\input amssym.def
\input amssym
%
%
\font\ninerm=cmr9 
\font\ninei=cmmi9
\font\nineit=cmti9
\font\ninesl=cmsl9
\font\ninebf=cmbx9
\font\ninesy=cmsy9
\def\rmnine{\fam0\ninerm}
\def\itnine{\fam\itfam\nineit}
\def\slnine{\fam\slfam\ninesl}
\def\bfnine{\fam\bffam\ninebf}
\def\ninepoint{\let\rm=\rmnine
\textfont0=\ninerm \scriptfont0=\sevenrm \scriptscriptfont0=\fiverm
\textfont1=\ninei\scriptfont1=\seveni \scriptscriptfont1=\fivei
\textfont2=\ninesy
\textfont\itfam=\nineit \let\it=\itnine
\textfont\slfam=\ninesl \let\sl=\slnine
\textfont\bffam=\ninebf \scriptfont\bffam=\sevenbf
\scriptscriptfont\bffam=\fivebf
\let\bf=\bfnine
\let\sc=\sevenrm
\normalbaselineskip=11pt\normalbaselines\rm}
\font\tenib=cmmib10
\font\tensc=cmcsc10
\def\rmten{\fam0\tenrm}
\def\itten{\fam\itfam\tenit}
\def\slten{\fam\slfam\tensl}
\def\bften{\fam\bffam\tenbf}
\def\tenpoint{\let\rm=\rmten
\textfont0=\tenrm\scriptfont0=\sevenrm\scriptscriptfont0=\fiverm
\textfont1=\teni\scriptfont1=\seveni\scriptscriptfont1=\fivei
\textfont2=\tensy
\textfont\itfam=\tenit \let\it=\itten
\textfont\slfam=\tensl \let\sl=\slten
\textfont\bffam=\tenbf
\scriptfont\bffam=\sevenbf
\scriptscriptfont\bffam=\fivebf
\let\bf=\bften
\let\sc=\tensc
\normalbaselineskip=12pt\normalbaselines\rm}
\font\twelmib=cmmib10 scaled \magstep1
\font\twelbf=cmbx10   scaled \magstep1
\font\twelsy=cmsy10   scaled \magstep1
\def\rmbftwel{\fam0\twelbf}
\def\itbftwel{\fam\itfam\twelmib}
\def\bftwel{\fam\bffam\twelbf}
\def\bftwelpoint{\let\rm=\rmbftwel
\textfont0=\twelbf  \scriptfont0=\tenbf \scriptscriptfont0=\ninebf
\textfont1=\twelmib \scriptfont1=\tenib \scriptscriptfont1=\tenib
\textfont2=\twelsy
\textfont\itfam=\twelmib  \let\it=\itbftwel
\textfont\slfam=\twelmib  \let\sl=\itbftwel 
\textfont\bffam=\twelbf   \scriptfont\bffam=\tenbf
\scriptscriptfont\bffam=\ninebf
\let\bf=\bftwel
\normalbaselineskip=14pt\normalbaselines\rm}
\font\fortmib=cmmib10 scaled \magstep2
\font\fortbf=cmbx10   scaled \magstep2
\font\fortsy=cmsy10   scaled \magstep2
\def\rmbffort{\fam0\fortbf}
\def\itbffort{\fam\itfam\fortmib}
\def\bffort{\fam\bffam\fortbf}
\def\bffortpoint{\let\rm=\rmbffort
\textfont0=\fortbf  \scriptfont0=\twelbf \scriptscriptfont0=\tenbf
\textfont1=\fortmib \scriptfont1=\twelmib \scriptscriptfont1=\tenib
\textfont2=\fortsy
\textfont\itfam=\fortmib  \let\it=\itbffort
\textfont\slfam=\fortmib  \let\sl=\itbffort 
\textfont\bffam=\fortbf   \scriptfont\bffam=\twelbf
\scriptscriptfont\bffam=\tenbf
\let\bf=\bffort
\normalbaselineskip=16pt\normalbaselines\rm}
%
\def\ul#1{$\setbox0=\hbox{#1}\dp0=0pt\mathsurround=0pt
\underline{\box0}$}
%
\pageno=1
\parindent=4mm
\hsize=120mm \hoffset=20mm
\vsize=190mm \voffset=20mm

%
\def\raggedcenter{\leftskip=4em plus 12em \rightskip=\leftskip
  \parindent=0pt \parfillskip=0pt \spaceskip=.3333em \xspaceskip=.5em
  \pretolerance=9999 \tolerance=9999
  \hyphenpenalty=9999 \exhyphenpenalty=9999 }
\def\\{\break}
\newtoks\LECTURE \LECTURE={ }
\long\def\title#1{\bgroup\vglue10mm\raggedcenter
	             {\tenpoint\bf\the\LECTURE\ }
	             \vskip10mm
	             {\bffortpoint #1}}

\long\def\author#1{\vskip5mm\tenpoint\rm #1}
\def\inst#1{$(^{#1})$}
\long\def\institute#1{\vskip5mm\tenpoint\sl#1}
\def\maketitle{\vskip5mm
               \vtop{\baselineskip=4pt
                     \vrule height.5pt width3cm\par
                     \vrule height.5pt width2cm}
               \vskip20mm\egroup\tenpoint
	          \let\lasttitle=Y\everypar={\let\lasttitle=N}}
%
\newbox\boxtitle
\newskip\beforesect \newskip\aftersect 
\newskip\beforesubsect \newskip\aftersubsect 
\newskip\beforesubsubsect \newskip\aftersubsubsect
%
\beforesect=7mm plus1mm minus1mm   \aftersect=5mm plus.5mm minus.5mm
\beforesubsect=5mm plus.5mm minus.5mm \aftersubsect=3mm plus.2mm minus.1mm 
\beforesubsubsect=3mm plus.2mm minus.1mm \aftersubsubsect=2mm plus.2mm minus.1mm 
\newcount\CNTa \CNTa=0
\newcount\CNTb \CNTb=0
\newcount\CNTc \CNTc=0
\newcount\CNTd \CNTd=0
\def\resetCN #1{\global\csname CN#1\endcsname =0}
\def\stepCN #1{\global
\expandafter\advance \csname CN#1\endcsname by 1}
\def\ArabicCN #1{\expandafter\number\csname CN#1\endcsname}
\def\RomanCN #1{\uppercase\expandafter{\romannumeral\csname CN#1\endcsname}}
\newcount\sectionpenalty  \sectionpenalty=0
\newcount\subsectionpenalty  \subsectionpenalty=0
\newcount\subsubsectionpenalty  \subsubsectionpenalty=0
\newdimen\indsect
\newdimen\dimensect
\indsect=1cm
\dimensect=\hsize\advance\dimensect by -\indsect
\def\section#1#2 {\par\resetCN{Tb}\resetCN{Tc}\resetCN{Td}%
              \if N\lasttitle\else\vskip-\beforesect\fi
              \bgroup
              \bf
              \pretolerance=20000
              \setbox0=\vbox{\vskip\beforesect
                       \noindent\ArabicCN{Ta}.\kern1em#1#2
                       \vskip\aftersect}
              \dimen0=\ht0\advance\dimen0 by\dp0 
              \advance\dimen0 by 2\baselineskip
              \advance\dimen0 by\pagetotal
              \ifdim\dimen0>\pagegoal
                 \ifdim\pagetotal>\pagegoal
                 \else\eject\fi\fi
              \vskip\beforesect
              \penalty\sectionpenalty \global\sectionpenalty=-200
              \global\subsectionpenalty=10007
              \ifx#1*\noindent #2\else\stepCN{Ta}
              \setbox0=\hbox{\noindent\ArabicCN{Ta}.}
              \indsect=\wd0\advance\indsect by 1em
              \parshape=2 0pt\hsize \indsect\dimensect
              \noindent\hbox to \indsect{\ArabicCN{Ta}.\hfil}#1\fi
              \vskip\aftersect
              \egroup
              \let\lasttitle=Y
              \nobreak\parindent=0pt
              \everypar={\parindent=4mm
                         \penalty0\let\lasttitle=N}\ignorespaces}
\def\subsection#1#2 {\par\resetCN{Tc}\resetCN{Td}%
              \if N\lasttitle\else\vskip-\beforesubsect\fi
              \bgroup\tenpoint\bf
                 \setbox0=\vbox{\vskip\beforesubsect
                 \noindent\ArabicCN{Ta}.\ArabicCN{Tb}.\kern1em#1#2
                 \vskip\aftersubsect}
              \dimen0=\ht0\advance\dimen0 by\dp0\advance\dimen0 by
                 2\baselineskip
              \advance\dimen0 by\pagetotal
              \ifdim\dimen0>\pagegoal
                 \ifdim\pagetotal>\pagegoal
                 \else \if N\lasttitle\eject\fi \fi\fi
              \vskip\beforesubsect
              \if N\lasttitle \penalty\subsectionpenalty \fi
              \global\subsectionpenalty=-100
              \global\subsubsectionpenalty=10007
              \ifx#1*\noindent#2\else\stepCN{Tb}
              \setbox0=\hbox{\noindent\ArabicCN{Ta}.\ArabicCN{Tb}.}
              \indsect=\wd0\advance\indsect by 1em
              \parshape=2 0pt\hsize \indsect\dimensect
              \noindent\hbox to \indsect{\ArabicCN{Ta}.\ArabicCN{Tb}.\hfil}#1\fi
              \vskip\aftersubsect
              \egroup\let\lasttitle=Y
              \nobreak\parindent=0pt
              \everypar={\parindent=4mm
                         \penalty0\let\lasttitle=N}\ignorespaces}
\def\subsubsection#1#2 {\par\resetCN{Td}%
              \if N\lasttitle\else\vskip-\beforesubsubsect\fi
              \bgroup\tenpoint\sl
                 \setbox0=\vbox{\vskip\beforesubsubsect\noindent
              {\ArabicCN{Ta}.\ArabicCN{Tb}.\ArabicCN{Tc}.\kern1em}#1#2
              \vskip\aftersubsubsect}
              \dimen0=\ht0\advance\dimen0 by\dp0\advance\dimen0 by
                 2\baselineskip
              \advance\dimen0 by\pagetotal
              \ifdim\dimen0>\pagegoal
                 \ifdim\pagetotal>\pagegoal
                 \else \if N\lasttitle\eject\fi \fi\fi
              \vskip\beforesubsubsect
              \if N\lasttitle \penalty\subsubsectionpenalty \fi
              \global\subsubsectionpenalty=-50
              \ifx#1*\noindent#2\else\stepCN{Tc}
              \setbox0=\hbox{\noindent
              	\ArabicCN{Ta}.\ArabicCN{Tb}.\ArabicCN{Tc}.}
              \indsect=\wd0\advance\indsect by 1em
              \parshape=2 0pt\hsize \indsect\dimensect
              \noindent\hbox to 
              \indsect{\ArabicCN{Ta}.\ArabicCN{Tb}.\ArabicCN{Tc}.\hfil}#1\/\fi
              \vskip\aftersubsubsect
              \egroup\let\lasttitle=Y
              \nobreak\parindent=0pt
              \everypar={\parindent=4mm
                         \penalty0\let\lasttitle=N}\ignorespaces}
\def\paragraph#1#2 {\par\if N\lasttitle\else\vskip-\aftersubsubsect\fi
    		    \bgroup\tenpoint\rm
         	    \setbox0=\vbox{\vskip\aftersubsubsect\noindent
         	    {\ArabicCN{Ta}.\ArabicCN{Tb}.\ArabicCN{Tc}.\ArabicCN{Td}}#1#2}
              \dimen0=\ht0\advance\dimen0 by\dp0\advance\dimen0 by
                 2\baselineskip
              \advance\dimen0 by\pagetotal
              \ifdim\dimen0>\pagegoal
              \ifdim\pagetotal>\pagegoal
              \else \if N\lasttitle\eject\fi \fi\fi
              \vskip\aftersubsubsect
              \if N\lasttitle \penalty-50 \fi
              \ifx#1*\noindent\ul{#2:}\ \else\stepCN{Td}
              \setbox0=\hbox{\noindent
              	\ArabicCN{Ta}.\ArabicCN{Tb}.\ArabicCN{Tc}.\ArabicCN{Td}.}
              \indsect=\wd0\advance\indsect by 1em
              \parshape=2 0pt\hsize \indsect\dimensect
              \noindent\hbox to 
              \indsect{\ArabicCN{Ta}.\ArabicCN{Tb}.\ArabicCN{Tc}.\ArabicCN{Td}.
              \hfil}\ul{#1:}\ \fi
              \egroup\let\lasttitle=N}
%
\newtoks\ACK \ACK={Acknowledgements}
\def\ack#1{\par\vskip\beforesect\goodbreak
\noindent{\tenpoint\bf\the\ACK }\par\vskip\aftersect\penalty500\noindent#1}
%
\newtoks\APPND \APPND={Appendix}
\def\appendix#1{\par\vskip\beforesect\goodbreak
\noindent{\tenpoint\bf\the\APPND \kern.5em#1}\par
\vskip\aftersect\penalty500\let\lasttitle=Y}
\def\titleapp #1{\if N\lasttitle\goodbreak
\else\vskip-\beforesubsect\penalty500\fi
\vskip\beforesubsect
{\tenpoint\bf\noindent\ignorespaces #1}
\vskip\aftersubsect\let\lasttitle=Y\noindent}
%
\newtoks\REFNAME \REFNAME={References}
\def\references{\begREF}
\def\begREF{\bgroup
             \setbox0=\vbox{\vskip\beforesect\noindent{\bf\the\REFNAME}
                       \vskip\aftersect}
              \dimen0=\ht0\advance\dimen0 by\dp0 
              \advance\dimen0 by 2\baselineskip
              \advance\dimen0 by\pagetotal
              \ifdim\dimen0>\pagegoal
                 \ifdim\pagetotal>\pagegoal
                 \else\eject\fi\fi
              \vskip\beforesect\noindent{\bf\the\REFNAME}
              \vskip\aftersect
               \frenchspacing \parindent=0pt \leftskip=1truecm
               \everypar{\hangindent=\parindent}}
\def\ref#1{ $[{\rm #1}]$}%
\gdef\refis#1{\item{$[$#1$]$\ }}   
\def\endreferences{\par\egroup}
%
\def\review#1, #2, 1#3#4#5, #6 {{\sl#1\/} {\bf#2} (1#3#4#5) #6}
\def\book#1, #2, #3, 1#4#5#6, #7 {#1 (#2, #3, 1#4#5#6) p. #7}
%
\newcount\foCN \foCN=0
\def\fonote{\global\advance\foCN by 1
$(^{\rm\number\foCN})$\vfootnote{$(^{\rm\number\foCN})$}}
%
\catcode`@=11
\def\vfootnote#1{\insert\footins\bgroup
  \ninepoint
  \interlinepenalty\interfootnotelinepenalty
  \splittopskip\ht\strutbox 
  \splitmaxdepth\dp\strutbox \floatingpenalty\@MM
  \leftskip\z@skip \rightskip\z@skip \spaceskip\z@skip \xspaceskip\z@skip
  \baselineskip=10pt\lineskip=10pt
  \noindent\kern10mm\llap{#1\enspace}\footstrut
  \futurelet\next\fo@t}
\def\@foot{\egroup}
\catcode`@=12
%
\newtoks\shorttitle
\newtoks\authors
\shorttitle={AUTOMATIC SEQUENCES}%
\authors={J.-P. Allouche and M. Mend\`es France}%
\def\firsthd{\hfill}
\def\hdleft{\tenpoint\folio\hfill{\sl \the\authors\/}\hfill}
\def\hdrigt{\tenpoint\hfill{\ninepoint\the\shorttitle}\hfill\folio}
\newif\ifbegpage
\headline={\ifbegpage\firsthd
\global\begpagefalse\else
\ifodd\pageno\hdrigt\else\hdleft\fi
\advance\pageno by 1\fi}
\footline={\hfil}
\def\makeheadline{\vbox to0pt{\vskip-10mm
\line{\vbox to8.5pt{}\the\headline}\vss}\nointerlineskip}
%
\begpagetrue
%
\newtoks\TABLE \TABLE={Table}
%
\newdimen\tableheight
\newskip\superskipamount \superskipamount=28pt plus 4pt minus 4pt
\def\superskip{\vskip\superskipamount}
\def\superbreak{\par\ifdim\lastskip<\superskipamount
  \removelastskip\penalty-200\superskip\fi}
\catcode`\@=11
\def\endinsert{\egroup 
  \if@mid \dimen@\ht\z@ \advance\dimen@\dp\z@
    \advance\dimen@24\p@ \advance\dimen@\pagetotal
    \ifdim\dimen@>\pagegoal\@midfalse\p@gefalse\fi\fi
  \if@mid \superskip\box\z@\superbreak
  \else\insert\topins{\penalty100 
    \splittopskip\z@skip
    \splitmaxdepth\maxdimen \floatingpenalty\z@
    \ifp@ge \dimen@\dp\z@
    \vbox to\vsize{\unvbox\z@\kern-\dimen@}
    \else \box\z@\nobreak\superskip\fi}\fi\endgroup}
\catcode`\@=12
\newcount\CNfig \CNfig=0
\let\captext=N
\def\begfig #1cm{\midinsert\tableheight=#1cm\advance\tableheight by 5mm
	\vglue\tableheight}
\def\topfig #1cm{\topinsert\tableheight=#1cm\advance\tableheight by 5mm
	\vglue\tableheight}
\long\def\caption#1{\let\captext=Y\stepCN{fig}\ninepoint\rm
	\noindent Fig.\ \number\CNfig.\kern.3em ---\kern.3em\ignorespaces
     \parindent=0pt#1\par}
\def\endfig{\if N\captext\stepCN{fig}
	\ninepoint\rm\noindent Figure \number\CNfig\else\let\figtext=N\fi
	\endinsert}
%
\newcount\CNtab \CNtab=0
\let\tabtext=N
\def\begtab #1cm{\midinsert\tableheight=#1cm}
\def\toptab #1cm{\topinsert\tableheight=#1cm}
\long\def\tabcap#1{\let\tabtext=Y\stepCN{tab}
	\ninepoint\rm\noindent Table\ \RomanCN{tab}.\kern.3em ---\kern.3em\ignorespaces
	\parindent=0pt#1\par}
\def\endtab{\if N\tabtext\stepCN{tab}
	\ninepoint\rm\noindent Table \RomanCN{tab} \else\let\tabtext=N\fi
	\ifdim\tableheight=0cm\vskip-\belowdisplayskip
     \else\advance\tableheight by 5mm\fi
     \vglue\tableheight\endinsert}
%

%
\def\(#1){(\call{#1})}

\catcode`\@=11
\def\logo@{}
\catcode`\@=\active

\document

\hbox{}
\catcode`\@ =11
\def\proclaimheadfont@{\smc}
\catcode`\@ = \active


\pageno=293

\ { }

\vskip -2truecm

\title{COURSE 11 \\  \vskip .18truecm  Automata and automatic sequences}

\vskip -.1truecm

\author{J.-P. Allouche \inst{1} and M. Mend\`es France \inst{2}}

\vskip -.15truecm

\institute{\inst{1} CNRS, LMD, Luminy Case 930, \\
F-13288 Marseille Cedex 9, France

\inst{2} Universit\'e Bordeaux I, Math\'ematiques, 351 cours
de la Lib\'eration,\\
F-33405 Talence Cedex, France}
\authors{J.-P. Allouche and M. Mend\`es France}
\shorttitle{AUTOMATIC SEQUENCES}

\vskip -.5truecm 

\maketitle

\vskip -1.5truecm

In the following pages we discuss infinite sequences defined on a 
finite alphabet, and more specially those which are generated by
finite automata. We have divided our paper into seven parts which
are more or less self-contained. Needless to say, we feel that the order 
we propose is the most natural one. References appear at the end of each
one of the parts which implies some redundancy. Extra references are
listed at the very end of our paper.

\medskip

\noindent
PART I. Introduction. Substitutions and finite automata

\medskip

1. The Fibonacci sequence 

\smallskip

2. The Prouhet-Thue-Morse sequence 

\smallskip

3. The paperfolding sequence 

\smallskip

4. Automatic sequences: definition; a zoo of examples.
A warning

\smallskip

5. Where finite automata enter the picture 

\smallskip

6. How random can an automatic sequence be? 

\smallskip

7. Miscellanea 

\smallskip

8. Appendix: automata generating the five examples of Paragraph 4 

\smallskip

References

\medskip

\noindent
Part II. Further properties of paperfolding

\medskip

1. Direct reading and reverse reading 

\smallskip

2. Words and diagrams 

\smallskip

3. The folding operators

\smallskip

4. The dimension of a curve 

\smallskip

5. Paperfolding and continued fractions 

\smallskip

References

\bigskip

\noindent
Part III. Complements

\medskip

1. Repetitions in infinite sequences

\smallskip

2. Multidimensional morphisms and (finite) automata

\smallskip

3. Links with cellular automata

\smallskip

References

\bigskip

\noindent
Part IV. Fourier analysis

\medskip

1. Fourier-Bohr coefficients

\smallskip

2. Bessel's inequality and Parseval's equality

\smallskip

3. The Wiener spectrum and the spectral measure

\smallskip

4. Analyzing the spectral measure

\smallskip

5. Appendix I: the spectral measure of the Thue-Morse sequence

\smallskip

6. Appendix II: the spectral measure of the paperfolding sequence

\smallskip

References

\bigskip

\noindent
Part V. Complexity of infinite sequences

\medskip

1. Sturmian sequences and generalizations

\smallskip

2. Complexity of infinite sequences

\smallskip

References

\bigskip

\noindent
Part VI. Opacity of an automaton

\medskip

1. Automata revisited

\smallskip

2. Computing the opacity

\smallskip

References

\bigskip

\noindent
Part VII. The Ising automaton

\medskip

1. The inhomogeneous Ising chain

\smallskip

2. The induced field

\smallskip

3. The Ising automaton

\smallskip

4. An ergodic property

\smallskip

5. Opacity of the Ising automaton

\smallskip

References

\bigskip

\noindent
Part VIII. Extra references

\newpage

\font\cyr=wncyr10


\document

\hbox{}
\catcode`\@ =11
\def\proclaimheadfont@{\smc}
\catcode`\@ = \active

\font\big= cmr10 scaled 1500
\pageno=295

\centerline{\big PART I}

\bigskip

\centerline{\big Introduction. Substitutions and finite automata}

\bigskip
 
The discovery of quasicrystals in 1984 by Shechtman et al [34] led
to renewed studies of the Penrose tiling with its fivefold symmetry.
In a regular Penrose tiling one can observe infinite arrays or
{\it worms} of short and long {\it ties} distributed according to the
Fibonacci sequence.
Similar sequences, which are generated by {\it substitutions} (or
{\it substitution rules}), became very popular among physicists.  
Instead of substitutions computer scientists rather speak of 
{\it morphisms} whereas physicists also use the term {\it inflation 
rules}.  

\bigskip

\noindent{\bf 1. THE FIBONACCI SEQUENCE}

\bigskip

Let us first recall what the Fibonacci morphism is and how it generates
the Fibonacci sequence. One starts with the {\it alphabet} (finite 
set) $\{0, 1\}$. One then considers the set $\{0, 1\}^*$ of all finite 
{\it words} on this alphabet, a finite word being a finite string
(for example $0101110$) which may be empty. The {\it concatenation} of words, 
consists in assembling words together one after the other, thus creating
a longer word. For instance 
$011.1010=0111010$. The set $\{0, 1\}^*$ endowed with the operation of 
concatenation is called the {\it free monoid} generated by $\{0, 1\}$.
 
We now define the map $\sigma$ on $\{0, 1\}$ by:
$$\sigma(0)= 01, \ \ \sigma(1) = 0.$$
This map can be extended ``morphically" to a map from $\{0, 1\}^*$ to
itself, for example:
$$\sigma(0101110) = \sigma(0)\sigma(1)\sigma(0)\sigma(1)\sigma(1)
\sigma(1)\sigma(0) = 0100100001.$$
The image of an infinite sequence is defined the same way.

\medskip

Let us compute now the iterates of the map $\sigma$ applied to the 
one-letter word $0$:
$$\alignedat2
\sigma(0) &= 01, \\
\sigma^2(0) &= \sigma(01) &&= 010, \\
\sigma^3(0) &= \sigma(010) &&=01001, \\
&\cdots &&
\endalignedat
$$
The reader will have noticed that each new word begins with the previous
one, which implies that the sequence of words $\sigma^n(0)$ converges
to an infinite word
$$w = 0 \ 1 \ 0 \ 0 \ 1 \ 0 \ 1 \ 0 \ 0 \ 1 \ 0 \ 0 \ 1 \ \cdots$$
called the Fibonacci sequence.
This infinite word is by construction a fixed point of $\sigma$:
$\sigma(w)=w$.

\medskip

Note that a rule as the one above is a {\it growth rule} in the
terminology of D. Gratias, and not a {\it matching rule}. A matching
rule is local, but both types of rules are {\it drawing rules}.

That such a rule gives - or may give - a quasicrystalline structure
will be discussed all along this volume. A quasicrystalline
structure is defined either via the notion of 
quasilattices, see the contribution of Y. Meyer [26] in this volume, or
via Fourier transforms. It is somewhere between periodicity and 
chaos (or randomness).

Note also that these morphisms are particular cases of formal grammars
which were invented as toy models for the study of languages. Needless
to say, computer scientists were among the first to develop the theory.

\bigskip

\noindent{\bf 2. THE PROUHET-THUE-MORSE SEQUENCE}

\bigskip

Let us make a (seemingly small) change in the definition of the Fibonacci
morphism, and let us consider the morphism defined by:
$$0 \longrightarrow 01, \ \ 1 \longrightarrow 10.$$
Iterating this morphism starting from $0$ yields:
$$
\aligned
&0 \\
&01 \\
&0110 \\
&01101001 \\
&0110100110010110 \\ 
&\cdots,
\endaligned
$$
hence, as previously noticed, we obtain an infinite sequence which is a 
fixed point of the morphism. This sequence is the celebrated
Prouhet-Thue-Morse sequence, (see [29], [36], [37], [27]). The 
sequence is also known as the Thue-Morse sequence or the Morse 
sequence. We shall freely use any of these denominations.

\medskip

What makes this sequence drastically different from the Fibonacci
sequence is that the images of $0$ and $1$ have the same length, 
namely $2$. The importance of this observation will
be emphasized in two ways:

\smallskip

- denoting by $(u_n)_{n \in \Bbb N}$ the Prouhet-Thue-Morse sequence, it is 
not difficult to prove that $u_n = 1$ if and only if
the number of $1$'s in the binary expansion of $n$ is odd. Such a simple
property is due to the underlying binary expansion in the morphisms
of length $2$. If one wants a similar arithmetic definition for the
Fibonacci sequence, one should introduce the Fibonacci expansion of 
integers which is not as straightforward;

\smallskip

- write down the Fibonacci sequence in a way which emphasizes the
definition as a fixed point of the morphism $\sigma$:
$$\alignedat9
&01 \ \, &&0 \ \, &&01 \ \, &&01 \ \, &&0 \ \, &&01 \ \, &&0 \ \, &&01 \ \,
&&\cdots
\\
&\,0  \ &&1 \ &&\,0  \ &&\,0  \ &&1 \ &&\,0  \ &&1 \ &&\,0 \ \, &&\cdots
\endalignedat
$$  
where each block $01$ or $0$ is written above the letter it originates
from.
This can be seen as a {\it renormalization} process. The constant length 
Prouhet-Thue-Morse morphism can be seen as a {\it constant renormalization}
process:
$$\alignedat9
&01 \ \, &&10 \ \, &&10 \ \, &&01 \ \, &&10 \ \, &&01 \ \, &&01 \ \, &&10 
\ \, &&\cdots
\\
&\,0   &&\,1       &&\,1     &&\,0     &&\,1     &&\,0     &&\,0   &&\,1
\ \, &&\cdots
\endalignedat
$$

\bigskip

\noindent{\bf 3. THE PAPERFOLDING SEQUENCE}

\bigskip

In this section we will discuss a slightly more general way of 
generating sequences: the morphisms will be followed
by a {\it literal map}.

Let us examplify this with the paperfolding sequence. This sequence can be 
generated by regularly folding a piece of paper onto itself and by 
considering, after unfolding, the sequence of folds (``mountains" and 
``valleys") created on its edge (see for instance [18] and [21], see also 
Part II of this paper). This sequence can also be defined as follows.

Consider the alphabet $A=\{a, b, c, d\}$ and the morphism on $A$
defined by
$$\aligned
a &\longrightarrow ab \\
b &\longrightarrow cb \\
c &\longrightarrow ad \\
d &\longrightarrow cd.
\endaligned
$$
This morphism admits a fixed point, obtained as usual by computing
the iterates of the morphism applied to the letter $a$:
$$a \ b \ c \ b \ a \ d \ c \ b \ a \ b \ c \ d \ \cdots$$
One then applies the {\it literal map}
(which can be also seen as a morphism of length $1$ onto a different
alphabet) defined by:
$$\aligned
a &\longrightarrow 1 \\
b &\longrightarrow 1 \\
c &\longrightarrow 0 \\
d &\longrightarrow 0
\endaligned
$$
thus obtaining the sequence
$$1 \ 1 \ 0 \ 1 \ 1 \ 0 \ 0 \ 1 \ 1 \ 1 \ 0 \ 0 \ \cdots$$
which is the paperfolding sequence on the alphabet $\{0,1\}$.
The sequence is obtained as the projection of a fixed point of a
morphism of constant length.

\bigskip

\newpage

\noindent{\bf 4. AUTOMATIC SEQUENCES: DEFINITION; A ZOO OF EXAMPLES.
A WARNING}

\bigskip

\proclaim{Definition}
Let $k$ be an integer greater than or equal to $2$.
A sequence $u = (u_n)_{n \in \Bbb N}$ with values in the alphabet $A$
is said $k$-automatic if there exists:

$\bullet$ an alphabet $B$, 

$\bullet$ a uniform morphism of $B^*$, of length $k$ (i.e. such that 
the image of each letter in $B$ is a word of length $k$ in $B^*$),

$\bullet$ an infinite sequence with values in $B$, say
$v = (v_n)_{n \in \Bbb N}$, which is a fixed point of this morphism,

$\bullet$ a map $\varphi$ from $B$ to $A$ such that \
$\forall n \in \Bbb N, \ \varphi(v_n) = u_n$.
\endproclaim

\medskip

\noindent {\bf Examples}

\medskip

- 1) The Prouhet-Thue-Morse sequence defined above is a $2$-automatic 
sequence.

\medskip

- 2) The paperfolding sequence defined above is a $2$-automatic sequence.

\medskip

- 3) The Rudin-Shapiro sequence (see [33], [31]) can be defined as follows.
Let $A=\{a, b, c, d\}$. Consider the morphism on $A$:
$$\aligned
a &\longrightarrow ab \\
b &\longrightarrow ac \\
c &\longrightarrow db \\
d &\longrightarrow dc
\endaligned
$$
Then define $\varphi : A \to \{-1, +1\}$ by:
$$\aligned
a &\longrightarrow +1 \\
b &\longrightarrow +1 \\
c &\longrightarrow -1 \\
d &\longrightarrow -1
\endaligned
$$
The image of the fixed point of the above morphism by $\varphi$ is the
Rudin-Shapiro sequence
$$+1 \ \ +1 \ \ +1 \ \ -1 \ \ +1 \ \ +1 \ \ -1 \ \ +1 \ \ +1 \ \
+1 \ \ +1 \ \ -1 \ \ \cdots$$

\smallskip

Prove that the Rudin-Shapiro sequence cannot be obtained as the image of
the fixed point of a morphism (of length $2$) on an alphabet with less than
$4$ elements.

\smallskip

This sequence is $2$-automatic. It has been considered both by Shapiro 
[33] and Rudin [31] in the hope of mimicking randomness in the following 
sense. Consider an arbitrary  sequence
of $+1$'s and $-1$'s, say $a=(a_n)_n$, and put
$$M_N(a) = \sup_{\theta \in [0, 1]} \left | \sum_{n=0}^{N-1} a_n 
e^{2i \pi n\theta} \right | = \left \Vert \sum_{n=0}^{N-1} a_n 
e^{2i \pi n\theta} \right \Vert_{\infty}.$$
Then one has: 
$$ \sqrt N = \left \Vert \sum_{n=0}^{N-1} a_n e^{2i \pi n\theta}
\right \Vert_2 \leq \left \Vert \sum_{n=0}^{N-1} a_n 
e^{2i \pi n\theta} \right \Vert_{\infty} = M_N(a) \leq N,$$
where the quadratic norm is defined by:
$$\Vert f \Vert_2 = \left ( \int_0^1 |f(\theta)|^2 d\theta 
\right )^{1/2}.$$
It is not hard to show that for a periodic (or even almost-periodic) sequence
$a$ the order of magnitude of $M_N(a)$ is actually $N$ (check for
instance this claim for the constant sequence $a_n = 1$ for all $n$).
On the other hand it can be shown that for Lebesgue-almost all
sequences $a$ one has, when $N$ goes to infinity:
$$M_N(a) \leq \sqrt{N \log N},$$
which can be reinterpreted by saying that, for a ``random" sequence, the
order of magnitude of $M_N(a)$ is between $\sqrt N$ and  $\sqrt{N \log N}$.

The Rudin-Shapiro sequence is an example of a clearly deterministic
sequence for which one can prove the existence of a positive constant
$C \leq 2 + \sqrt 2$ such that:
$$M_N(a) \leq C \sqrt N. \tag{$*$}$$

\smallskip

This property of the Rudin-Shapiro sequence to simulate randomness will 
also be seen later on (see the contribution of M. Queff\'elec [30] in this 
volume).

\medskip

Let us see how the proof of $(*)$ works. We first show that the 
Rudin-Shapiro sequence $(a_n)_n$ satisfies:
$$\forall n \geq 0, \ a_{2n} = a_n, \ a_{4n+3} = - a_{2n+1}.$$
If $(u_n)_n$ is the fixed point of the above morphism on the
four-letter alphabet $\{a, b, c, d \}$ the very definition of
being a fixed point implies that:
$$\forall n \geq 0, \ u_{2n} = \alpha(u_n), \ u_{2n+1} = \beta(u_n),$$
where the maps $\alpha$ and $\beta$ are defined by:
$$\alignedat8
&\alpha(a) &&= a, \ &&\alpha(b) &&= a, \ 
&&\alpha(c) &&= d, \ &&\alpha(d) &&= d, \\
&\beta(a) &&=b, \ &&\beta(b) &&=c, \ 
&&\beta(c) &&= b, \ &&\beta(d) &&=c.
\endalignedat
$$
One then notices that:
$$
\varphi \circ \alpha = \varphi, \ 
\varphi \circ \beta^2 = - \varphi \circ \beta, \
\varphi \circ \beta \circ \alpha = \varphi.
$$
Hence, $\forall n \geq 0$:
$$\alignedat2
&a_{2n} &&= \varphi(u_{2n}) = \varphi\alpha(u_n) = \varphi(u_n) = a_n, 
\\
&a_{4n+1} &&= \varphi(u_{4n+1}) = \varphi\beta\alpha(u_n) = 
\varphi(u_n) = a_n, \\
&a_{4n+3} &&= \varphi(u_{4n+3}) = \varphi \beta^2(u_n) = 
- \varphi\beta(u_n) = - \varphi(u_{2n+1}) = - a_{2n+1}.
\endalignedat
$$
For all $n \geq 0$ define the two-dimensional vector 
$V_n = \left (\matrix a_n \\ a_{2n+1} \endmatrix \right )$.
Then:
$$\forall n \geq 0, \ V_{2n} = M_0 V_n, \ V_{2n+1} = M_1 V_n,$$
where $M_0$ and $M_1$ are the $2 \times 2$-matrices:
$$M_0 = \left ( \matrix 1 & 0 \\ 1 & 0 \endmatrix \right ), \ \
  M_1 = \left ( \matrix 0 & 1 \\ 0 & -1 \endmatrix \right ).$$
Now let
$$S(K, \theta):= \sum_{n=0}^{2^K -1} V_n e^{2i\pi n \theta}.$$
One has:
$$\aligned
S(K+1, \theta) &= \sum_{n=0}^{2^{K+1} -1} V_n e^{2i\pi n \theta} \\
&= \sum_{n=0}^{2^K -1} V_{2n} e^{2i\pi (2n) \theta}
 + \sum_{n=0}^{2^K -1} V_{2n+1} e^{2i\pi(2n+1) \theta} \\
&=M(\theta) \sum_{n=0}^{2^K -1} V_n e^{2i\pi (2n) \theta} \\
&=M(\theta) S(K,2\theta),
\endaligned
$$
where
$$M(\theta) = \left (\matrix 1 & e^{2i\pi \theta} \\ 1 & -e^{2i\pi \theta}
\endmatrix \right ).
$$
Therefore:
$$\aligned
S(K, \theta) &= M(\theta) M(2\theta) \cdots M(2^{K-1}\theta) S(0, 2^K\theta) \\
&= M(\theta) M(2\theta) \cdots M(2^{K-1}\theta) \left ( \matrix 1 \\ 1 
\endmatrix \right ).
\endaligned
$$
Hence, taking the $L^2$-norm:
$$\Vert S(K, \theta) \Vert_2 \leq 
\sqrt 2 \prod_{j=0}^{K-1} \Vert M(2^j \theta) \Vert_2.$$
But the $L^2$-norm of a matrix 
$M = \left ( \matrix 1 &z \\ 1 &-z \endmatrix \right )$,
where $z$ is a complex number of modulus $1$ is easy to compute: this is the
square root of the spectral radius of the matrix 
$\overline{^tM}M = \left( \matrix 2 &0 \\ 0 &2 \endmatrix \right )$, 
hence $\Vert M \Vert_2 = \sqrt 2$. Thus for all $\theta$:
$$\left | \sum_{n=0}^{2^K-1}a_n e^{2i \pi n \theta} \right | \leq 
\Vert S(K, \theta) \Vert_2 \leq \sqrt 2\ 2^{K/2}.$$
In other words the inequality $M_N(\theta) \leq C \sqrt N$ holds if $N$ is a
power of $2$ with $C = \sqrt 2$. It is then possible to extend the
inequality to arbitrary $N$ provided the value $C=\sqrt 2$ is replaced
by $C=2 + \sqrt 2$.

\medskip

- 4) The period-doubling sequence occurs when iterating a map of the unit
interval. Consider a one-parameter family of unimodal maps showing a Feigenbaum 
cascade. Choose the value of the parameter where the ``chaos" begins.
Then the kneading sequence of the point where the function has its maximum is
universal (even in cases where the Feigenbaum constant is not obtained) and 
equal to the period-doubling sequence. For a general reference on this subject 
see [15] and the included bibliography. This sequence can be defined as the 
fixed point of the morphism
$$\aligned
0 &\longrightarrow 01 \\
1 &\longrightarrow 00;
\endaligned
$$
it is therefore a $2$-automatic sequence. Note that this sequence is 
actually connected with the Thue-Morse sequence (see [5], this seems
also to be implicit from reading between the lines in [24]).

\medskip

- 5) The Hanoi sequence underlies the game known as the Towers of Hanoi. 
In this puzzle there are three pegs and $N$ disks of increasing diameters. 
At the beginning all the disks are stacked on the first peg in increasing
order, the topmost being the one with the smallest diameter. Then, at 
each step, one may take a disk on the top of a peg and put it on another
peg provided that it is never stacked on a smaller disk. The game ends 
when all the disks are on a new peg (necessarily in increasing order). 
Denote by $a$, $b$, $c$ the moves consisting in taking the topmost disk 
from peg I, (resp. peg II, III), and putting it on peg II, (resp. III, 
I), and by $\overline a$, $\overline b$, $\overline c$ the inverse moves.
It can be shown that there exists an infinite sequence on the six-letter
alphabet $A = \{a, b, c, \overline{a}, \overline{b}, \overline{c}\}$
such that its prefixes of length $2^N-1$ give for each $N$ the minimal
strategy to move $N$ disks from peg I to peg II if $N$ is odd and
from peg I to peg III if $N$ is even. Furthermore this sequence can be
obtained (see [4]) as the fixed point of the morphism defined on $A$
by:
$$\aligned
a &\longrightarrow a \overline{c} \\
b &\longrightarrow c \overline{b} \\ 
c &\longrightarrow b \overline{a} \\
\overline{a} &\longrightarrow ac \\
\overline{b} &\longrightarrow cb \\
\overline{c} &\longrightarrow ba.
\endaligned
$$
It is hence also a $2$-automatic sequence. For a survey and a ``direct" proof
of automaticity, see [6]. Another general survey on the towers of Hanoi
is [23].

\medskip

\noindent {\bf Warning}

\medskip

The definition of the Fibonacci sequence ($0 \longrightarrow 01$,
$1 \longrightarrow 0$) uses a morphism of non-constant length. Hence
the question arises whether this sequence can be generated in another
way (constant length morphism followed by a literal map) i.e. whether
it is $k$-automatic for some $k \geq 2$. The answer is {\bf NO}: it can
be shown that, for all $k \geq 2$, this sequence is not $k$-automatic.

In the same vein consider the ``cyclic" towers of Hanoi, where the
towers are disposed on a circle and where only clockwise moves
are permitted (using the aboves notations this means that one may
use only the moves $a$, $b$ and $c$). Then there exists, as previously,
an infinite sequence which is the limit of the finite sequences of moves 
for $N$ disks.
This sequence is given (see [6]) by the following morphism (of 
non-constant length) on the alphabet $A = \{ f, g, h, u, v, w\}$:
$$\aligned
f &\longrightarrow fvf \\
g &\longrightarrow gwg \\
h &\longrightarrow huh \\
u &\longrightarrow fg \\
v &\longrightarrow gh \\
w &\longrightarrow hf,
\endaligned
$$
followed by the projection:
$$\aligned
f &\longrightarrow a \\
g &\longrightarrow c \\
h &\longrightarrow b \\
u &\longrightarrow c \\
v &\longrightarrow b \\
w &\longrightarrow a.
\endaligned
$$
But it can be shown (see [1]) that, for all $k \geq 2$, this sequence is 
not $k$-automatic.

\medskip

Many other sequences discussed by physicists are not $k$-automatic:
for instance the so-called generalized Fibonacci sequences, generated by
$0 \longrightarrow 01^{a+1}$, $1 \longrightarrow 01^a$, are $k$-automatic 
for no value of $k \geq 2$. 
Another example of a sequence which is $k$-automatic for no value of 
$k \geq 2$ is the ``circle sequence": this sequence is associated to the 
substitution $a \longrightarrow cac$, $b \longrightarrow accac$, 
$c \longrightarrow abcac$;
(actually the substitution has no fixed point, but its square has two fixed 
points).

\bigskip

\noindent{\bf 5. WHERE FINITE AUTOMATA ENTER THE PICTURE}

\bigskip

In this section we give the reason why the terminology ``automatic"
is used. A finite automaton (with output function) - also called 
``uniform tag system" by Cobham in [14] - consists of states, transition
functions and an output function. It reads words written with an input 
alphabet and computes an output for any given word. More precisely, if 
$k\geq 2$ is an integer, a $k$-automaton consists of:

\smallskip

$\bullet$ a finite set $S$ of states; one of the states is distinguished
and is called the initial state. $S$ can be thought of as the set of
vertices of a graph;

\smallskip

$\bullet$ $k$ (transition-) maps from $S$ to $S$ labelled $0$, $1$, $\cdots$,
$k-1$, represented as oriented edges (arrows) of the above graph;

\smallskip

$\bullet$ an output function $\varphi$, i.e. a map from the set of states
$S$ to some set $U$.

\smallskip

Such a machine works as follows: the input alphabet is $\{0,1, \cdots
k-1\}$. One chooses to read words on this alphabet either 
from left to right (direct automaton) or from right to left (reverse
automaton). Given an input word one reads it and
feeds the automaton with its letters interpreted as transition maps,
starting from the initial state and following the arrows in order. Having
finished to read the given word, one finds oneself on a state. Take
the image of this state by the output function. Hence to each word on 
$\{0,1, \cdots k-1\}$ one associates an element of $U$.

The machine generates an infinite sequence $(u_n)_{n \geq 0}$
as follows: to each input $n \in \Bbb N$, or rather to its $k$-ary expansion,
the automaton associates an element $u_n \in U$.

\bigskip

\noindent {\bf Example}

\medskip

Take $k=2$, hence the input alphabet is $\{0, 1\}$. Let $S=\{A,B,C\}$ and
let $A$ be the initial state. The arrows of the graph below define the 
transition maps. Put $\varphi(A)=\varphi(B)=0$, $\varphi(C)=1$. 

\bigskip

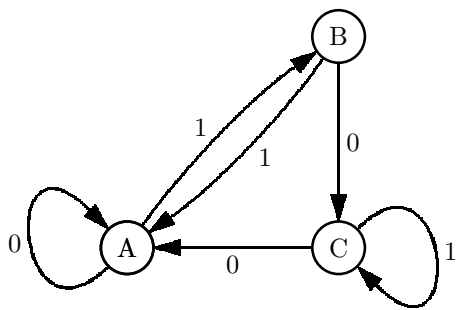
\input DRATEX.STY 
\input ALDRATEX.STY

\Draw
\NewCIRCNode(\StateNode,103,)
\Define\StateAt(3){ \MoveTo(#2,#3) \StateNode(#1)(--#1--)}
\CycleEdgeSpec(45,50)
\DiagramSpec(\StateAt&\TransEdge)
\ArrowHeads(1) 
\SaveAll
\Diagram
(A,0,0 & B,80,80 & C,80,0) (A,B,1,1 & A,C, ,0 & B,C,0,  
& C,C,0,1) 
\RecallAll
\CycleEdgeSpec(225,50)
\Diagram  
(A,0,0) (A,A,0,0)
\EndDraw

\bigskip

Choose to read words from right to left. 
The input word $001110$ yields:

\noindent
$001110.A = 00111.A = 0011.B = 001.A = 00.B = 0.C = A$; and finally 
$\varphi(A) = 0$.

\noindent
To the sequence of binary integers corresponds the sequence of outputs:
$$\aligned
0 &\longrightarrow 0 \\
1 &\longrightarrow 0 \\
10 &\longrightarrow 0 \\
11 &\longrightarrow 0 \\
100 &\longrightarrow 0 \\
101 &\longrightarrow 1 \\
110 &\longrightarrow 0 \\
111 &\longrightarrow 0 \\
1000 &\longrightarrow 0 \\
1001 &\longrightarrow 0 \\
1010 &\longrightarrow 1 \\
1011 &\longrightarrow 0 \\
&\cdots
\endaligned
$$

We shall give in the last paragraph of this chapter more examples of
automata which actually generate the five automatic sequences discussed
in Paragraph 4.

Finally, and to conclude the section, we mention a theorem of Cobham's [14] 
which asserts that a sequence is generated by a 
$k$-automaton as above if and only if it is the literal image of a fixed
point of a morphism of length $k$. Furthermore {\it automatic sequences 
can be generated either by direct or reverse reading automata}.

\bigskip
 
\noindent{\bf 6. HOW RANDOM CAN AN AUTOMATIC SEQUENCE BE?}

\bigskip

The automatic sequences occur in many mathematical fields (number theory,
harmonic analysis, fractals ...) but also in theoretical computer science,
or even in music (see [7] for example). The ``philosophical" reason, if
we may say, is that they lie between order (periodicity)
and chaos (randomness), but that they are however easy to construct and ...
theorems can actually be proved.

\medskip

Now, what happens if one perturbs an automatic sequence?

\smallskip

Firstly one can change the name of the letters without changing
the structure of the sequence: for instance the Rudin-Shapiro sequence
above has been defined on the alphabet $\{-1, +1\}$. As far as automaticity 
is concerned it is perfectly irrelevant to take $\{-1, +1\}$, $\{0, 1\}$ or 
say $\{a, b\}$.

\smallskip

Secondly the family of $k$-automatic sequences is closed under simple
operations such as shifting, changing finitely many terms, adding two
sequences, etc.

\medskip

How ``random" can an automatic sequence be? Of course we should exclude 
the periodic or ultimately periodic sequences. The remaining automatic
sequences:

- have zero entropy (whatever the definition of entropy is, see the 
contribution of V. Berth\'e [9] in this volume), hence are 
deterministic;

- have a low factor complexity (see Part III of this paper), more
precisely the number of blocks of length $n$ in an automatic sequence 
is $O(n)$, whereas for a ``random" binary sequence one has $2^n$ blocks;

- form a countable set, (hence there are very few of them, contrarily 
to what would be expected from ``random" sequences);

- form a set of measure zero, (of course almost all the sequences on a
finite alphabet are expected to be ``random");

- BUT they may have an absolutely continuous spectrum, such as the 
Rudin-Shapiro sequence for which the spectral measure is actually
equal to the Le\-besgue measure, just as would be expected for a random
sequence (see the contribution of M. Queff\'elec [30] in this volume).

\bigskip

\noindent{\bf 7. MISCELLANEA}

\bigskip

\noindent $\bullet$ Define a sequence of words on the alphabet $\{0,1\}$
by $w_0 = 0$, $w_{n+1} = w_n \overline{w_n}$, $\forall n \geq 0$, where
$\overline{w}$ is obtained from $w$ by replacing the $0$'s by $1$'s
and vice versa.

\noindent
Hence $w_1 = 01$, $w_2 = 0110$, $w_3 = 01101001$, $\cdots$

Prove that the sequence of words $(w_n)_n$ converges to the
Prouhet-Thue-Morse sequence, (all these words are prefixes of
our Prouhet-Thue-Morse sequence). For other such examples see
[4], and for general results see [32].

\medskip

\noindent $\bullet$ Define a sequence of words on the alphabet $\{0,1\}$
by $w_0 = 1$, $w_{n+1} = w_n 1 f(w_n)$, $\forall n \geq 0$, where
$f(w)$ is obtained from $w$ by reading $w$ backwards then replacing the 
$0$'s by $1$'s and vice versa. Prove that the sequence of words one
obtains converges to the paperfolding sequence. This construction is
known as ``perturbed symmetry", (see [10]).

\medskip

\noindent $\bullet$ A deep theorem of Cobham (see [14], see also [11])
asserts that the only sequences which are both $k$- and $\ell$-automatic
for two multiplicatively independent integers $k$ and $\ell$ (i.e.
such that the ratio $\frac{\log k}{\log \ell}$ is irrational) are the
ultimately periodic sequences. There is yet no general result for
the non-constant length substitutions. Try to guess such a result
(there is a conjecture by Hansel [22], and a partial solution has 
been given quite recently by F. Durand). 
Before guessing, consider the 
sequence $u=(u_n)_n$ which is the fixed point of the morphism
$$\aligned
0 &\longrightarrow 12 \\
1 &\longrightarrow 102 \\
2 &\longrightarrow 0.
\endaligned
$$
Now prove that this sequence can be obtained by taking the fixed point
beginning by $1$ of the following morphism (this result is due to Berstel [8]):
$$\aligned
0 &\longrightarrow 12 \\
1 &\longrightarrow 13 \\
2 &\longrightarrow 20 \\
3 &\longrightarrow 21
\endaligned
$$
and projecting it modulo $3$
$$
0 \longrightarrow 0, \ \
1 \longrightarrow 1, \ \
2 \longrightarrow 2, \ \
3 \longrightarrow 0.
$$

The same sequence
$$1 \ \ 0 \ \ 2 \ \ 1 \ \ 2 \ \ 0 \ \ 1 \ \ 0 \ \ 2 \ \ 0 \ \ \cdots$$
has a remarkable property which is worth mentioning: {\it it has no squares in 
it (i.e. no two consecutive identical blocks)}, see Part III of this paper. 
For surveys on similar topics see also [8] and [3].

\medskip

Here is yet another property of the sequence which links it to the 
Prouhet-Thue-Morse sequence. Add $1$ modulo $3$ to each term, obtaining:
$$2 \ \ 1 \ \ 0 \ \ 2 \ \ 0 \ \ 1 \ \ 2 \ \ 1 \ \ 0 \ \ 1 \ \ \cdots.$$
This is the sequence of lengths of the strings of $1$'s between two
consecutive $0$'s in the Morse sequence 
$$0 \ 1 \ 1 \ 0 \ 1 \ 0 \ 0 \ 1 \ 1 \ 0 \ 0 \ 1 \ 0 \ 1 \ 1 \ 0 \ 1 \ 0
\ 0 \ 1 \ 0 \ \cdots.$$

\medskip

\noindent $\bullet$ Let $\Cal A$ be defined as the
smallest (for the lexicographical order) subset of $\Bbb N$ which
contains $1$ and such that $n \in \Cal A \implies 2n \notin \Cal A$.
Hence the elements of $\Cal A$ in increasing order are:
$$1 \ \ 3 \ \ 4 \ \ 5 \ \ 7 \ \ 9 \ \ 11 \ \ 12 \ \ 13 \ \ 15 \ \ 16 \ \
17 \ \ 19 \ \ \cdots.$$
Now take the first differences to get:
$$2 \ \ \ 1 \ \  \ 1 \ \ \ 2 \ \ \ 2 \ \ \ 2 \ \ \ 1 \ \ \ 1 \ \ \ 
2 \ \ \ 1 \ \ \ 1 \ \ \ 2 \ \ \cdots.$$
The reader might now consider again his favourite Prouhet-Thue-Morse 
sequence:
$$0 \ 1 \ 1 \ 0 \ 1 \ 0 \ 0 \ 1 \ 1 \ 0 \ 0 \ 1 \ 0 \ 1 \ 1 \ 0 \ 1 \ 0
\ 0 \ \cdots$$
and write down the lengths of the blocks composed only with $0$'s or
only with $1$'s. He will obtain:
$$1 \ \ \ 2 \ \ \ 1 \ \ \ 1 \ \ \ 2 \ \ \ 2 \ \ \ 2 \ \ \ 1 \ \ \ 1
\ \ \ 2 \ \ \ 1 \ \ \ 1 \ \ \ 2 \ \ \ \cdots,$$
which is - up to the first term - the sequence of first differences above.
To finish with these two sequences of $1$'s and $2$'s, the patient reader
can prove that the first one is the fixed point of:
$$\aligned
2 &\longrightarrow 211 \\
1 &\longrightarrow 2
\endaligned
$$
whereas the second one is the fixed point of:
$$\aligned
1 &\longrightarrow 121 \\
2 &\longrightarrow 12221.
\endaligned
$$

To read more on these sequences see [2] and [35].
Note that it has been proved independently by Mkaouar and Shallit that
both sequences above are not $2$-automatic.

\medskip

\noindent $\bullet$ Another intriguing sequence is the self-running
Kolakoski sequence. Consider any infinite sequence of $1$'s and $2$'s,
say for example:
$$u : 2 \ 1 1 1 \ 2 \ 1 \ 2 2 \ 1 \ \cdots.$$
Let $K(u)$ be the sequence formed by the lengths of the strings of $1$'s 
and the strings of $2$'s. In our example:
$$K(u) : 1 \ 3 \ 1 \ 1 \ 2 \ \cdots.$$
The Kolakoski sequence is the only sequence beginning by $2$ which
satisfies the remarkable property $K(u) = u$.
$$
\underset{K(u):} \to {\underset{ } \to {u:}} \ \
\underset{2} \to {\underbrace{22}} \
\underset{2} \to {\underbrace{11}} \
\underset{1} \to {\underbrace{2}} \
\underset{1} \to {\underbrace{1}} \
\underset{2} \to {\underbrace{22}} \
1 \
\cdots.
$$
It can be shown that this sequence can be obtained by iterating the 
following map:
$$\aligned
11 &\longrightarrow 21 \\
12 & \longrightarrow 211 \\
22 &\longrightarrow 2211 \\
21 &\longrightarrow 221.
\endaligned
$$
Note that this map is not a morphism (there is some kind of 
``context-depen\-dence") and that it is not known whether this sequence
can be obtained as the projection of the fixed point of a morphism
(even with non-constant length). For more properties of this sequence
see [25], [19], [20], [38], [16], [17], [12] and [28].

\medskip

This last example may open the way to new constructions of sequences
which are still algorithmically (easy) computable but not ``too simple",
hence again between order and chaos. An interesting paper 
of Wen and Wen [39] discusses the composite iteration of two
different morphisms ruled by a third morphism,
(see also [16] and [17]).

\bigskip

\noindent{\bf 8. APPENDIX: AUTOMATA GENERATING THE FIVE EXAMPLES OF
SECTION 4}

\bigskip

{\bf 1)} A reverse automaton generating the Prouhet-Thue-Morse sequence:

\bigskip

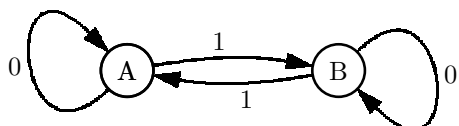
\Draw
\NewCIRCNode(\StateNode,103,)
\Define\StateAt(3){ \MoveTo(#2,#3) \StateNode(#1)(--#1--)}
\CycleEdgeSpec(45,50)
\DiagramSpec(\StateAt&\TransEdge)
\ArrowHeads(1) 
\SaveAll
\Diagram
(A,0,0 & B,80,0) (A,B,1,1 & B,B,0,0) 
\RecallAll
\CycleEdgeSpec(225,50)
\Diagram  
(A,0,0) (A,A,0,0)
\EndDraw

\smallskip

\centerline{Output function: $\varphi(A) = 0$, $\varphi(B) = 1$.}

\bigskip

Can you find what the direct automaton is?

\bigskip

{\bf 2)} A reverse automaton generating the paperfolding sequence:

\bigskip

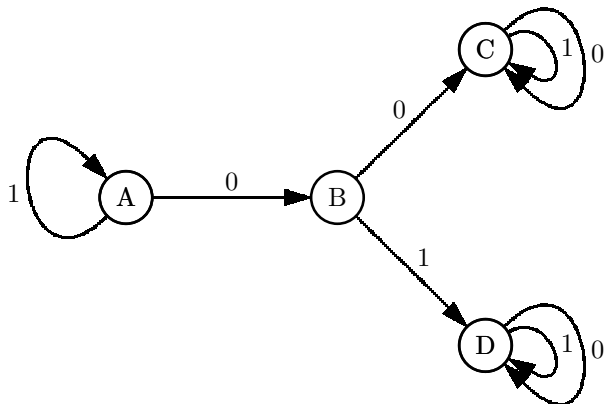
\Draw
\NewCIRCNode(\StateNode,103,)
\Define\StateAt(3){ \MoveTo(#2,#3) \StateNode(#1)(--#1--)}
\CycleEdgeSpec(45,50)
\DiagramSpec(\StateAt&\TransEdge)
\ArrowHeads(1) 
\SaveAll
\Diagram
(A,0,0 & B,80,0 & C,136,56 & D,136,-56) 
(A,B,0, & B,C,0, & B,D,1, & C,C,0,0 & D,D,0,0)
\RecallAll
\CycleEdgeSpec(225,50)
\SaveAll
\Diagram  
(A,0,0) (A,A,0,1)
\RecallAll
\CycleEdgeSpec(30,20)
\Diagram
(C,136,56 & D,136,-56) (C,C,0,1 & D,D,0,1)
\EndDraw

\smallskip

\centerline{Output function: $\varphi(A) = 1$, $\varphi(B) = 1$, 
$\varphi(C) = 1$, $\varphi(D) = 0$.}

\bigskip

The direct automaton is given in Part II.

\bigskip

{\bf 3)} A reverse automaton generating the Rudin-Shapiro sequence:

\bigskip

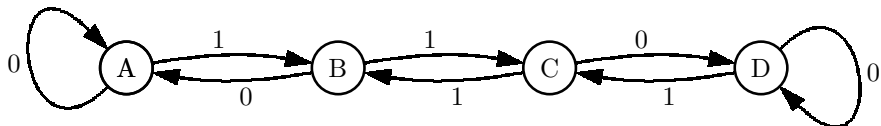
\Draw
\NewCIRCNode(\StateNode,103,)
\Define\StateAt(3){ \MoveTo(#2,#3) \StateNode(#1)(--#1--)}
\CycleEdgeSpec(45,50)
\DiagramSpec(\StateAt&\TransEdge)
\ArrowHeads(1) 
\SaveAll
\Diagram
(A,0,0 & B,80,0 & C,160,0 & D,240,0) (A,B,1,0 & B,C,1,1
& C,D,0,1 & D,D,0,0) 
\RecallAll
\CycleEdgeSpec(225,50)
\Diagram  
(A,0,0) (A,A,0,0)
\EndDraw

\smallskip

\centerline{Output function: $\varphi(A) = +1$, $\varphi(B) = +1$, 
$\varphi(C) = -1$.}

\bigskip

The value $\varphi(D)$ is irrelevant. What is the direct automaton?

\bigskip

{\bf 4)} A reverse automaton generating the period-doubling 
sequence:

\bigskip

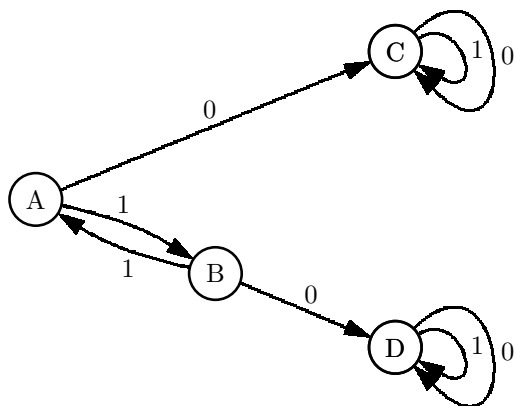
\Draw
\NewCIRCNode(\StateNode,103,)
\Define\StateAt(3){ \MoveTo(#2,#3) \StateNode(#1)(--#1--)}
\CycleEdgeSpec(45,50)
\DiagramSpec(\StateAt&\TransEdge)
\ArrowHeads(1) 
\SaveAll
\Diagram
(A,0,0 & B,68,-28 & C,136,56 & D,136,-56) 
(A,B,1,1 & A,C,0, & B,D,0, & C,C,0,0 & D,D,0,0) 
\RecallAll
\CycleEdgeSpec(30,20)
\Diagram  
(C,136,56 & D,136,-56) (C,C,0,1 & D,D,0,1)
\EndDraw

\smallskip

\centerline{Output function: $\varphi(A) = 0$, $\varphi(B) = 1$, 
$\varphi(C) = 0$, $\varphi(D) = 1$.}

\bigskip

{\bf 5)} A reverse automaton generating the Hanoi sequence:

\bigskip

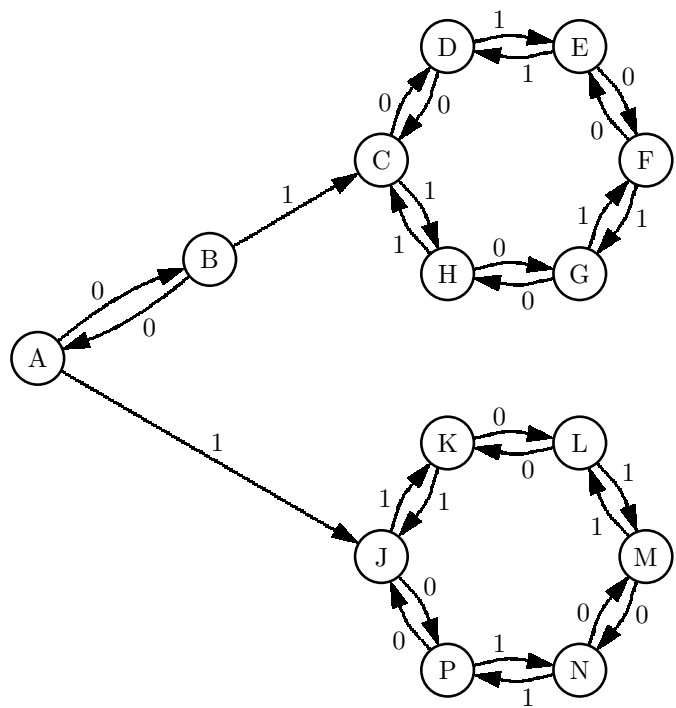
\Draw
\NewCIRCNode(\StateNode,103,)
\Define\StateAt(3){ \MoveTo(#2,#3) \StateNode(#1)(--#1--)}
\CycleEdgeSpec(45,50)
\DiagramSpec(\StateAt&\TransEdge)
\ArrowHeads(1) 
\Diagram
(A,0,0 & B,65,37.5 
& C,130,75 & D,155,118 & E,205,118 
& F,230,75 & G,205,32 & H,155,32
& J,130,-75 & K,155,-32 & L,205,-32
& M,230,-75 & N,205,-118 & P,155,-118) 
(A,B,0,0 & B,C,1, & C,D,0,0 & D,E,1,1 & E,F,0,0 & F,G,1,1 & G,H,0,0
& H,C,1,1
& A,J,1,
& J,K,1,1 & K,L,0,0 & L,M,1,1 & M,N,0,0 & N,P,1,1 & P,J,0,0) 
\EndDraw
$$\alignedat6
\ \ \ \ \text{Output function: } \ &\varphi(A) &&= \text{any value}, \ 
 &&\varphi(B) &&= \text{any value},  && &&\\
&\varphi(C) &&= \overline{c}, \ &&\varphi(D) &&=  \overline{c}, \ 
 &&\varphi(E) &&= \overline{a}, \\ 
&\varphi(F) &&=  \overline{a}, \ &&\varphi(G) &&= \overline{b}, \
 &&\varphi(H) &&= \overline{b}, \\
&\varphi(J) &&= a, \ &&\varphi(K) &&= b, \ &&\varphi(L) &&= b, \\
&\varphi(M) &&= c, \ &&\varphi(N) &&= c, \ &&\varphi(P) &&= a.
\endalignedat
$$
 
\bigskip
\bigskip

\references

\refis{1} 
Allouche J.-P., Note on the cyclic towers of Hanoi, \review
Theoret. Comput. Sci., 123, 1994, 3--7.

\refis{2}
Allouche J.-P., Arnold A., Berstel J., Brlek S., Jockusch W., Plouffe S.,
Sagan B. E.,  A relative of the Thue-Morse sequence,
\review Discrete Math.,  , 1994, to \ appear.

\refis{3}
Allouche J.-P., Astoorian D., Randall J., Shallit J.,
Morphisms, squarefree strings and the tower of Hanoi puzzle,
\review Amer. Math. Monthly, 101, 1994, 651--658.

\refis{4}
Allouche J.-P., B\'etr\'ema J., Shallit J., Sur des points
fixes de morphismes d'un mono\"{\i}de libre, \review {RAIRO, Informatique
Th\'eorique et Applications}, 23, 1989, 235--249.

\refis{5} 
Allouche J.-P., Cosnard M., It\'eration de fonctions unimodales
et suites engendr\'ees par automates, \review {C. R. Acad. Sci., Paris, 
S\'erie A}, 296, 1983, 159--162.

\refis{6}
Allouche J.-P., Dress F., Tours de Hano\"{\i} et automates,
\review {RAIRO, Informatique Th\'eorique et Applications}, 24, 1990, 
1--15.
 
\refis{7}
Allouche J.-P., Johnson T., Finite automata and morphisms
in assisted musical composition, \review J. New Music Research, 24, 1995,
to \ appear.

\refis{8}
Berstel J., Sur la construction de mots sans carr\'e,
\review {S\'eminaire de Th\'eorie des Nombres de Bordeaux, Expos\'e 18}, 
\! , 1978--1979, 18-01--18-15.

\refis{9}
Berth\'e V., {\it  this volume}.

\refis{10}
Blanchard A., Mend\`es France M., Sym\'etrie et transcendance,
\review Bull. Sci. Math., 106, 1982, 325--335.

\refis{11}
Bruy\`ere V., Hansel G., Michaux C., Villemaire R.,
Logic and $p$-recogni\-zable sets of integers, \review Bull. Belg. Math.
Soc., 1, 1994, 191--238.

\refis{12}
Carpi A., Repetitions in the Kolakoski sequence, \review Bull.
EATCS, 50, 1993, 194--196.

\refis{13}
Cobham A., On the base-dependence of sets of
numbers recognizable by finite automata, \review Math. Systems Theory,
3, 1969, 186--192.

\refis{14}
Cobham A., Uniform tag sequences, \review Math. Systems
Theory, 6, 1972, 164--192.

\refis{15}
Collet P., Eckmann J.-P., Iterated maps on the interval as 
dynamical systems (Progress in Physics, Birkh\"auser, 1980).
\refis{16} Culik II K., Karhum\"aki J., Lepis\"o A., Alternating 
iteration of morphisms and the Kolakoski sequence, in: Lindenmayer 
systems: impacts in theoretical computer science, computer graphics and 
developmental biology,  G. Rozenberg and A. Salomaa Eds 
(Springer Verlag, 1992) p. 93--103.

\refis{17}
Culik II K., Karhum\"aki J., Iterative devices generating
infinite words,  in: STACS 1992: 9th Annual Symposium on Theoretical 
Aspects of Computer Science, A. Finkel and M. Jantzen Eds (Lecture Notes in 
Computer Science, {\bf 577},  1992) p. 531--543.

\refis{18}
Davis C., Knuth D. E., Number representations and dragon curves, \review 
J. Recr. Math., 3, 1970, 161--181; \ 133--149.

\refis{19}
Dekking F. M., Regularity and irregularity of sequences
generated by automata, \review {S\'eminaire de Th\'eorie des Nombres de 
Bordeaux, Expos\'e 9}, \! , 1979--1980, 9-01--9-10.

\refis{20}
Dekking F. M., On the structure of self-generating sequences,
\review {S\'eminaire de Th\'eorie des Nombres de Bordeaux, Expos\'e 31}, 
\! , 1980--1981, 31-01--31-06.

\refis{21}
Dekking F. M., Mend\`es France M., van der Poorten A.,
FOLDS!, \review Math. Intell., 4, 1982, {130--138, \ \!
173--181 \ \!  and \ \! 190--195}. 

\refis{22}
Hansel G., private communication.

\refis{23}
Hinz A. M., The tower of Hanoi, \review Enseign. Math., 35,  1989,
289--321.

\refis{24}
Jonker L., Periodic orbits and kneading invariants,
\review Proc. London Math. Soc., 39, 1979, 428--450.

\refis{25}
Kolakoski W.,  Self-generating runs, Prob. 5304, sol. in
\review Amer. Math. Monthly, 73, 1966, 681--682.

\refis{26}
Meyer Y., {\it this volume}.

\refis{27}
Morse M., Recurrent geodesics on a surface of negative curvature,
\review Trans. Amer. Math. Soc., 2, 1921, 84--100.

\refis{28}
P\v{a}un G., How much Thue is Kolakoski?, \review Bull. EATCS, 49, 1993, 
183--185.

\refis{29}
Prouhet E., \review  {C. R. Acad. Sci., Paris}, 33, 1851, 225.

\refis{30}
Queff\'elec M., {\it this volume}.

\refis{31}
Rudin W., Some theorems on Fourier coefficients,
\review Proc. Amer. Math. Soc., 10, 1959, 855--859.

\refis{32}
Shallit J., A generalization of automatic sequences, \review
Theoret. Comput. Sci., 61, 1988, 1--16.

\refis{33}
Shapiro H. S., Extremal problems for polynomial and power series
(Thesis, M.I.T., 1951).

\refis{34}
Shechtman D., Blech I. A., Gratias D., Cahn J. W.,
Metallic phase with long-range orientational order and no 
translational symmetry, \review Phys. Rev. Letter, 53, 1984, 1951--1953.

\refis{35}
Tamura J., Some problems having their origin in the power series
\linebreak $\sum_{n=1}^{\infty}z^{[\alpha n]}$, in:  Reports of the 
meeting on analytic theory of numbers and related topics, (Gakushuin 
University, 1992), p. 190--212.

\refis{36}
Thue A., \"Uber unendliche Zeichenreihen, \review Norske
vid. Selsk. Skr. I. Mat. Kl. Christiana, 7, 1906, 1--22.

\refis{37}
Thue A., Lage gleicher Teile gewisse Zeichenreihen, \review
Norske vid. Selsk. Skr. I. Mat. Kl. Christiana, 1, 1912, 1--67.

\refis{38}
Weakley W. D., On the number of $C^{\infty}$-words of each length,
\review J. Combin. Theory Ser. A, 51, 1989, 55--62.

\refis{39}
Wen Z.-X., Wen Z.-Y.,  Remarques sur la suite engendr\'ee 
par des substitutions compos\'ees, \review Ann. Fac. Sci. Toulouse Math.,
9, 1988, 55--63.

\endreferences

\newpage
 
\centerline{\big PART II}
 
\bigskip
 
\centerline{\big Further properties of paperfolding}
 
\bigskip

In this part we shall discuss some unrelated properties of the paperfolding
sequence and its generalizations. We shall be concerned among other
things with dimensions and continued fractions.

\bigskip

\noindent
{\bf 1. DIRECT READING AND REVERSE READING}

\bigskip

Consider the two automata $\Cal A'$ and $\Cal A''$ where $A'$ (resp. $A''$)
is the initial state.

\bigskip

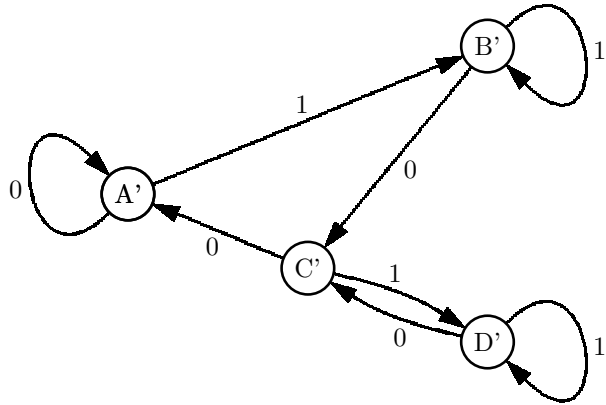
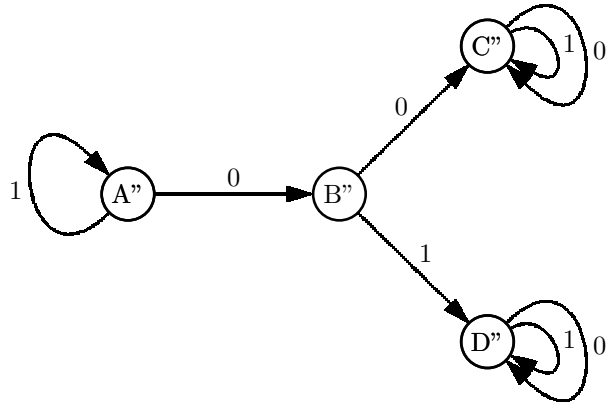
\Draw
\NewCIRCNode(\StateNode,103,)
\Define\StateAt(3){ \MoveTo(#2,#3) \StateNode(#1)(--#1--)}
\CycleEdgeSpec(45,50)
\DiagramSpec(\StateAt&\TransEdge)
\ArrowHeads(1) 
\SaveAll
\Diagram
(A',0,0 & C',68,-28 & B',136,56 & D',136,-56) 
(A',C', ,0 & A',B',1, & B',C',0, & C',D',1,0 
& B',B',0,1 & D',D',0,1) 
\RecallAll
\CycleEdgeSpec(225,50)
\Diagram  
(A',0,0) (A',A',0,0)
\EndDraw

\smallskip

\centerline{Output function: $\varphi'(A') = \varphi'(B') = a$, 
$\varphi'(C') = \varphi'(D') = b$.}

\bigskip

\Draw
\NewCIRCNode(\StateNode,103,)
\Define\StateAt(3){ \MoveTo(#2,#3) \StateNode(#1)(--#1--)}
\CycleEdgeSpec(45,50)
\DiagramSpec(\StateAt&\TransEdge)
\ArrowHeads(1) 
\SaveAll
\Diagram
(A",0,0 & B",80,0 & C",136,56 & D",136,-56) 
(A",B",0, & B",C",0, & B",D",1, & C",C",0,0 & D",D",0,0)
\RecallAll
\CycleEdgeSpec(225,50)
\SaveAll
\Diagram  
(A",0,0) (A",A",0,1)
\RecallAll
\CycleEdgeSpec(30,20)
\Diagram
(C",136,56 & D",136,-56) (C",C",0,1 & D",D",0,1)
\EndDraw

\smallskip

\centerline{Output function: $\varphi"(A") = \varphi"(B") = 
\varphi"(C") = a$, $\varphi"(D") = b$.}

\bigskip

The input instructions are to be read from left to right for the
automaton $\Cal A'$ (direct reading) whereas for the automaton $\Cal A''$ we
are to read from right to left (reverse reading). For example
``nineteen''=10011 puts $\Cal A'$ onto state $B'$ and $\Cal A''$ onto state
$C''$. In both cases the output is
$a=\varphi'(B')=\varphi''(C'')$. We leave it as an exercise to
verify that for all $n=0,1,2\cdots$
$$\varphi'(A'_n)=\varphi''(A''_n)$$
where $A'_n$ (resp. $A''_n$) is the state of the automaton $\Cal A'$ (resp.
$\Cal A''$) at step $n$.

\medskip

This illustrates a general result that we had already mentioned in 
Part I: {\it automatic sequences can be generated either by direct or
reverse reading automata}.

\bigskip

\noindent
{\bf 2. WORDS AND DIAGRAMS}

\bigskip

Let $\{L, R\}^*$ be the set of words on two symbols $L$ (left) and $R$
(right). To each word $w\in \{L, R\}^*$ corresponds a diagram on the
lattice $\Bbb Z^2$ in the following way. Starting from the origin follow
the horizontal unit interval. At the endpoint turn left or right
according to the instruction given by the first letter of the word $w$.
Having then followed the vertical segment (upward or downward) read out
the second letter of $w$ which instructs you to turn left or right on a
horizontal segment, etc... For example the diagram corresponding to
$w=L\,R\,L\,L$ is 

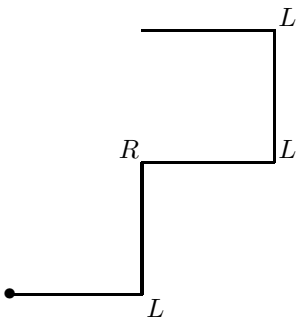
\bigskip

\Draw
\Text(--$\bullet$--)
\LineTo(50,0)
\LineTo(50,50)
\LineTo(100,50)
\LineTo(100,100)
\LineTo(50,100)

\MoveTo(55,-5) \Text(--$L$--)
\MoveTo(45,55) \Text(--$R$--)
\MoveTo(105,55) \Text(--$L$--)
\MoveTo(105,105) \Text(--$L$--)
\EndDraw

\bigskip

\noindent
The length of the word is $|w|=4$ and the length of the diagram is
$|w|+1=5$.

\smallskip

Similarly $L^4=LLLL$ represents

\bigskip

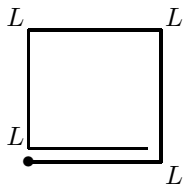
\Draw
\Text(--$\bullet$--)
\LineTo(50,0)
\LineTo(50,50)
\LineTo(0,50)
\LineTo(0,5)
\LineTo(45,5)

\MoveTo(55,-5) \Text(--$L$--)
\MoveTo(55,55) \Text(--$L$--)
\MoveTo(-5,55) \Text(--$L$--)
\MoveTo(-5,10) \Text(--$L$--)
\EndDraw

\bigskip

\noindent and $LRL^3RL$ represents 

\bigskip

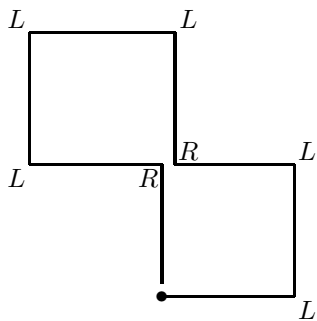
\Draw
\Text(--$\bullet$--)
\LineTo(50,0)
\LineTo(50,50)
\LineTo(100,50)
\LineTo(100,100)
\LineTo(50,100)
\LineTo(50,55)
\LineTo(0,55)
\LineTo(0,5)

\MoveTo(55,-5) \Text(--$L$--)
\MoveTo(55,45) \Text(--$R$--)
\MoveTo(105,45) \Text(--$L$--)
\MoveTo(105,105) \Text(--$L$--)
\MoveTo(45,105) \Text(--$L$--)
\MoveTo(45,60) \Text(--$R$--)
\MoveTo(-5,50) \Text(--$L$--)
\EndDraw

\bigskip

The first and the third examples show {\it self-avoiding} curves:
edges are described only once. The associated words are said to be
self-avoiding.
The third example illustrates a closed curve or {\it cyclic} curve:
the two ends coincide. We then speak of cyclic words.\medskip
Later we shall encounter infinite length diagrams which correspond to
infinite sequences of $R$'s and $L$'s.

\bigskip

\noindent
{\bf 3. THE FOLDING OPERATORS} [10]

\bigskip

Folding a word or its associated diagram is done in three steps.

\medskip

{\bf Step 1.} Fold the polygon $w$ onto the unit segment by
shrinking all angles to $0$. Hence $L^3$ becomes

\bigskip

\Draw(.8pt, .8pt)
\Text(--$\bullet$--)
\LineTo(50,0)
\LineTo(50,50)
\LineTo(0,50)
\LineTo(0,5)

\MoveTo(100,0)
\Text(--$\bullet$--)
\LineTo(150,0)
\LineTo(150,50)
\LineTo(100,50)
\MoveTo(100,47.5)
\DrawOvalArc(2.5,2.5)(90,270)
\MoveTo(100,45)
\LineTo(145,45)

\MoveTo(200,0)
\Text(--$\bullet$--)
\LineTo(250,0)
\LineTo(250,50)
\MoveTo(246,50)
\DrawOvalArc(4,4)(0,180)
\MoveTo(242,50)
\LineTo(242,4)
\MoveTo(244,4)
\DrawOvalArc(2,2)(180,0)
\MoveTo(246,4)
\LineTo(246,46)

\MoveTo(300,0)
\Text(--$\bullet$--)
\LineTo(350,0)
\MoveTo(350,6)
\DrawOvalArc(6,6)(-90,90)
\MoveTo(350,12)
\LineTo(304,12)
\MoveTo(304,8)
\DrawOvalArc(4,4)(90,270)
\MoveTo(304,4)
\LineTo(350,4)
\MoveTo(350,6)
\DrawOvalArc(2,2)(-90,90)
\MoveTo(350,8)
\LineTo(306,8)

\EndDraw

\bigskip

{\bf Step 2.} Multiply the length of the edges by 2 and fold the
collapsed diagram onto itself either in the positive direction (operator
$F_+$) or in the negative direction $(F_-)$

\bigskip

\Draw

\Text(--$\bullet$--)
\LineTo(60,0)
\MoveTo(50,6)
\Node(a)(--{}--)
\MoveTo(10,6)
\Node(b)(--{}--)
\CurvedEdgeSpec(-45,.3,45,.3)
\ArrowSpec(V)
\ArrowHeads(1)
\CurvedEdge(a,b)
\MoveTo(30,30)
\Text(--$F_+$--)

\MoveTo(90,0)
\Text(--$\bullet$--)
\LineTo(150,0)
\MoveTo(140,-6)
\Node(a)(--{}--)
\MoveTo(100,-6)
\Node(b)(--{}--)
\CurvedEdgeSpec(45,.3,-45,.3)
\ArrowSpec(V)
\ArrowHeads(1)
\CurvedEdge(a,b)
\MoveTo(120,-30)
\Text(--$F_-$--)

\EndDraw

\bigskip

{\bf Step 3.} Unfold to 90$^\circ$. We thus obtain a new diagram $w'$ twice
as long as the original one which we denote $F_+(w)$ or $F_-(w)$. Clearly
$|F_{\pm}(w)|=2|w|-1$.
\smallskip
As an example $F_+(L^3)$ is the diagram

\bigskip

\Draw
\Text(--$\bullet$--)
\LineTo(50,0)
\LineTo(50,50)
\LineTo(5,50)
\LineTo(5,100)
\LineTo(-50,100)
\LineTo(-50,50)
\LineTo(0,50)
\LineTo(0,5)

\MoveTo(55,-5) \Text(--$L$--)
\MoveTo(55,55) \Text(--$L$--)
\MoveTo(10,55) \Text(--$R$--)
\MoveTo(10,105) \Text(--$L$--)
\MoveTo(-55,105) \Text(--$L$--)
\MoveTo(-55,45) \Text(--$L$--)
\MoveTo(-5,45) \Text(--$R$--)
\EndDraw

\bigskip

\noindent so that
$$F_+(L^3)=L^2\,R\,L^3\,R.$$

The following result can easily be established.
\proclaim{Theorem 1} Let $w=A_0\,A_1\,\cdots\,A_{n-1}\in \{L,R\}^*$. Then
$$F_+(w)=L\,A_0\, R\,A_1\,L\,A_2\,R\cdots\,A_{n-1}\, X$$
where
$$
X=\left\{\alignedat3
&L\ &&\text{ if }\ &&n\equiv 0 \pmod 2 , \\
&R\ &&\text {if }\ &&n\equiv 1 \pmod 2 . 
\endalignedat
\right.
$$
Similarly
$$F_-(w)=R\,A_0\,L\,A_1\,R\, A_2 \,L\,\cdots\,A_{n-1}\,Y$$
where
$$Y=\left\{\alignedat3
&R\ &&\text{ if }\ &&n\equiv 0 \pmod 2 , \\
&L\ &&\text{ if }\ &&n\equiv 1 \pmod 2 . 
\endalignedat
\right.
$$
\endproclaim

The operation which is involved in this theorem is the so-called shuffle
product {\cyr X} (Russian letter ``Sha''): if
$w=A_0\,A_1\,\cdots\,A_{n-1}\,A_n$ and $w'=B_0\,B_1\cdots B_{n-1}$ (it may
happen that $A_n$ is the empty letter), then
$$w\,\text{\cyr X}\, w'=A_0\,B_0\,A_1\,B_1\, A_2\,B_2\,\cdots
A_{n-1}\,B_{n-1}\,A_n.$$
The theorem asserts
$$\aligned
&F_+(w)=(L\,R)^\times \text{\cyr X}\,w \\
&F_-(w)=(R\,L)^\times \text{\cyr X}\,w
\endaligned
$$
where it is to be understood that $(L\,R)^\times$ (resp. $(R\,L)^\times$) is the
alternating word $L\,R\,L\,R$... (resp. $R\,L\,R\,L$) of length $|w|+1$.
\medskip
Let $S$ be the set of self-avoiding words and let $C$ be the set of cyclic
words.
\proclaim{Theorem 2 }(Davis and Knuth).
$$F_{\pm}(S)\subset S,\ \ F_{\pm}(C)\subset C.$$
\endproclaim
We refer to [3] for the proof.
\medskip
Suppose we start out with the empty word $\emptyset$. Then, in accordance
with Theorem~1
$$F_+(\emptyset)=L,\ \ \ \ F_-(\emptyset)=R.$$
As $\emptyset$ is obviously self-avoiding (it is the unit interval!) so is
$F_{\pm}\ (L$ or $R)$ and by induction, given any sequence
$\varepsilon_1\,\varepsilon_2\cdots \varepsilon_n$ of $(\pm)$ signs
$$F_{\varepsilon_n\,\varepsilon_{n-1}\cdots\varepsilon_1}
(\emptyset)=F_{\varepsilon_n}\, F_{\varepsilon_{n-1}}\cdots F_{\varepsilon_1}
(\emptyset)$$
is self-avoiding.\medskip
To an infinite sequence
$\varepsilon=(\varepsilon_1\,\varepsilon_2\,\cdots)$ of signs correspond an
infinite diagram called a {\it dragon curve} $F_\varepsilon(\emptyset)$.
The above discussion shows that dragon curves are self-avoiding.\medskip
The figures below represent respectively $F_\varepsilon(\emptyset)$ where
$\varepsilon=(+ -)^\infty$ and $\varepsilon=(+)^\infty$. (There are two
dragon curves corresponding to $\varepsilon=(+ -)^\infty$; we only show one
of them.)

\bigskip
\bigskip
\bigskip

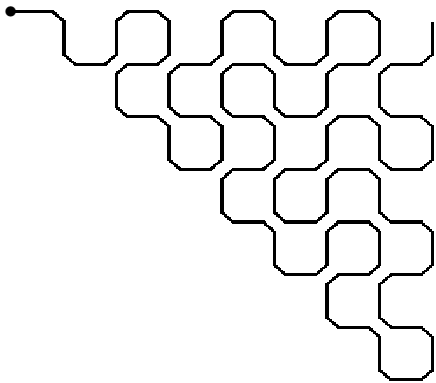
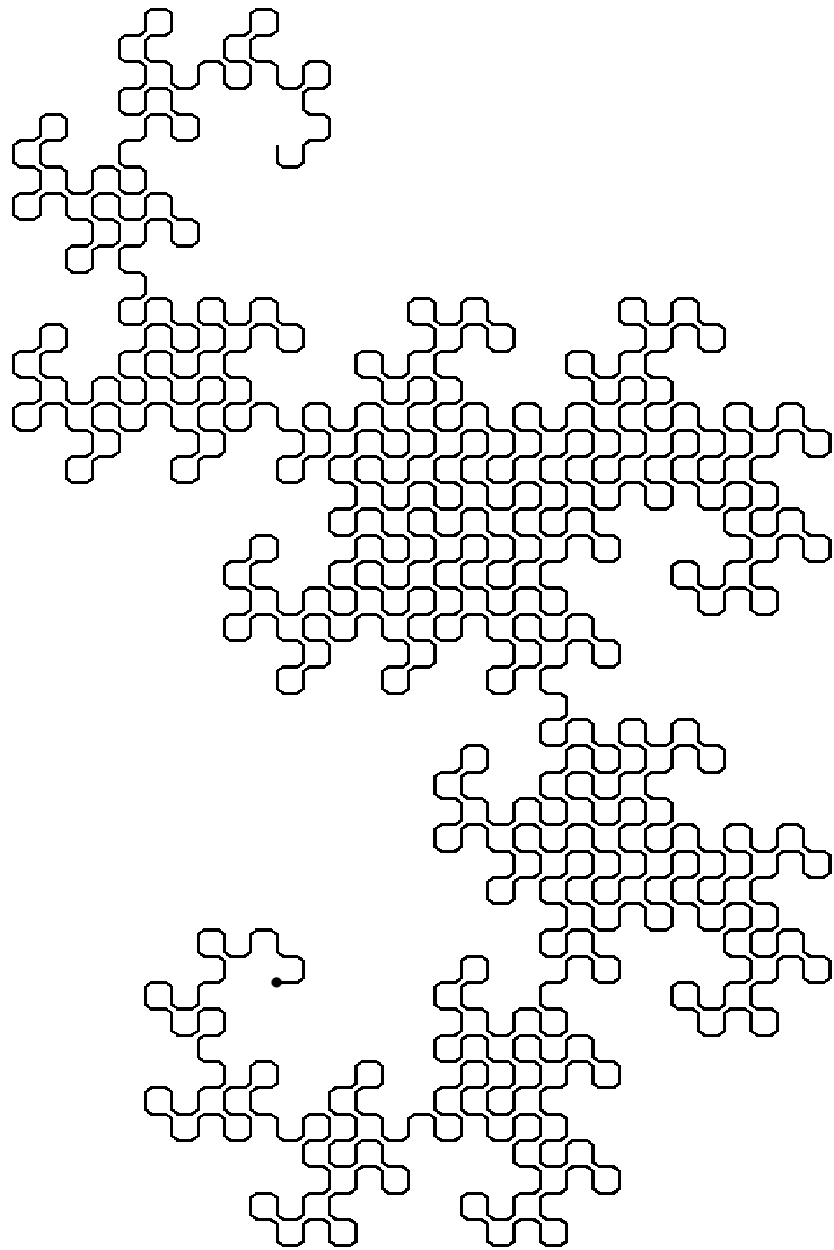
\Draw(2pt,2pt)
 
\Define\0{
\Rotate(45)
\LineF(2.8)
\Rotate(45)
\LineF(6)}

\Define\1{
\Rotate(-45)
\LineF(2.8)
\Rotate(-45)
\LineF(6)}
 
\Define\ra{\1 \0 \0}
\Define\sa{\1 \1 \0}
 
\Define\rb{\ra \1 \sa \0 \ra \0 \sa}
\Define\sb{\ra \1 \sa \1 \ra \0 \sa}
 
\Define\rc{\rb \1 \sb \0 \rb \0 \sb}
\Define\sc{\rb \1 \sb \1 \rb \0 \sb}
 
\Text(--$\bullet$--)
\LineF(8)
\rc
 
\EndDraw

\newpage

\Draw

\Define\0{
\Rotate(45)
\LineF(2.8)
\Rotate(45)
\LineF(6)}

\Define\1{
\Rotate(-45)
\LineF(2.8)
\Rotate(-45)
\LineF(6)}

\Define\pa{\0 \0 \1}
\Define\qa{\0 \1 \1}

\Define\pb{\pa \0 \qa}
\Define\qb{\pa \1 \qa}

\Define\pc{\pb \0 \qb}
\Define\qc{\pb \1 \qb}

\Define\pd{\pc \0 \qc}
\Define\qd{\pc \1 \qc}

\Define\pe{\pd \0 \qd}
\Define\qe{\pd \1 \qd}

\Define\pf{\pe \0 \qe}
\Define\qf{\pe \1 \qe}

\Define\pg{\pf \0 \qf}
\Define\qg{\pf \1 \qf}

\Define\ph{\pg \0 \qg}
\Define\qh{\pg \1 \qg}

\Define\pj{\ph \0 \qh}
\Define\qj{\ph \1 \qh}

\Text(--$\bullet$--)
\LineF(8)
\pj

\EndDraw

\newpage

The reader will no doubt have noticed that the infinite sequence
$F_{(+)^\infty}(\emptyset)$ is the paperfolding sequence discussed earlier in
Paragraph 1 and in Part I. If $\varepsilon$  is an arbitrary $(\pm)$
sequence, we shall say that $F_\varepsilon(\emptyset)$ is a generalized
paperfolding sequence. It is easily seen that if $A_0\,A_1\,A_2\cdots $ is
the original paperfolding sequence (the regular paperfolding) then
$$\left\{\aligned
&A_{2n}\ \text{ is the alternating sequence } (L\,R)^\infty, \\
&A_{2n+1}=A_n.
\endaligned
\right.
$$
More generally it $B_0\,B_1\,B_2\cdots$ is any paperfolding sequence then
$$\left\{\aligned
&B_{2n}\, \text{ is an alternating sequence}, \\
&B_{2n+1}\, \text{ is a paperfolding sequence}.
\endaligned
\right.
$$
This property characterizes the family of paperfolding
sequences.

\medskip

\noindent
{\bf Exercise.} The generalized paperfolding sequence
$F_\varepsilon(\emptyset)$ is automatic if and only if $\varepsilon$ is
ultimately periodic.\medskip By simple inspection one feels that dragon
curves are somehow space-filling (space $=\Bbb Z^2$), or that at least they
fill portions of $\Bbb Z^2$. This is indeed the case and we shall show in
the next paragraph that all dragon curves are 2-dimensional. First of all we
must define the concept of dimension.

\bigskip

\noindent
{\bf  4. THE DIMENSION OF A CURVE} [4], [11].

\bigskip

Let $\Gamma$ be a locally rectifiable unbounded curve in the plane. All
arcs of $\Gamma$ have finite length unlike fractal Weierstrass
curves, Peano curves or Brownian curves. (See Mandelbrot [8] for the
properties of the last mentioned curves.)

\medskip

Consider a portion $\Gamma_L$ of $\Gamma$ of length $L$. Let
$\varepsilon>0$ and let $\Gamma(L,\varepsilon)$ be the area of the
$\varepsilon$-sausage of $\Gamma_L$.

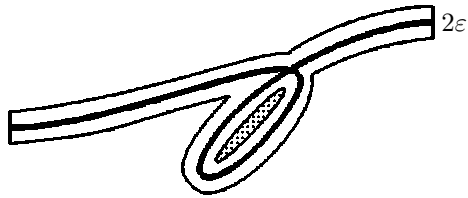
\bigskip

\Draw

\MoveTo(0,0)    \MarkLoc(a)  \MoveTo(50,0)   \MarkLoc(r)
\MoveTo(100,0)  \MarkLoc(b)  \MoveTo(140,50) \MarkLoc(s)
                             \MoveTo(80,-25) \MarkLoc(t)
\MoveTo(80,0)   \MarkLoc(c)  \MoveTo(60,-25) \MarkLoc(u)
                             \MoveTo(100,25) \MarkLoc(v) 
\MoveTo(160,40) \MarkLoc(d)  \MoveTo(140,40) \MarkLoc(w)

\PenSize(2pt)
\Curve(a,r,s,b)
\Curve(b,t,u,c)
\Curve(c,v,w,d)
\PenSize(.75pt)

\MoveTo(0,6)    \MarkLoc(aa)  \MoveTo(50,11)  \MarkLoc(rr)
\MoveTo(105,28) \MarkLoc(bb)  \MoveTo(95,30)  \MarkLoc(ss)
                              \MoveTo(115,36) \MarkLoc(tt)
\MoveTo(160,46) \MarkLoc(dd)  \MoveTo(140,46) \MarkLoc(uu)

\Curve(aa,rr,ss,bb)
\Curve(bb,tt,uu,dd)

\MoveTo(0,-6)   \MarkLoc(aaa) \MoveTo(50,-1)  \MarkLoc(rrr)
\MoveTo(83,12)  \MarkLoc(bbb) \MoveTo(30,-1)  \MarkLoc(sss)
                              \MoveTo(60,-12) \MarkLoc(ttt)
\MoveTo(70,-23) \MarkLoc(ccc) \MoveTo(64,-18) \MarkLoc(uuu)
                              \MoveTo(82,-33) \MarkLoc(vvv)
\MoveTo(114,20) \MarkLoc(ddd) \MoveTo(118,4)  \MarkLoc(www)
                              \MoveTo(125,26) \MarkLoc(xxx)
\MoveTo(160,34) \MarkLoc(eee) \MoveTo(140,32) \MarkLoc(yyy)

\Curve(aaa,rrr,sss,bbb)
\Curve(bbb,ttt,uuu,ccc)
\Curve(ccc,vvv,www,ddd)
\Curve(ddd,xxx,yyy,eee)

\LineAt(0,6,0,-6)
\LineAt(160,46,160,34)

\SetBrush(2,2){\PaintRect(.5,.5)}

\MoveTo(91,1)
\RotatedAxes(47,137)
\DrawOval(19,2.5)
\PaintOval(19,2.5)
\EndRotatedAxes

\MoveTo(168,40)
\Text(--$2\varepsilon$--)

\EndDraw

\bigskip

Denote by $D(L)$ the diameter of $\Gamma_L$:
$$D(L)=\max\{\text{dist} (M,N) \ | \ M\in\Gamma_L,\ N\in\Gamma_L\}.$$
By definition the dimension of $\Gamma$ is
$$\text{dim}\ (\Gamma)=\liminf\limits_{L \to \infty}
\dfrac{\log \Gamma(L, \varepsilon)}{\log D(L)}.$$
It can be shown that dim $(\Gamma)$ is independent of $\varepsilon>0$ and
that
$$1\le \text{dim}\ (\Gamma)\le 2.$$

A straight line, an infinite branch of an algebraic curve have dimension 1.
The spiral $\rho=\exp\,\theta$ is also one-dimensional. Let $\alpha>0$. The
spiral $\rho=\theta^\alpha$ has dimension $1+1/\alpha$ if $\alpha\ge1$ and
dimension 2 if $\alpha \leq 1$.
\proclaim{Theorem 3} (Mend\`es France and Tenenbaum).

All dragon curves are two-dimensional.
\endproclaim

Let us sketch the proof which seems quite relevant here since it involves
sequences related to the Rudin-Shapiro sequence. (For details see [12].)

\medskip

The $N^{\text{th}}$ vertex of a given dragon curve $\Gamma$ has coordinates
$$x_N=\sum\limits_{k<N/2} s_k,\ \ \ y_N=\sum\limits_{k<N/2} t_k,$$
where $s_k$ and $t_k$ are equal to $\pm 1$; $x_N$ is the algebraic sum of
the right and left moves whereas $y_N$ is the sum of ups and downs. Both
sequences $(s_n)$ and $(t_n)$ are closely linked to the Rudin-Shapiro
sequence. In particular for the dragon curve
$F_{(+-)^\infty}(\emptyset)=L\,R\,R\,L\,R\,R\cdots$ the sequence $(s_n)$ is
exactly the Rudin-Shapiro sequence. The sequence $(t_n)$ is closely related.
\smallskip
It can be shown that for all dragon curves the Rudin-Shapiro property
holds, namely
$$\aligned
&\left\vert\sum_{k<N/2} s_k\right\vert\le (2+\sqrt{2})\sqrt{N/2}, \\
& \\
&\left\vert\sum_{k<N/2} t_k\right\vert\le (2+\sqrt{2})\sqrt{N/2}. 
\endaligned
$$

Therefore the diameter $D(N)$ satisfies
$$D(N)\le \underset n<N\to\max (x^2_n + y^2_n)^{\frac{1}{2}}\le a\sqrt N$$
for some constant $a$.\smallskip
On the other hand, as the dragon curve is self-avoiding,
$$\Gamma(N,\varepsilon) \sim 2\varepsilon N$$
so that
$$\text{dim} (\Gamma)\ge\lim\limits_{N\to\infty}\ \dfrac{\log
(2\varepsilon N)}{\log a\sqrt N}=2.$$
But the dimension is at most 2 therefore
$$\text{dim}(\Gamma)=2,\qquad\qquad\qquad \text{QED.}$$

The entropy of a curve (see V. Berth\'e [1]) is linked to the dimension.
The complete relationship is discussed in Mend\`es France [11]. The higher
the entropy, the higher the dimension, so in some sense the dimension
measures the complexity of a curve (this is one of Mandelbrot's
principles). In that respect dragon curves are among the most complex
self-avoiding curves on $\Bbb Z^2$.

\bigskip

\noindent
{\bf 5. PAPERFOLDING AND CONTINUED FRACTIONS}

\bigskip

The continued fraction expansion of a real number $x$ reads
$$x=a_0+\dfrac{1}{a_1+\dfrac{1}{a_2+\ddots}}=[a_0, a_1, a_2, \cdots]$$
where the so-called partial quotients $a_n\ge1$ are integers; $a_0$ may be
negative. If $x$ is irrational the expansion is infinite whereas if $x$ is
rational the expansion terminates. Due to the trivial equality
$$a=a-1+\frac{1}{1},\ a\ge2,$$
rational numbers have two expansions which only differ by the last two
partial quotients. The rank $n$ of the last partial quotient $a_n$ is
called the depth of $x$. The above remark allows us to assume with no loss
of generality that the depth is always even.
\medskip
In this paragraph we shall discuss the continued fraction expansion of the
number
$$x=\sum\limits_{n\ge0} g^{-2^n}$$
where $g\ge2$ is a given integer (the case $g=2$ is slightly more
complicated so we shall suppose $g\ge3$). The following analysis is
essentially due to Kmo\v sek [6] and Shallit [13]. See also the papers of
Mend\`es France [9], Blanchard, Mend\`es France [2] or 
Dekking, Mend\`es France, van der Poorten [5].

\medskip

Let $p/q$ be a rational number in the interval $(0,1)$. We assume $p$ and
$q$ coprime. It can be shown that if
$$
\frac{p}{q}=[0, a_1, a_2, \cdots, a_n], \  (n\text{
even}),$$
then
$$\frac{p}{q}+\frac{1}{q^2}=[0, a_1, a_2, \cdots, a_n+1, a_n-1, a_{n-1},
\cdots, a_1].$$
If $\overrightarrow{w}=a_1\,a_2\cdots a_n$ is the word associated to $p/q$
then the word corresponding to $p/q+1/q^2$ is
$$S_{\overrightarrow{p}}(\overrightarrow{w})=
\overrightarrow{w}\,\overrightarrow{p}\,\overleftarrow{w}$$
where $\overleftarrow{w}$ is the word $\overrightarrow{w}$ reversed and
where $\overrightarrow{p}$ denotes the ``operator'' which perturbs the
last letter of $\overrightarrow{w}$ and the first letter of 
$\overleftarrow{w}$. $\overleftarrow{p}$ is the inverse operator: it adds
$-1$ (resp. $+1$) to the last letter of the first word (resp. first letter
of the last word).\medskip
Consider
$$\aligned
x_2&=\frac{1}{g}+\frac{1}{g^2}= [0, g-1, g+1], \\
x_3&=\frac{1}{g}+\frac{1}{g^2}+\frac{1}{g^{2^2}}=
S_{\overrightarrow{p}} (g-1, g+1) \\
&=[0, g-1, g+2, g, g-1].
\endaligned
$$
Then
$$\aligned
x_4&= S_{\overrightarrow{p}}\ [g-1, g+2, g, g-1] \\
&=[0, g-1, g+2, g, g, g-2, g, g+2, g-1]
\endaligned
$$
and
$$\aligned
x=x_\infty &= S^\infty_{\overrightarrow{p}} (g-1, g+1) \\
&=[0, g-1, g+2, g, g, g-2, g, g+2, g, g-2, g+2, g, \cdots].
\endaligned
$$
The partial quotients of $x$ are completely determined by the algorithm.
Their values are $g-2,\,g-1,\,g,\,g+2$. Mahler [7] proved that $x$ is
transcendental and this is quite an achievement since many of the easier
proofs in transcendental number theory are based on the speed with which
the partial quotients tend to infinity (think of Liouville numbers).\medskip
Let us analyse the structure of the sequence of the partial quotients of $x$.
Put $\overrightarrow{w}=g-1,\,g+1$. Then $x=S^\infty_{\overrightarrow{p}}\
(\overrightarrow{w})$
$$x=\overrightarrow{w}\,
\overrightarrow{p}\,\overleftarrow{w}\,\overrightarrow{p}
\,\overrightarrow{w}\,\overleftarrow{p}\,\overleftarrow{w}
\,\overrightarrow{p}\,\overrightarrow{w}\,\overrightarrow{p}
\,\overleftarrow{w}\,\overleftarrow{p}\,\overrightarrow{w}\,\cdots$$
The sequence of $w$'s and $p$'s is obviously periodic but more surprisingly
(?) the sequence of arrows is the regular paper-folding sequence. As it is
automatic one can correctly guess that the sequence of partial quotients is
indeed automatic. We show below the direct reading automaton which
generates the sequence of partial quotients.

\bigskip

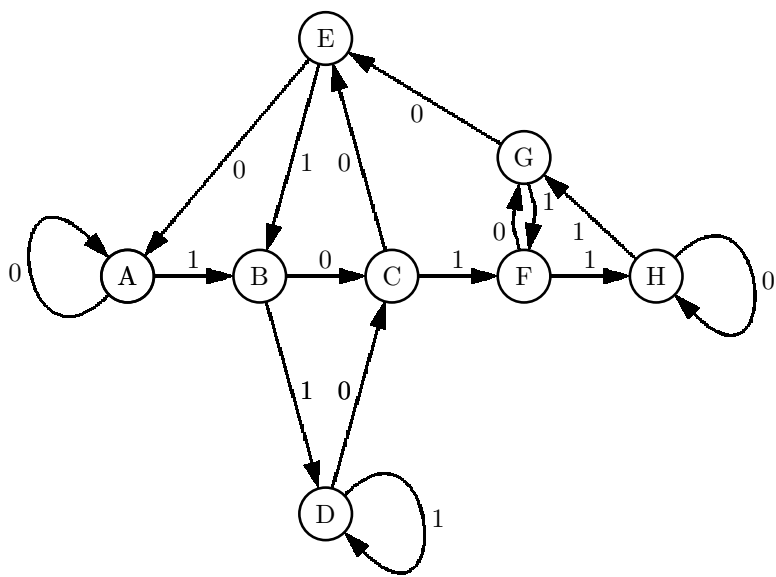
\Draw
\NewCIRCNode(\StateNode,103,)
\Define\StateAt(3){ \MoveTo(#2,#3) \StateNode(#1)(--#1--)}
\CycleEdgeSpec(45,50)
\DiagramSpec(\StateAt&\TransEdge)
\ArrowHeads(1) 
\SaveAll
\Diagram
(A,0,0 & B,50,0 & C,100,0 & D,75,-90 & E,75,90 & F,150,0 
& G,150,45 & H,200,0) (A,B,1, & B,C,0, & C,F,1, & F,H,1, 
& H,G,1, & G,E,0, & E,A,0, & B,D,1, & D,C,0, & B,D,1, & D,C,0,
& E,B,1, & C,E,0, & F,G,0,1 & D,D,0,1 & H,H,0,0) 
\RecallAll
\CycleEdgeSpec(225,50)
\Diagram  
(A,0,0) (A,A,0,0)

\EndDraw

\smallskip

$$\alignedat4
\text{Output function: } \ 
&\varphi(A) &&= g+2, \ \ \ \ &&\varphi(E) &&= g,   \\
&\varphi(B) &&= g,   \ \ \ \ &&\varphi(F) &&= g+2, \\
&\varphi(C) &&= g,   \ \ \ \ &&\varphi(G) &&= g-2, \\
&\varphi(D) &&= g-2, \ \ \ \ &&\varphi(H) &&= g.
\endalignedat
$$

\bigskip
\bigskip

\references

\refis{1}
Berth\'e V., {\it this volume}.

\refis{2}
Blanchard A., Mend\`es France M., Sym\'etrie et transcendance,
\review Bull. Sci. Math., 106, 1982, 325--335.

\refis{3}
Davis C., Knuth D. E., Number representations and dragon curves, \review
J. Recr. Math., 3, 1970, 161--181; \ 133--149.

\refis{4} 
Dekking F. M., Mend\`es France M., Uniform distribution modulo one: 
a geometrical viewpoint, \review J. reine angew. Math., 129, 1981, 143--153.

\refis{5}
Dekking F. M., Mend\`es France M., van der Poorten A.,
FOLDS!, \review Math. Intell., 4, 1982, {130--138, \ \!
173--181 \ \!  and \ \! 190--195}.

\refis{6}
Kmo\v{s}ek M., Report by A. Schinzel, Oberwolfach 28 may - 2 june 1979.

\refis{7}
Mahler K., Ein Analog zu einem Schneiderschen Satz, \review
Proc. Akad. \linebreak Wetensch. Amsterdam, 39, 1936, {633--640 \ \! and \ \! 
729--737}.

\refis{8}
Mandelbrot B., Les objets fractals (Nouvelle Biblioth\`eque Scient. 
Flammarion, Paris, 1975).

\refis{9}
Mend\`es France M., Sur les fractions continues limit\'ees, \review
Acta Arith., 2, 1973, 207--215.

\refis{10}
Mend\`es France M., Paper folding, space-filling curves and
Rudin-Shapiro sequences, \review Contempory Mathematics, 9, 1982, 85--95.

\refis{11}
Mend\`es France M., The Planck constant of a curve, in: Fractal
Geometry and Analysis J. Belair and S. Dubuc Eds (Kluwer Acad. Publ., 1991)
p. 325--366

\refis{12}
Mend\`es France M., Tenenbaum G., Dimension des courbes planes, 
papiers pli\'es et suites de Rudin-Shapiro, \review Bull. Soc. math.
France, 109, 1981, 207--215.

\refis{13}
Shallit J., Simple continued fractions for some irrational numbers 
\review J. Number Theory, 11, 1979, 209--217.

\endreferences

\newpage

\centerline{\big PART III}

\bigskip

\centerline{\big Complements}

\bigskip

In this chapter we will give some complements on finite automata:
a quick survey of the notion of ``repetitions" in a finite or infinite word,
the multidimensional generalization of morphisms of constant length and of
finite automata and finally a short account on differences and 
similarities between finite automata and cellular automata.

\bigskip

\noindent {\bf 1. REPETITIONS IN INFINITE SEQUENCES} 

\bigskip

\noindent
{\bf 1.1. The beginning of the story}

\medskip

The pioneering work of Thue (see [18] and [19]) stems from the following
easy observation:

\proclaim{Observation} Every infinite word on a two-letter alphabet contains
at least a (non-empty) square, (i.e. a block $ww$ where $w$ is a non empty
word).
\endproclaim 

\noindent {\it Proof}

\medskip

As the  proof is very easy we will give it: let $\{a,b\}$ be the two-letter
alphabet. Let us try to construct a ``long" word with no square in it.
With no loss of generality we can suppose that the infinite word
begins with an $a$. The following letter cannot be an $a$ (since $aa$ is a square),
hence our word begins with $ab$. The next letter cannot be a $b$ (since the word
would contain the block $bb$), hence the word begins with $aba$. Now if we add
a letter to $aba$ the new word would necessarily contain a square, as it would 
be either $abaa$ or $abab$.

\medskip

Note that the above reasoning shows that every word of length $\geq 4$ on a 
two-letter alphabet contains a square.

\medskip

A natural question hence arises: is it possible to find an infinite sequence 
on a three-letter alphabet without squares in it? One can also ask whether
there exists an infinite sequence on a two-letter alphabet without cubes,
(i.e. without blocks $www$ where $w$ is a non-empty word).
The answer to both questions is YES, see [18], [19] and the survey [8] with
its bibliography:

\proclaim{Theorem} 

\smallskip

- The fixed point of the morphism
$$\aligned
0 &\longrightarrow 12 \\
1 &\longrightarrow 102 \\
2 &\longrightarrow 0
\endaligned
$$
has no square. This sequence is $2$-automatic as mentioned in Part I.

\smallskip

- The Thue-Morse sequence has no cube. More precisely this sequence has no 
overlap, i.e. no word $wwx$, where $x$ is the first letter of the word $w$. 
This sequence is linked to the preceding one as we have seen in Part I.

\endproclaim

\newpage

\noindent {\bf Remark}

Note that the Thue-Morse sequence contains arbitrarily long squares: indeed
it is a fixed point of the morphism $\sigma$ defined on $\{0, 1\}$ by:
$$\sigma(0)= 01, \ \ \sigma(1) = 10.$$
This sequence begins by $0110 \cdots$ hence by $\sigma^k(0) \sigma^k(1) 
\sigma^k(1) \sigma^k(0)$ for every $k \geq 2$. Hence it contains the square 
$\sigma^k(1)\sigma^k(1)$.

\bigskip

\noindent
{\bf 1.2. Why study repetitions?}

\medskip
 
When he asked the above questions concerning repetitions in sequences, Thue
added that he had no particular applications in mind but that the problem
seemed interesting and not trivial; there would certainly be applications 
in the future.

And indeed this question has been studied very precisely by computer scientists,
see for example [11], see also [8]. The notion of morphism was again an 
essential tool: our leitmotiv is that {\it sequences generated by morphism are 
both easy to construct and not necessarily trivial}.

The existence of repetitions is somehow linked to randomness: 
a random sequence is expected to be {\it normal}, (i.e. all possible blocks
of given length occur with the same frequency), hence a sequence without squares 
(or cubes or $\cdots$) cannot be random.

\medskip

We finally mention a paper of Bovier and Ghez [9] on the discrete Schr\"odinger 
equation with automatic potential where, curiously enough, one needs the 
automatic sequence of potentials to begin with a square. For more information on 
``automatic Schr\"odinger equations" one can read the contribution of 
S\"ut\H{o} [17] in this volume, one can also read [6], [7].

\bigskip

\noindent
{\bf 1.3. More examples}

\medskip

We quote here two results which are interesting in that they lead to other
developments that we will not describe here.

\smallskip

- {\it The paperfolding sequence has no square $ww$ with $|w| \geq 5$}. 
This result is announced in [1] and more details can be found in [2].
This implies that this sequence does not contain any $12^{\text th}$ power,
i.e. that no block has the form $w^{12}$, but a more precise result is proved
in [2]: {\it the paperfolding sequence has no fourth power}.
The result holds for {\it any} paperfolding sequence. For the
Rudin-Shapiro sequence(s) some results of the same kind are also given in [2].

\smallskip

- {\it The Fibonacci sequence has no fourth power}. It does however contain cubes 
since it begins by:
$$ 0 1 0 0 1 0 1 0 0 (1 0 0 1 0) (1 0 0 1 0) (1 0 0 1 0) \cdots$$
Considering this sequence one might also define fractional powers: for
example $10010010$ is a power $2 + \frac{2}{3} = \frac{8}{3}$, as it can be 
written $(100)(100)(10)$. Considering the supremum of the fractional powers which 
occur in an infinite word permits us to define a real maximal (positive) power 
occurring in a sequence. For the Fibonacci sequence one can prove that this 
maximal power is equal to $2+\tau = 3.618 \cdots$, see [13].

\bigskip

\noindent 
{\bf 2. MULTIDIMENSIONAL MORPHISMS AND (FINITE) AUTO\-MA\-TA}

\bigskip

Going back to the Penrose tiling, remember that the underlying morphism
is the Fibonacci morphism:
$$0 \longrightarrow 01, \ \ 1 \longrightarrow 0.$$
This morphism is a one-dimensional morphism and it might seem odd to
describe a two-dimensional pattern by means of a one-dimensional tool.
Among the possible generalizations of morphisms we will describe the 
two-dimensional substitutions and the associated two-dimensional finite 
automata. Although these morphisms are not suitable to describe the Penrose 
tiling, they are a natural generalization of the uniform morphisms and have
many interesting properties.

\bigskip

\noindent
{\bf 2.1. An example}

\medskip

Consider the alphabet $\{0, 1\}$ and the following map:
$$0 \ \ \  \longrightarrow 
\matrix &0 &0 &0 \\
        &0 &0 &0 \\
        &0 &0 &0 
\endmatrix \ \ \ \ \ \ \ \ \ 
1 \ \ \ \longrightarrow 
\matrix &1 &0 &1 \\
        &0 &1 &0 \\ 
        &1 &0 &1  
\endmatrix.   
$$
Starting from the letter $1$ and iterating this map by replacing each letter by a
$3 \times 3$ square gives:
$$1 \ \ \ \longrightarrow   
\matrix &1 &0 &1 \\  
        &0 &1 &0 \\ 
        &1 &0 &1   
\endmatrix \ \ \ 
\longrightarrow
\matrix &{\matrix &1 &0 &1 \\  
                  &0 &1 &0 \\ 
                  &1 &0 &1   
          \endmatrix}
        &{\matrix &0 &0 &0 \\
                  &0 &0 &0 \\ 
                  &0 &0 &0 
          \endmatrix} 
        &{\matrix &1 &0 &1 \\   
                  &0 &1 &0 \\  
                  &1 &0 &1    
          \endmatrix} \\
        & \ & \ & \ \\
        & \ & \ & \ \\
        &{\matrix &0 &0 &0 \\ 
                  &0 &0 &0 \\  
                  &0 &0 &0  
          \endmatrix} 
        &{\matrix &1 &0 &1 \\   
                  &0 &1 &0 \\  
                  &1 &0 &1    
          \endmatrix} 
        &{\matrix &0 &0 &0 \\ 
                  &0 &0 &0 \\  
                  &0 &0 &0  
          \endmatrix} \\
        & \ & \ & \ \\ 
        & \ & \ & \ \\
        &{\matrix &1 &0 &1 \\   
                  &0 &1 &0 \\  
                  &1 &0 &1    
          \endmatrix} 
        &{\matrix &0 &0 &0 \\ 
                  &0 &0 &0 \\  
                  &0 &0 &0  
          \endmatrix} 
        &{\matrix &1 &0 &1 \\    
                  &0 &1 &0 \\   
                  &1 &0 &1     
          \endmatrix} 
\endmatrix \ \ \ \longrightarrow \ \ \ \cdots 
$$  
An infinite number of iterations leads to the double infinite sequence:
$$\matrix         &1 &0 &1 &0 &0 &0 &1 &0 &1 &\cdots \\
                  &0 &1 &0 &0 &0 &0 &0 &1 &0 &\cdots \\
                  &1 &0 &1 &0 &0 &0 &1 &0 &1 &\cdots \\
                  &0 &0 &0 &1 &0 &1 &0 &0 &0 &\cdots \\
                  &0 &0 &0 &0 &1 &0 &0 &0 &0 &\cdots \\
                  &0 &0 &0 &1 &0 &1 &0 &0 &0 &\cdots \\
                  &1 &0 &1 &0 &0 &0 &1 &0 &1 &\cdots \\
                  &0 &1 &0 &0 &0 &0 &0 &1 &0 &\cdots \\
                  &1 &0 &1 &0 &0 &0 &1 &0 &1 &\cdots \\
                  &\vdots  &\vdots   &\vdots   &\vdots   &\vdots   &\vdots 
                  &\vdots   &\vdots   &\vdots 
\endmatrix  
$$ 
which is by construction a fixed point of this morphism.

\bigskip

\noindent
{\bf 2.2. Properties}

\medskip

This construction, which can be extended to any dimension, gives
rise to multi-indexed sequences with many arithmetic properties (see [14] and 
[15]). Note in particular that all lines, columns and diagonals of the above 
sequence are $3$-automatic. Quite obviously the same construction can be done 
with any integer $d \geq 2$ instead of $3$.

\medskip

These sequences can be used in Physics: an application to the
Robinson tiling is given in [5]. One could also imagine a two-dimensional
discrete Schr\"odinger equation with potentials generated this way, thus
permitting to study the eigen-frequencies of, say, a two-dimensional 
quasicrystalline mattress of springs or of a quasicrystalline {\it trempolino}.
To our knowledge this has not yet been studied.

\medskip

Another application to Physics is the construction of classical {\it fractal}
figures: first note that the {\it constant renormalization} we alluded to in
Part I, paragraph 2, can be interpreted as {\it self-similarity} for 
one-dimensional automatic sequences; in the same way double automatic sequences
are self-similar. Take the above example, replace the $1$'s by
black squares and the $0$'s by white squares; after each iteration renormalize
the pattern of black and white squares. This gives the well-known Sierpinski
triangle. Compare with the classical construction which can be seen as ``dual". 
Many such fractals can be found in the Appendix by Shallit to the paper [14], see 
also [16].

A last example is given by ... the handkerchief-folding, [14].

\bigskip  
            
\noindent {\bf 3. LINKS WITH CELLULAR AUTOMATA}

\bigskip

Cellular automata are often used in Physics. They have been 
introduced by von Neumann and they are discrete tools to mimick phenomena
which can be continuous. From the point of view of their computing power they
are equivalent to Turing machines, hence they are strictly more powerful in 
general than finite automata. Many books on cellular automata have been
published. See for example [10]. For an example of a two-dimensional cellular 
automaton as a model of a chemical reaction one can read [4], where
the Greenberg and Hastings model of the Belouzov-Zhabotinski-Zaikin 
oscillating reaction is discussed in detail.

\medskip

Let us give a quick definition of cellular automata. A cellular automaton 
consists of:

\smallskip

- a lattice, $\Bbb Z^d$. Each vertex is called a site or a cell;

\smallskip

- a (usually finite) neighbourhood of sites $V_i$ for each site $i \in \Bbb Z^d$,
such that $V_i = V_j + i - j$, (translation invariance). The most frequently used
neighbourhoods in two dimensions are the von Neumann and the Moore neighbourhood,
(see below); 

\smallskip

- a set of states $Q$, (usually finite);

\smallskip

- a map $f$, called the transition function, from $Q^v$ to $Q$, where $v$ is the
cardinality of each $V_i$.

\bigskip

\Draw

\MoveTo(0,0) \Text(--$i$--)

\LineAt(-35,0,-5,0)    \LineAt(5,0,35,0)
\LineAt(-35,30,35,30)
\LineAt(-35,-30,35,-30)

\LineAt(-30,35,-30,-35)
\LineAt(0,35,0,5)     \LineAt(0,-5,0,-35)
\LineAt(30,35,30,-35)

\MoveTo(30,0)  \Text(--$\bullet$--)
\MoveTo(-30,0) \Text(--$\bullet$--)
\MoveTo(0,30)  \Text(--$\bullet$--)
\MoveTo(0,-30) \Text(--$\bullet$--)

\MoveTo(110,0) \Text(--$i$--)

\LineAt(75,0,105,0)    \LineAt(115,0,145,0)
\LineAt(75,30,145,30)
\LineAt(75,-30,145,-30)

\LineAt(80,35,80,-35)
\LineAt(110,35,110,5)     \LineAt(110,-5,110,-35)
\LineAt(140,35,140,-35)
 
\MoveTo(140,0)   \Text(--$\bullet$--)
\MoveTo(80,0)    \Text(--$\bullet$--)
\MoveTo(110,30)   \Text(--$\bullet$--)
\MoveTo(110,-30)  \Text(--$\bullet$--)

\MoveTo(80,30)   \Text(--$\bullet$--)
\MoveTo(80,-30)  \Text(--$\bullet$--)
\MoveTo(140,30)  \Text(--$\bullet$--)
\MoveTo(140,-30) \Text(--$\bullet$--)

\MoveTo(0,-55) 
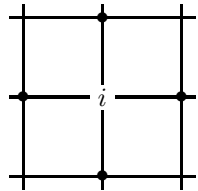
\Text(--von Neumann's 
neighbourhood of
$i$ in dimension $2$--)

\MoveTo(110,-55)
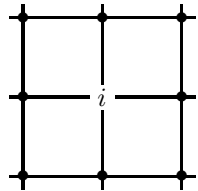
\Text(--Moore's
neighbourhood of
$i$ in dimension $2$--)

\EndDraw

\bigskip
 
The evolution of the cellular automaton is the following:
one assigns to each site $i$ an element $x_i^0$ in $Q$. The set 
$$\Cal C_0 = \{x_i^0, \ i \in \Bbb Z^d\}$$ 
is called the initial configuration. 
For each $i$ one then defines $x_i^1$ as the image by $f$ of 
$(x_j^0)_{j \in V_i}$. Hence the set 
$$\Cal C_1 = \{x_i^1, \ i \in \Bbb Z^d\}$$ 
is constructed by parallel updating. 
One writes $\Cal C_1 = f(\Cal C_0)$. Similarly $\Cal C_2 = f(\Cal C_1)$, $\cdots$, 
$\Cal C_{n+1} = f(\Cal C_n)$. The index of iteration  $n$ is sometimes called the
(discrete) time-scale.

\medskip

For many results on the dynamical behaviour of cellular automata we refer to [12]
and its bibliography. 

\medskip

In the case where the lattice is one-dimensional, one usually draws a 
two-dimensional pattern each line of which is a configuration $\Cal C_i$.
A simple class in one dimension is the class of linear cellular automata:
the values of the sites are taken in $\Bbb Z /d \Bbb Z$
the ring of integers modulo $d$ and the transition rules are linear.

For example the cellular automaton given by the Pascal's triangle rule
modulo $d$ is linear. The two-dimensional pattern it generates is exactly
the double sequence $u = ((u_{m,n})_{m,n})$ where $u_{m,n} = {m \choose n}
\mod d$, hence the Sierpinski triangle modulo $d$ (up to rotation).
It has been proved in [3] that {\it the sequence $u$ is a double 
automatic sequence if and only if $d$ is equal to a prime power $p^k$. The
sequence $u$ is then $p$-automatic}.

\bigskip
\bigskip

\references

\refis{1} 
Allouche J.-P., Suites infinies \`a r\'ep\'etitions born\'ees, 
\review {S\'eminaire de Th\'eorie des Nombres de Bordeaux, Expos\'e 20},
\!, 1983--1984, 20-01--20-11. 

\refis{2} 
Allouche J.-P., Bousquet-M\'elou M., Facteurs des suites de Rudin-Shapiro 
g\'en\'eralis\'ees, \review Bull. Belg. Math. Soc., 1, 1994, 145--164.

\refis{3} 
Allouche J.-P., von Haeseler F., Peitgen H.-O., Skordev G., Linear 
cellular automata, finite automata and Pascal's triangle, \review 
Discrete Appl. Math., , 1995, to \ appear.

\refis{4}
Allouche J.-P., Reder C., Oscillations spatio-temporelles engendr\'ees 
par un automate cellulaire, \review Discrete Appl. Math., 8, 1984, 215--254. 
 
\refis{5} 
Allouche J.-P., Salon O., Robinson tilings and $2$-dimensional automata, in:
Quasicrystals, Networks and Molecules of Fivefold Symmetry, I. Hargittai 
Ed (VCH Publishers Inc., 1990) p. 97--105. 

\refis{6}
Axel F., Allouche J.-P., Kl\'eman M., Mend\`es France M., Peyri\`ere J.,
Vibrational modes in a one dimensional ``quasi-alloy", the Morse case,  
\review {J. Phys. Colloques, 1986, Workshop on aperiodic crystals, Les 
Houches}, , 1986, 11--20.

\refis{7}
Axel F., Peyri\`ere J.,  Spectrum and extended states in a harmonic
chain with controlled disorder: effects of the Thue-Morse symmetry, 
\review J. Stat. Phys., 57, 1989, 1013--1047.

\refis{8}
Berstel J., Sur la construction de mots sans carr\'e, \review
{S\'eminaire de Th\'eorie des Nombres de Bordeaux, Expos\'e 18}, \!, 
1978--1979, 18-01--18-15.

\refis{9}
Bovier A., Ghez J.-M., Spectral properties of one-dimensional Schr\"o\-dinger
operators with potentials generated by substitutions, 
\review Comm. \linebreak Math. Phys., 158, 1993, 45--66.

\refis{10}
Burks A. W., Essays on cellular automata (University of Illinois
Press, 1970).

\refis{11}
Cassaigne J., Motifs \'evitables et r\'egularit\'es dans les mots
(Th\`ese, Universit\'e Paris 6/7, 1994).

\refis{12}
Goles E., Mart\'{\i}nez S., Neural and automata networks
(Mathematics and its applications, {\bf 58}, 1991).

\refis{13}
Mignosi F., Pirillo G., Repetitions in the Fibonacci infinite word, 
\linebreak \review {RAIRO, Informatique Th\'eorique et Applications}, 26, 1992, 
199--204.

\refis{14}
Salon O., Suites automatiques \`a multi-indices, {\it S\'eminaire 
de Th\'eorie des Nombres de Bordeaux, Expos\'e 4} (1986--1987)
4-01--4-27; followed by an appendix by Shallit J., 4-29A--4-36A.

\refis{15}
Salon O., Suites automatiques \`a multi-indices et alg\'ebricit\'e,
\review {C. R. Acad. Sci. Paris, S\'erie I}, 305, 1987, 501--504.

\refis{16}
Shallit J., Stolfi J., Two methods for generating fractals,
\review Comp. and Graphics, 13, 1989, 185--191.

\refis{17}
S\"ut\H{o} A., {\it this volume}.

\refis{18}
Thue A., \"Uber unendliche Zeichenreihen, \review Norske vid. Selsk. 
Skr. I. Mat. Kl. Christiana, 7, 1906, 1--22.

\refis{19}
Thue A., Lage gleicher Teile gewisse Zeichenreihen, \review Norske vid. 
Selsk. Skr. I. Mat. Kl. Christiana, 1, 1912, 1--67.

\endreferences

\newpage

\centerline{\big PART IV}

\bigskip

\centerline{\big Fourier analysis}

\bigskip

We present a rather naive point of view on the Fourier analysis of complex
valued sequences. For a deeper understanding the reader is referred to
Martine Queff\'elec's contribution to this volume [8].

\bigskip

\noindent {\bf 1. FOURIER-BOHR COEFFICIENTS}

\bigskip

Let $f=(f(n))$ be an infinite sequence of complex numbers such that for all
$\lambda\in\Bbb R$ the limit
$$\widehat f(\lambda)=\lim\limits_{N\rightarrow\infty}\ \dfrac{1}{N}\
\sum\limits_{n=0}^{N-1} f(n) \exp (-2i\pi\lambda n)$$
exists. This hypothesis is rather strong and eliminates such sequences as
$$f(n)=\exp (2i\pi \log n), \ (n\ge1),$$
for which the limit does not exist for $\lambda=0$.\medskip
Observe that whenever the hypothesis is fulfilled $\widehat
f(\lambda+1)=\widehat f(\lambda)$ so that it is enough to restrict $\lambda$ to
the interval $[0,1[$.

\medskip

\indent From now on we shall always assume the existence of 
$\widehat f(\lambda)$ for all $\lambda\in [0,1[$.
Under this assumption it can be shown that the {\it spectrum} of $f$
(the Fourier-Bohr spectrum)
$$S(f)=\{\lambda\ |\ \widehat f(\lambda)\ne0\}$$
is {\it at most countable}.
We use the notation
$$f(n)\sim \sum\limits_{\lambda\in S(f)}\ \widehat f(\lambda) \exp
2i\pi\lambda n.$$

\smallskip

\noindent
{\bf Warning}: The series may well diverge. We do not specify in which
order the terms are to be written down. Even if the the series converges
(whatever we mean by convergence!) its sum may be different from
$f$.

\medskip

We now present some examples which can be considered as exercises.

\medskip

\noindent
{\bf Example 1.} $f(n)=\exp(2i\pi\log n)$.

\medskip

\noindent As we already mentioned, $\widehat f(0)$ does not exist.

\medskip

\noindent
{\bf Example 2.} Let $\alpha\in\Bbb R$ and define
$f(n)=(-1)^{[\alpha n]}$ where $[\cdots]$ denotes the integer part. One
can show that the spectrum is
$$S(f)=\left\{\dfrac{\alpha}{2}+k\,\alpha\ |\ k\in\Bbb Z\right\} \pmod 1.$$
Observe that when $\alpha=p/q$, $(p,q)=1$, is rational, the spectrum boils
down to the finite set
$$\left\{\dfrac{\alpha}{2}+\dfrac{k}{q}\ |\ k=0,1, \cdots,
q-1\right\} \pmod 1.$$
When $\alpha$ is irrational
$$(-1)^{[\alpha n]}\sim \frac{2}{i\pi} \sum\limits_{k\in\Bbb Z}
\frac{1}{2k+1} \exp 2i\pi \left(k+\frac{1}{2}\right)\alpha n.$$
Exercise: what is the corresponding formula when $\alpha$ is rational? 

\medskip

\noindent
{\bf Example 3.} Let $f(n)=(-1)^{s(n)}$ be the $\pm$ Thue-Morse sequence ($s(n)$
is the sum of the binary digits of $n$). Then $\widehat f(\lambda)=0$ for
all $\lambda$ hence
$$f(n) \sim 0.$$

\medskip

\noindent
This result may be seen as a consequence of our Appendix 1. However a direct
computation is possible.

\medskip 

\noindent
{\bf Example 4.} The Rudin-Shapiro $(\pm)$ sequence has empty spectrum: 
$$f(n)\sim 0.$$

\medskip

\noindent
{\bf Example 5.} Let $\alpha$ be irrational. Consider $f(n)=(-1)^{[\alpha
n^2]}$. Then as in the preceding cases, the spectrum is empty and therefore
$$f(n) \sim 0.$$

\medskip

\noindent
This example is worked out in the paper of J. Bass [1].

\medskip

\noindent
{\bf Example 6.} Let $f(n)$ be the regular paperfolding sequence with
values $\pm 1$. Then
$$\widehat f(\lambda)=\left\{
\aligned
&\dfrac{i(-1)^{a-1}}{2^{\nu+1}}\qquad\ \text{ if }\ \ \ \ \ \ 
\lambda=\dfrac{2a+1}{2^{\nu+2}},\ \nu\ge0,\ 0\le a<2^{\nu+1} \\
& \\
&0\qquad\qquad\qquad \text{ if not }
\endaligned
\right.
$$
and
$$f(n-1)\sim\ \sum\limits_{\nu=0}^\infty \ \sum\limits_{a=0}^{2^{\nu+1}-1}
\dfrac{i(-1)^{a-1}}{2^{\nu+1}} \exp 2i\pi \dfrac{2a+1}{2^{\nu+2}} n.$$

\bigskip

Examples 3 and 6 will be worked out explicity in the appendices (Paragraphs 5
and 6).

\newpage\par\noindent
{\bf 2. BESSEL'S INEQUALITY AND PARSEVAL'S EQUALITY}

\bigskip

Define the norm of a sequence $f=(f(n))$ by
$$\Vert f\Vert = \limsup\limits_{N \to \infty}
\left(\dfrac{1}{N} \sum\limits_{n=0}^{N-1} |f(n)|^2\right)^{1/2}.$$
This is really a semi-norm since it may well happen that $\Vert f\Vert=0$
and yet\par\noindent $f\ne(0,0,0,\cdots)$.\smallskip
Here are two such examples:
$$f=\left(1, \dfrac{1}{2}, \dfrac{1}{3}, \dfrac{1}{4}, \dfrac{1}{5},
\cdots, \dfrac{1}{n}, \cdots\right)$$
and
$$f=(1,\ \overset 2^0\to{\oversetbrace\to 0},\ 1,\ \overset
2^1\to{\oversetbrace\to{0,0}}\ ,\ 1,\ \overset
2^2\to{\oversetbrace\to{0,0,0,0}}\ ,\ 1,\ \overset
2^3\to{\oversetbrace\to{0,\cdots,0}}\ ,\ 1,\ \overset
2^4\to{\oversetbrace\to{0,\cdots}}\ )$$
In this last example, the density of $1$'s  vanishes.
\medskip
From now on we identify two sequences $f$ and $g$ as soon as $\Vert
f-g\Vert=0$, so that the two above examples are considered as null
sequences.\medskip
Bessel's inequality asserts that in all cases
$$\sum\limits_{\lambda\in S(f)}\ |\widehat f(\lambda)|^2 \le \Vert
f\Vert^2.$$

A series of sequences $u_k(\,.\,)$ defined on $\Bbb N$ is said to converge
to a sequence $s(\,.\,)$ with respect to the norm $\Vert\cdots\Vert$ if
$$\lim\limits_{K\rightarrow\infty} \left\Vert
s(\,.\,)-\sum\limits_{k=0}^K\ u_k(\,.\,)\right\Vert =0.$$
In other words
$$\lim\limits_{K\rightarrow\infty}
\limsup\limits_{N \to \infty} \left(\dfrac{1}{N}
\sum\limits_{n=0}^{N-1} \left| s(n)-\sum\limits_{k=0}^{K}
u_k(n)\right|^2\right)^{\frac{1}{2}}=0.$$
We then write
$$s(n)\cong \sum\limits_{k=0}^\infty u_k(n).$$
This must not be confused with a pointwise convergence.

\proclaim{Theorem 1} (Parseval [9])
$$\sum\limits_{\lambda\in S(f)} \widehat f(\lambda) \exp 2i\pi\lambda n
\cong f(n)$$
if and only if
$$\sum\limits_{\lambda\in S(f)} |\widehat f(\lambda)|^2= \Vert f\Vert^2.$$
\endproclaim

\noindent
In this case we say that $f$ is an almost-periodic sequence in the sense
of Besicovitch; $f$ is represented by a sum of exponentials.
\medskip
We leave it to the reader to verify that $(-1)^{[\alpha n]}$ and that the
paperfolding sequence are Besicovitch almost-periodic. Obviously the
Thue-Morse sequence, the Rudin-Shapiro sequence and $(-1)^{[\alpha n^2]}$
($\alpha$ irrational) are not since Parseval's equality does not hold.

\bigskip

\noindent
{\bf 3. THE WIENER SPECTRUM AND THE SPECTRAL MEASURE}

\bigskip

The correlation function $\gamma^f$ which we assume to exist for all
$h\in\Bbb Z$ is defined by
$$\gamma^f(h)=\lim\limits_{N\rightarrow\infty} \dfrac{1}{N}
\sum\limits_{n=0}^{N-1} \overline{f}(n) f(n+h).$$
A deep result of Bochner's and Herglotz's asserts the existence of a bounded 
increasing function $\sigma^f$ such that
$$\gamma^f(h)=\int^1_0 \exp 2i\pi h x \  d\sigma^f(x).$$
$d\sigma^f$ is known as the spectral measure of $f$.

\medskip

Let $E$ be a measurable subset of the unit interval. Its $\sigma^f$
measure is denoted
$$\sigma^f\{E\}=\int_E d\sigma^f(x).$$
In particular, let $\lambda\in]0,1[$. The measure of the set $\{\lambda\}$
is
$$\sigma^f\{\lambda\}=\lim\limits_{\varepsilon\searrow 0}
\int^{\lambda+\varepsilon}_{\lambda-\varepsilon} d\sigma^f(x)=
\lim\limits_{\varepsilon\searrow 0}
(\sigma^f(\lambda+\varepsilon)-\sigma^f(\lambda-\varepsilon)).$$
Similarly
$$\sigma^f\{1\}= \lim\limits_{\varepsilon\searrow0}
\int^1_{1-\varepsilon}$$
and
$$\sigma^f\{0\}= \lim\limits_{\varepsilon\searrow0} \int^\varepsilon_0.$$

\newpage

One should not confuse the measure $\sigma^f\{\lambda\}$ of the set
$\{\lambda\}$ with the value $\sigma^f(\lambda)$ of the function
$\sigma^f$ at $\lambda$. There is of course a trivial relationship between
both
$$\sigma^f\{\lambda\}=\sigma^f(\lambda+0)-\sigma^f(\lambda-0);$$
$\sigma^f\{\lambda\}$ is the length of the discontinuity at $\lambda$. In
particular if $\sigma^f$ is continuous at $\lambda$ then
$\sigma^f\{\lambda\}=0$ and conversely.

\medskip

The set of $\lambda$'s such that $\sigma^f\{\lambda\}\ne0$ is called the
{\it Wiener spectrum} and we denote it $W(f)$. It is obviously at most
countable, (indeed the number of steps is at most countable).

\medskip

Lebesgue's decomposition theorem asserts that $\sigma^f$ is the sum of
three increasing functions
$$\sigma^f=\sigma^f_{st}+\sigma^f_{ac}+ \sigma^f_{sg}$$
or
$$d\sigma^f=d\sigma^f_{st}+d\sigma^f_{ac}+d\sigma^f_{sg},$$
where $\sigma^f_{st}$ is a step function, $\sigma^f_{ac}$ is absolutely 
continuous (it is the primitive of its derivative) and $\sigma^f_{sg}$ is
continuous  singular (its derivative vanishes almost everywhere with
respect to the Lebesgue measure).
\proclaim{Theorem 2} (Bertrandias [2])
$$|\widehat f(\lambda)|\le \sqrt{\sigma^f\{\lambda\}}.$$
\endproclaim

This important result shows that the Fourier-Bohr spectrum of $f$ is a
subset of the Wiener spectrum: $S(f)\subseteq W(f)$. In particular if
$\sigma^f$ is continuous, then $\widehat f(\lambda)=0$ for all
$\lambda$, but the converse is not true.

\medskip

Here are some examples:
$$\alignedat3
&f(n)=\exp(2i\pi\alpha n^2), \ \alpha \text{ irrational} &&: \ \ 
&&d\sigma^f(x)=dx, 
\text{ the Lebesgue measure,} \\
& \  &&\ &&\text{(see [1])}; \\
&f(n) \text{ is the Thue-Morse sequence} &&: \ \ &&d\sigma^f \text{ is
purely singular,} \\
& \ && \ &&\text{(see Appendix 1)}; \\
&f(n) \text{ is the Rudin-Shapiro sequence} &&: \ \ &&d\sigma^f(x)=dx, 
\text{ the Lebesgue measure};\\
&f(n)=\exp 2i\pi\alpha n &&: \ \
&&d\sigma^f(x)=\delta_\alpha, \text{ the Dirac mass at } \alpha;\\
&f(n)=\exp 2i\pi\sqrt n &&: \ \ &&d\sigma^f(x)=\delta_0.
\endalignedat
$$
If $f$ is a Besicovitch almost-periodic function
$$f(n)\cong\sum\limits_{\lambda\in S(f)} \widehat f(\lambda)
\exp 2i\pi\lambda n,$$
then
$$d\sigma^f(x)=\sum\limits_{\lambda\in S(f)} |\widehat f(\lambda)|^2
\delta_\lambda.$$

By definition a mean-periodic sequence $f$ is a sequence whose spectral
measure is a sum of Dirac masses. A Besicovitch almost-periodic function
is therefore mean-periodic. The converse is false as the example
$f(n)=\exp 2\pi i\sqrt n$ shows.

\medskip

In 1958 J. Bass [1] defined the family of pseudo-random functions and
sequences. His definition, modified by J.-P. Bertrandias [2] is that
$f(n)$ is pseudo-random if $\sigma^f$ is continuous. This is known to be
equivalent to the condition
$$\lim\limits_{N\rightarrow\infty} \dfrac{1}{N} \sum\limits_{k\le N}
|\gamma^f(h)|^2=0,$$
i.e. $\gamma^f(h)$ tends in average to 0 as $h$ increases to
infinity.

\medskip

We conclude this paragraph by a diagram which illustrates the relationship
between the different classes we discussed.

\bigskip

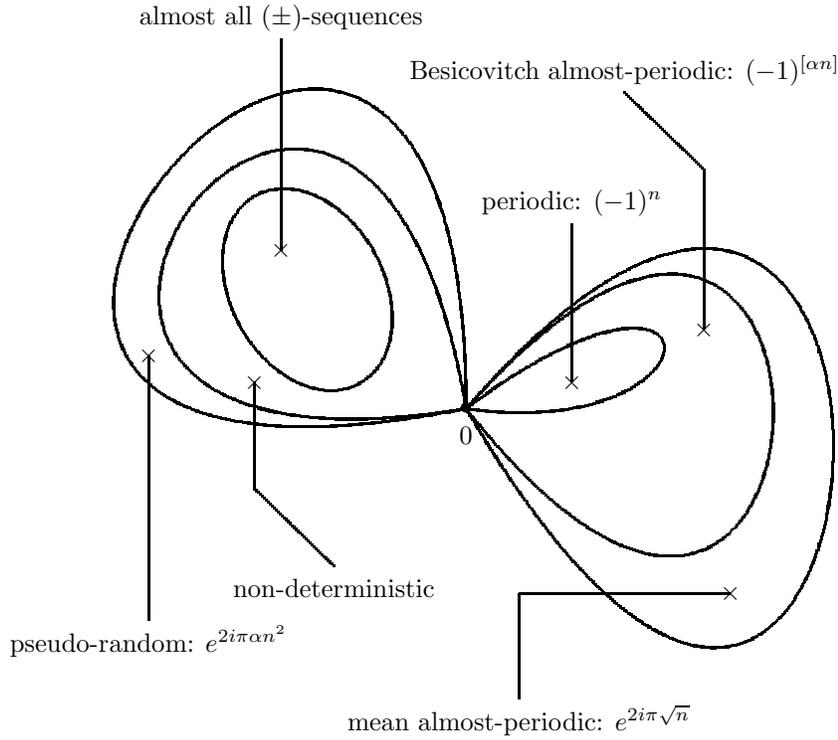
\Draw

\MoveTo(0,0)     \MarkLoc(a) \Text(--$\bullet$--)
\MoveTo(0,-10)               \Text(--$0$--)
\MoveTo(-40,240) \MarkLoc(r) \MoveTo(-240,-40) \MarkLoc(s)
\MoveTo(0,300)   \MarkLoc(t) \MoveTo(-300,-60) \MarkLoc(u)

\Curve(a,r,s,a)
\Curve(a,t,u,a)

\MoveTo(-60,45)
\RotatedAxes(-50,50)
\DrawOval(40,30)
\EndRotatedAxes

\MoveTo(100,75)  \MarkLoc(v)
\MoveTo(100,-14) \MarkLoc(w)
\Curve(a,v,w,a)

\MoveTo(160,180)  \MarkLoc(x)
\MoveTo(150,-190) \MarkLoc(y)
\Curve(a,x,y,a)

\MoveTo(200,230)  \MarkLoc(z)
\MoveTo(170,-290) \MarkLoc(q)
\Curve(a,z,q,a)

\MoveTo(-70,60) \Text(--$\times$--)
\LineAt(-70,60,-70,140)
\MoveTo(-70,148)
\Text(--almost all ($\pm$)-sequences--)

\MoveTo(-80,10) \Text(--$\times$--)
\LineAt(-80,10,-80,-30)
\LineAt(-80,-30,-50,-60)
\MoveTo(-50,-68) \Text(--non-deterministic--)

\MoveTo(-120,20) \Text(--$\times$--)
\LineAt(-120,20,-120,-80)
\MoveTo(-120,-88) 
\Text(--pseudo-random: $e^{2i\pi \alpha n^2}$--)

\MoveTo(40,10) \Text(--$\times$--)
\LineAt(40,10,40,70)
\MoveTo(40,78) \Text(--periodic: $(-1)^n$--)

\MoveTo(90,30) \Text(--$\times$--)
\LineAt(90,30,90,90)
\LineAt(90,90,60,120)
\MoveTo(60,128) 
\Text(--Besicovitch almost-periodic: $(-1)^{[\alpha n]}$--)

\MoveTo(100,-70) \Text(--$\times$--)
\LineAt(100,-70,20,-70)
\LineAt(20,-70,20,-110) 
\MoveTo(20,-118) 
\Text(--mean almost-periodic: $e^{2i\pi \sqrt n}$--)

\EndDraw

\bigskip
\bigskip

\newpage

\noindent
{\bf 4. ANALYZING THE SPECTRAL MEASURE}

\bigskip

We are given the sequence $f=(f(n))$ and we assume that the Fourier-Bohr
coefficients and the correlation exist.

\medskip

{\bf Case 1.} Parseval's equality holds. Then the spectral measure
$d\sigma^f$ is discrete (i.e. a sum of Dirac masses or Bragg peaks,
$\sigma^f$ is a step function):
$$d\sigma^f=\sum\limits_{\lambda\in S(f)} |\widehat f(\lambda)|^2
\delta_\lambda.$$
$f$ is Besicovitch almost-periodic.

\medskip

{\bf Case 2.} Suppose $S(f)=\emptyset$. A theorem of Wiener's asserts that
$$\lim\limits_{N\rightarrow\infty} \dfrac{1}{N} \sum\limits_{n<N}
|\gamma^f(n)|^2= \sum\limits_{\lambda\in W(f)} (\sigma^f\{\lambda\})^2,$$
where $\sigma^f\{\lambda\}$ is the mass at $\lambda$. Hence we have a first
important result which we already mentioned in Paragraph 3:

\medskip

\noindent
{\it If
$$\lim\limits_{N\rightarrow\infty} \dfrac{1}{N}\ \sum\limits_{n<N}
|\gamma^f(n)|^2=0,$$
then $d\sigma^f$ is continuous, and conversely.}

\medskip

The question arises to decide whether $d\sigma^f$ is absolutely continuous or
purely singular (or sum of both). A well known result of Riemann and Lebesgue
shows that if $d\sigma^f$ is absolutely continuous then
$$\gamma^f(n)=\int^1_0 e^{2i\pi n x} d\sigma^f(x)$$
tends to zero as $n$ increases to infinity. Therefore if
$$\limsup\limits_{n \to \infty} |\gamma^f(n)|>0,$$
there is a nonzero singular component to $d\sigma^f$.

\medskip 

We shall illustrate these ideas in the following appendices.

\bigskip

\noindent
{\bf 5. APPENDIX I: THE SPECTRAL MEASURE OF THE THUE-MORSE SEQUENCE}

\bigskip

Let $f=(f(n))$ be the $\pm$ Thue-Morse sequence. We show that its spectral 
measure is singular continuous. 

We first notice that $f(2n)=f(n)$
and $f(2n+1)=-f(n)$, $(f(0)=1)$. We shall compute the correlation function
$$\gamma(h)=\lim\limits_{N\rightarrow\infty} \dfrac{1}{N} \sum\limits_{n\in
N}\ \overline f(n)\ f(n+h).$$
One of the difficulties is to prove that $\gamma(h)$ actually exists. Obviously
$\gamma(0)=1$.

\medskip

Define
$$
\aligned
S(N, h)
&=\sum\limits_{n<N} \overline f(n)\ f(n+h) \\
& \ \\
&=\left\{
\aligned &\sum\limits_{m<\frac{N}{2}} \overline f(2m)\ f(2m+h) \\
+ &\sum\limits_{m<\frac{N}{2}} \overline f(2m+1) f(2m+1+h)+O(1).
\endaligned \right. \\
& \ \\
S(N, 2k+1) &= \left\{
\aligned &\sum\limits_{m<\frac{N}{2}} \overline f(2m) f(2m+2k+1) \\
+ &\sum\limits_{m<\frac{N}{2}} \overline f(2m+1) f(2m+2k+2)+ O(1)
\endaligned \right.  \\
& \ \\
&=-\sum\limits_{m<\frac{N}{2}} \overline f(m) f(m+k) - 
\sum\limits_{m<\frac{N}{2}} \overline f(m) f(m+k+1) + O(1).
\endaligned
$$
Hence
$$S(N, 2k+1) = -S\left(\frac{N}{2}, k \right)- 
S\left(\frac{N}{2}, k+1 \right)+O(1). \tag{1}$$
Similarly
$$\aligned
S(N, 2k)&= S\left(\frac{N}{2}, k\right) + S\left(\frac{N}{2}, k\right) + 
O(1) \\ 
&=2 S\left(\frac{N}{2}, k\right)+ O(1).
\endaligned
\tag{2}
$$

In our first recurrence relation choose $k=0$, divide by $N$ and let $N$ go to
infinity. Define

$$\left\{
\aligned
&\gamma^*(h)=\limsup\limits_{N \to \infty} \dfrac{1}{N}
\sum\limits_{n < N} f(n) f(n+h), \\
&\gamma_*(h)=\liminf\limits_{N \to \infty} \dfrac{1}{N}
\sum\limits_{n < N}\ f(n) f(n+h).
\endaligned
\right.
$$
Then

$$\left\{
\aligned
&\gamma^*(1)= -\frac{1}{2} -\frac{1}{2} \gamma_*(1), \\
&\gamma_*(1)= -\frac{1}{2} -\frac{1}{2} \gamma^*(1). 
\endaligned
\right.
$$

\noindent
These two equations with two unknowns show that
$$\gamma_*(1)= \gamma^*(1)=-\frac{1}{3}.$$ 
Therefore $\gamma(1)$ exists and its value is $-1/3$.  

\medskip

We now assume $\gamma(k)$ exists for all $k<2 K$. Then equation (2) implies that
$\gamma(2K)$ exists and
$$\gamma(2K)=\gamma(K).$$
Now equation (1) shows that $\gamma(2K+1)$ exists and
$$\gamma(2K+1)=-\dfrac{1}{2}(\gamma(K)+\gamma(K+1)).$$

We know there exists a measure $d\sigma$ such that
$$\gamma(K)=\int^1_0 e^{2i\pi K x} d\sigma(x).$$
Let us show that $d\sigma$ is a continuous measure. It suffices to show that
$$\lim\limits_{N\rightarrow\infty} \frac{1}{N} \sum\limits_{k<N}
|\gamma(k)|^2=0.$$
Put 
$$\Gamma(h)=\lim\limits_{N\rightarrow\infty} \frac{1}{N} \sum\limits_{n<N}
\overline\gamma (n) \gamma(n+h)$$
(it can easily be established that the limit exists for all $h\in\Bbb Z$).
Define
$$\aligned
T(N,h)
&= \sum\limits_{n<N} \overline\gamma(n) \gamma(n+h) \\
&=\sum\limits_{n<\frac{N}{2}} \overline\gamma(2n) \gamma(2n+h)
+\sum\limits_{n<\frac{N}{2}} \overline\gamma(2n+1) \gamma(2n+1+h)+ O(1), \\
T(N, 2k)
&=\sum\limits_{n<\frac{N}{2}} \overline\gamma(2n) \gamma(2n+2k)+
\sum\limits_{n<\frac{N}{2}} \overline\gamma (2n+1) \gamma(2n+1+2k) + O(1) \\
&=
\left\{
\aligned &T\left(\frac{N}{2}, k\right) \\
+ &\frac{1}{4} \sum\limits_{n<\frac{N}{2}}
(\gamma(n)+\gamma(n+1))(\gamma(n+k)+\gamma(n+k+1))+ O(1) 
\endaligned \right. \\
&= \left\{
\aligned &T\left(\frac{N}{2}, k\right) \\
+ &\frac{1}{4}\left(2T\left(\frac{N}{2},
k\right)+T\left(\frac{N}{2}, k+1\right)+ T\left(\frac{N}{2}, k-1\right)\right)+
O(1).
\endaligned \right. 
\endaligned
$$ 
Therefore $$\aligned \Gamma(2k)&
=\frac{1}{2}\ \Gamma(k)+\frac{1}{4} \Gamma(k)+ \frac{1}{8}
\Gamma(k+1)+ \frac{1}{8} \Gamma (k-1) \\
&=\frac{3}{4} \Gamma(k)+ \frac{1}{8} \Gamma(k+1) + \frac{1}{8}
\Gamma(k-1).
\endaligned
$$
For $k=0$,
$$\Gamma(0)=\frac{3}{4} \Gamma(0) + \frac{1}{8} \Gamma (1)+ \frac{1}{8} \Gamma
(-1).$$
But $\gamma$ is real valued, so that $\Gamma$ is real valued hence
$\Gamma(-1)=\Gamma(1)$. Therefore
$$\Gamma(0)=\Gamma(1).\tag{3}$$

On the other hand
$$T(N, 2k+1)=
\left\{
\aligned &-\frac{1}{2} T\left(\frac{N}{2}, k\right)-\frac{1}{2}
T\left(\frac{N}{2}, k+1\right) \\
&-\frac{1}{2} T\left(\frac{N}{2}, k\right)-\frac{1}{2} 
T\left(\frac{N}{2}, k-1\right)+O(1),
\endaligned
\right.
$$
hence
$$\Gamma(2k+1)=-\frac{1}{2} \Gamma(k)- \frac{1}{4} \Gamma (k+1) -\frac{1}{4}
\Gamma(k-1).$$
For $k=0$,
$$\Gamma(0)= -3 \Gamma(1).$$
This, compared with equality (3), implies that $\Gamma(0)=0$ and this in turn
implies that $d\sigma$ is a continuous measure.

\medskip

Remember that $|\widehat f(\lambda)|\le \sqrt{\sigma\{\lambda\}}=0$ so that the
Fourier-Bohr spectrum is empty.
\medskip
So far we have thus established that the spectral measure $d\sigma$ is
continuous and that the correlation $\gamma$ satisfies the equations
$$\left\{\aligned 
&\gamma(0)=1 \\
&\gamma(2h)=\gamma(h) \\
&\gamma(2h+1)=-\frac{1}{2} [\gamma(h)+\gamma(h+1)].
\endaligned
\right.
\tag{4}
$$

Observe that $\gamma(2^n)=\gamma(2^{n-1})=\cdots = \gamma(1)=-\frac{1}{3}$ hence
$$\limsup\limits_{n \to \infty} |\gamma(n)|\ge\frac{1}{3}.$$
This shows that $d\sigma$ has a nonzero singular component $d\sigma_{sg}$
$$d\sigma= d\sigma_{ac} + d\sigma_{sg}.$$

\smallskip

We shall now prove that $d\sigma_{ac}=0$. To simplify notations put
$$\sigma_{ac} = \sigma_1\ \ \ \text{ and }\ \ \sigma_{sg}=\sigma_2,$$
and
$$\gamma_1(n)= \int_0^1 e^{2i\pi n x}\ d\sigma_1(x),\ \ \ \ \ \ \gamma_2(n)=
\int^1_0 e^{2i\pi n x} d\sigma_2(x).$$
The following equalities hold
$$\aligned
\gamma(2n)
&= \int^1_0 e^{2i\pi 2n x} d\sigma(x) \\
&=\int^2_0 e^{2i\pi n t} d\sigma\left(\frac{t}{2}\right) \\
&=\int^1_0 e^{2i\pi n t} d\sigma\left(\frac{t}{2}\right) + \int^2_1
e^{2i\pi n t} d\sigma\left(\frac{t}{2}\right)
\endaligned
$$
In the last integral put $t=u+1$
$$\aligned
\gamma(2n)
&=\int^1_0 e^{2i\pi n u }d\sigma\left(\frac{t}{2}\right) + \int^1_0
e^{2i\pi n t} d\sigma\left(\frac{u+1}{2}\right) \\
&=\int^1_0 e^{2i\pi n t} d\left[\sigma\left(\frac{t}{2}\right) +
\sigma\left(\frac{t+1}{2}\right)\right].
\endaligned
$$
But $\gamma(2n)=\gamma(n)$, hence
$$d\sigma(t)=d\sigma\left(\frac{t}{2}\right) +
d\sigma\left(\frac{t+1}{2}\right),$$

$$d\sigma_1(t)+d\sigma_2(t)= d\sigma_1\left(\frac{t}{2}\right) + d\sigma_2 
\left(\frac{t}{2}\right)+ d\sigma_1\left(\frac{t+1}{2}\right) + d\sigma_2 
\left(\frac{t+1}{2}\right),$$

$$d\sigma_1(t)-d\sigma_1\left(\frac{t}{2}\right) -
d\sigma_1\left(\frac{t+1}{2}\right) = d\sigma_2 \left(\frac{t}{2}\right) +
d\sigma_2\left(\frac{t+1}{2}\right) - d\sigma_2(t).$$
The left hand side represents an absolutely continuous measure and the right
hand side a purely continuous singular measure. Therefore both vanish.
In particular
$$d\sigma_1(t)= d\sigma_1 \left(\frac{t}{2}\right) + d\sigma_1
\left(\frac{t+1}{2}\right).$$
Reading the above calculation backwards, we see that
$$\gamma_1(2n)=\gamma_1(n)$$
and in the same fashion we can prove that
$$\gamma_1(2n+1)=-\frac{1}{2} (\gamma_1(n)+ \gamma_1(n+1)). \tag{5}$$
Remember that $d\sigma_1$ is absolutely continuous hence
$$\gamma_1(n)\rightarrow 0\ \ \ \text{ as }\ \ \ \ n\rightarrow\infty,$$
so that in particular
$$\gamma_1(2^n)\rightarrow 0\ \ \ \ \text{ as }\ \ \ \ n\rightarrow\infty.$$
But $\gamma_1(2^n)=\gamma_1(1)$ hence $\gamma_1(1)=0$. Equation (5) then
implies with $n=0$ that 
$$\gamma_1(0)=0 = \int^1_0 d\sigma_1 (x).$$
Therefore $d\sigma_1=0$. We conclude that $d\sigma$ is indeed a purely
continuous singular measure.

\medskip

These results were known to Kakutani [4] and Mahler [5]. See also [3].

\bigskip

\newpage

\noindent
{\bf 6. APPENDIX 2: THE SPECTRAL MEASURE OF THE PAPERFOLDING SEQUENCE}

\bigskip

Let $f=(f(n))$ be the $(\pm)$ paperfolding sequence:
$$\left\{\aligned
&f(2n)=(-1)^n \\
&f(2n+1)=f(n).
\endaligned
\right.
$$
We shall prove that the sequence is Besicovitch almost-periodic.

To do so we compute the Fourier-Bohr coefficients
$$\widehat f(\lambda)=\lim\limits_{N\rightarrow\infty} \frac{1}{N}
\sum\limits_{n<N} f(n) \exp (-2i\pi n \lambda).$$
Define 
$$\aligned
S(N,\lambda)
&=\sum\limits_{n<N} f(n) \exp (-2i\pi n \lambda) \\
& \ \\
&=\left\{
\aligned &\sum\limits_{m<N/2} f(2m) \exp (-2i\pi 2m \lambda) \\
+&\sum\limits_{m<N/2} f(2m+1) \exp (-2i\pi (2m+1)\lambda)
\endaligned \right. \\
& \ \\
&= \left\{
\aligned &\sum\limits_{m<\frac{N}{2}} (-1)^m \exp (-4i\pi m \lambda) \\
+&e^{-2i\pi \lambda} \sum\limits_{m<\frac{N}{2}} f(m) \exp
(-2i\pi 2m \lambda).
\endaligned \right.
\endaligned
$$
The first sum is easily seen to be equal to
$$\sum\limits_{m<N/2} \exp 2i\pi m \left(\frac{1}{2}-2\lambda\right)=
\frac{N}{2}\ \goth X(\lambda)+O(1),$$
where
$$\goth X(\lambda)=\left\{\aligned
&1 \ \text{ if } \ 2\lambda\in \frac{1}{2}+\Bbb Z, \\ 
&0 \ \text{ if not. }
\endaligned
\right.
$$
To simplify notations we put $e(x)=\exp 2i\pi x$. The second sum is
$$e(-\lambda) S\left(\frac{N}{2}, 2\lambda\right),$$
so that
$$S(N, \lambda)= e(-\lambda) S\left(\frac{N}{2}, 2\lambda\right) +
\frac{N}{2} \goth X(\lambda)+O(1).$$
By induction
$$\aligned
S(N,\lambda)&=e(-\lambda-2\lambda) S\left(\frac{N}{2^2}, 2^2\lambda\right) +
\frac{N}{2} \goth X(\lambda) +\frac{N}{2^2} \goth X(2\lambda) e(-\lambda) 
+ O(1), \\
&=N \sum\limits_{j=0}^{k-1} \frac{1}{2^{j+1}} \goth X(2^j\lambda)
e((1-2^j)\lambda) 
+ e((1-2^k)\lambda) S\left(\frac{N}{2^k}, 2^k\lambda\right) \\
& \ \ + k.O(1).
\endaligned
$$
Choose $k=1+[\log N/\log 2]$. Then the sum $S(N 2^{-k}, 2^k \lambda)$ is
empty.
$$S(N,\lambda)=N \sum\limits_{j=0}^{k-1} \frac{1}{2^{j+1}} \goth X
(2^j\lambda) e((1-2^j)\lambda)+ O(\log N).$$
Divide by $N$ and let $N$ go to infinity:
$$\widehat f(\lambda)= \sum\limits_{j=0}^{\infty} 
\frac{\goth X(2^j\lambda)}{2^{j+1}} e((1-2^j)\lambda).$$
If $\lambda$ is irrational or rational with denominator other than a power of
$2$, $\widehat f(\lambda)=0$.

\medskip

Suppose $\lambda=\frac{2a+1}{2^\ell}$ where $a$ and $\ell$ are positive
integers. Then
$$\goth X\left(2^j \frac{2a+1}{2^\ell}\right)= \goth X
\left(\frac{2a+1}{2^{\ell-j}}\right)=\left\{
\aligned
&1 \ \text{ if }\ j=\ell-2, \\
&0 \ \text{ if not.}
\endaligned
\right.
$$
The Fourier-Bohr coefficient boils down to
$$\aligned
\widehat f\left(\frac{2a+1}{2^\ell}\right)
&=\frac{1}{2^{\ell-1}} e\left((1-2^{\ell-2}) \frac{2a+1}{2^\ell}\right), 
\ \ell\ge 2 \\
&=i \frac{(-1)^{a-1}}{2^{\ell-1}} 
\exp \left(2i\pi\frac{2a+1}{2^\ell}\right). 
\endaligned
$$
Then
$$\aligned
f(n)&\sim \sum\limits_{\ell=2}^\infty\,\sum\limits_{a=0}^{2^{\ell-1}-1}
i\frac{(-1)^{a-1}}{2^{\ell-1}} \exp 2i\pi \frac{2a+1}{2^\ell} \exp 2i\pi
\frac{2a+1}{2^\ell} n \\
&\sim \sum\limits_{\ell\ge2}\ \sum\limits_{a=0}^{2^{\ell-1}-1}
\frac{i(-1)^{a-1}}{2^{\ell-1}} \exp\left(2i\pi \frac{2a+1}{2^\ell}
(n+1)\right)
\endaligned
$$
It is now easy to see that Parseval's equality holds.
Indeed
$$\aligned
\sum\limits_{\lambda} |\widehat f(\lambda)|^2
&=\sum\limits_{\ell\ge2}\ \sum\limits_{a=0}^{2^{\ell-1}-1}
\frac{1}{2^{2(\ell-1)}} \\
&=\sum\limits_{\ell\ge2} \frac{2^{\ell-1}}{2^{2(\ell-1)}}=
\sum\limits_{\ell\ge2} \frac{1}{2^{\ell-1}}=1 =\Vert f\Vert^2.
\endaligned
$$
This proves that the regular paperfolding sequence is Besicovitch
almost-perio\-dic. Its spectral measure is
$$d\sigma=\sum\limits_{\ell=2}^\infty \frac{1}{4^{\ell-1}}
\sum\limits_{a=0}^{2^{\ell-1}-1} \delta_{\frac{2a+1}{2^\ell}}.$$
\smallskip
This result was first published in [6] where it is proved that all paperfolding
sequences have the same spectral measure (but the Fourier-Bohr coefficients
differ).

\bigskip
\bigskip

\references

\refis{1}
Bass J., Suites uniform\'ement denses, moyennes trigonom\'etriques,
fonctions pseudo-al\'ea\-toi\-res, \review Bull. Soc. math. France, 
87, 1959, 1--69.

\refis{2}
Bertrandias J.-P., Espaces de fonctions born\'ees et continues en
moyenne asymptotique d'ord\-re $p$, (Th\`ese), \review {Bull. Soc. math. 
France, M\'emoire}, 5, 1966, 3--106.

\refis{3}
Coquet J., Kamae T., Mend\`es France M., Sur la mesure spectrale de
certaines suites arithm\'etiques, \review Bull. Soc. math. France, 105, 
1977, 369--384.

\refis{4}
Kakutani S., Strictly ergodic symbolic dynamical systems,
in: Proceedings of the $6^{\text{th}}$ Berkeley symposium on mathematical 
statistics and probability [1970 Berkeley] vol. 2  
(Berkeley, University of California Press, 1972) p. 319--326.

\refis{5} 
Mahler K., On the translation properties of a simple class of arithmetical 
functions, \review Jour. Math. and Phys., 6, 1927, 158--163.

\refis{6} 
Mend\`es France M., van der Poorten A. J., Arithmetic and analytic
properties of paperfolding sequences (dedicated to Kurt Mahler),
\review Bull. Austr. Math. Soc., 24, 1981, 123--131.

\refis{7} 
Meyer Y., Le spectre de Wiener, \review Studia Mathematica, 27, 
1966, 189--201.

\refis{8} 
Queff\'elec M., {\it this volume}.

\refis{9} 
Wiener N., Generalized harmonic analysis, \review Acta Math., 55, 
1930, 117--258.

\endreferences

\newpage

\centerline{\big PART V}

\bigskip

\centerline{\big Complexity of infinite sequences}

\bigskip

Another approach to the order versus chaos paradigm consists of the study
of the factor complexity of sequences. A factor of a (finite or infinite)
sequence is a finite block of consecutive letters.

\proclaim{Definition} The complexity function (or for short the complexity)
of an infinite sequence $u$ with values in a finite alphabet $A$ is the function
$n \rightarrow p_u(n)$, where $p_u(n)$ denotes the number of factors of length 
$n$ in the sequence $u$.
\endproclaim

\noindent
{\bf Remarks}

\medskip

- Let $u$ be any infinite sequence on the (finite) alphabet $A$. Then obviously
$\forall n \geq 1, \ 1 \leq p_u(n) \leq (\sharp A)^n$. The function $p_u$ is 
clearly increasing.

\smallskip

- If $u$ is a ``random" sequence, one expects it to be {\it normal}, i.e.
every possible block of length $n$ occurs with frequency $1/(\sharp A)^n$. 
In particular the complexity of such a sequence is maximal and
is equal to $(\sharp A)^n$. On the other end of the spectrum it is not difficult 
to check that an ultimately periodic sequence has an ultimately constant complexity, 
(the converse holds with a weaker hypothesis, see below).
Complexity distinguishes between normality and periodicity, and much more.

Note however that there exist algorithmically constructed sequences which are normal:
the Champernowne sequence defined, say in base $2$, by writing down
consecutively the binary expansions of the integers,
$$0 \ 1 \ 10 \ 11 \ 100 \ 101 \ 110 \ 111 \ 1000 \ 1001 \ 1010 \ \cdots$$
is known to be normal.

\smallskip

- Using the above definition of complexity one defines the {\it 
topological entropy of a sequence $u$ as 
$$h(u) = \lim\limits_{n \to \infty}\frac{\log p_u(n)}{n \log \sharp A}.$$
The limit always exists}. Clearly $0 \leq h(u) \leq 1$. 
If the structure of the sequence is simple, one expects $h(u) = 0$. On the 
contrary  a complex structured sequence will have maximal entropy $h(u) = 1$. 
Needless to say for an automatic sequence $h(u) = 0$ and for a normal sequence 
$h(u) = 1$.
For any $\alpha \in [0,1]$ there exist sequences $u$ for which $h(u) = \alpha$.
The notion of entropy will be discussed in more details by V. Berth\'e in this 
volume [12].

\smallskip

- Many other notions of complexity exist for a sequence, a
very interesting one being the Kolmogorov-Chaitin-Solomonoff complexity
which compares a given finite or infinite sequence to the length of the
smallest possible ``program" which generates it, see [21].

\bigskip

\newpage
\noindent {\bf 1. STURMIAN SEQUENCES AND GENERALIZATIONS}

\bigskip

We begin this paragraph with a theorem which shows that, if the complexity of
a sequence is ``not too large", then the sequence is periodic from some point
on, and this implies in turn that the complexity is ultimately constant.

\proclaim{Theorem} [24], [25] and [15] 

Let $u$ be a sequence such that $\exists n \geq 1, \ p_u(n) \leq n$. Then the
sequence $u$ is ultimately periodic, (hence $p_u$ is ultimately constant).
\endproclaim

In view of this theorem the minimal possible complexity among the
sequences which are not ultimately periodic is given by $p(n) = n+1$, $\forall
n \geq 1$. With $n = 1$ one sees that, if there exist such sequences, the alphabet
must have cardinality $2$. Such sequences actually exist, as shown by the following
result. They are called {\it Sturmian sequences}.

\proclaim{Theorem} To each Sturmian sequence $(u_n)_n$ on the alphabet $\{0, 1\}$
corresponds a unique couple $(\alpha, \beta) \in [0,1[^2$, $\alpha$ irrational
such that
$$\alignedat2
&\text{- either} \ \  &&\forall n, \ u_n = \lfloor (n+1)\alpha + \beta \rfloor
- \lfloor n\alpha + \beta \rfloor, \\
&\text{- or} &&\forall n, \ u_n = \lceil (n+1)\alpha + \beta \rceil
- \lceil n\alpha + \beta \rceil,
\endalignedat
$$
where $\lfloor x \rfloor$ is the usual integer part and $\lceil x \rceil = 
- \lfloor -x \rfloor$.
\endproclaim

A nice survey on Sturmian sequences is [11].

\bigskip

\noindent {\bf Remarks}

\medskip

- The above theorem shows that Sturmian sequences can also be
constructed by intersecting a two-dimensional lattice by a straight line
with irrational slope. This implies that they can also be generated as
billiard trajectories. In particular our friend {\it the Fibonacci sequence is
a Sturmian sequence}. One should remark that very few Sturmian sequences can
be generated by morphisms: the set of morphic sequences is countable, whereas the set
of Sturmian sequences is not, since each of them is determined by a couple
$(\alpha, \beta)$. To know precisely which Sturmian sequences are fixed point of 
morphisms, one can read [16]. 

\medskip

- One can study billiard trajectories in more than two dimensions. Some results for the 
complexity are known: for instance the three-dimensional irrational
trajectories of billiard balls have complexity $n^2 + n + 1$, see [8] and [9], 
where interesting conjectures are given.

\medskip

- One can play billiard in a triangle or in a convex polygon instead of playing
in a square. The following has been proved in [19]: consider a polygonal billard 
with angles $\frac{a_1}{r}\pi$, $\frac{a_2}{r}\pi$, $\cdots$, $\frac{a_d}{r}\pi$,
where $\gcd(a_1,a_2, \cdots, a_d,r)=1$, and start with any angle $\alpha$
(with a countable number of exceptions), then the complexity is:
$p(n) = nr + (d-1)r$, for $n \geq n_0(\alpha)$. In particular $p(n) = 4n+8$
for a right-angle triangle with two equal sides and $p(n) = 3n+6$ for an equilateral 
triangle.

\medskip

- Other sequences with geometric properties have been studied: sequences
with complexity $2n+1$, see [29], [10], [18] and [31], sequences with complexity
$2n$, see [30] and [1]. Another class is defined as follows. Let $\Delta$ be
the first difference operator: for a sequence $u$, the sequence $\Delta u$
is defined by $(\Delta u)_n = u_{n+1} - u_n$. Now let $\alpha$ be an irrational
number. Instead of considering the Sturmian sequence $\left( [(n+1)\alpha] -
[n\alpha] \right) = (\Delta u)_n$, where $u_n = [n\alpha]$, consider
$\Delta^2 v$, where $v_n = [n^2 \alpha]$. 
The complexity of this new sequence is given by [7]:
$$p(n) = \frac{(n+1)(n+2)(n+3)}{6}.$$

\bigskip

\noindent {\bf 2. COMPLEXITY OF AUTOMATIC SEQUENCES}

\bigskip

\noindent
{\bf 2.1. An upper bound}

\medskip

The first result was given by Cobham [14]. It states that the complexity
of an automatic sequence is not too large.

\proclaim{Theorem} [14]

If $u$ is an automatic sequence, there exists a constant depending on $u$
such that:
$$\forall n \geq 1, \ p_u(n) \leq Cn.$$
\endproclaim

We list below some special results.

\bigskip

\noindent
{\bf 2.2. Complexity function of some automatic sequences}

\medskip

- If $u$ is the Prouhet-Thue-Morse sequence, its complexity $p_u$ is given by
a ... rather complicated expression, see [22], [23] and [13], from which we shall
only retain that the sequence $n \rightarrow p_u(n+1) - p_u(n)$ is a 
$2$-automatic sequence. This result will be revisited below.

\medskip

- If $u$ is the (regular) paperfolding sequence, then its complexity function
satisfies $\forall n \geq 7, \ p_u(n) = 4n$. Actually this result holds for
{\it any} paperfolding sequence [2], and one can also prove that there is
a kind of synchronization between the factors of different paperfolding
sequences [5]. Note that the paperfolding sequences can be obtained as
``Toeplitz sequences", see [3] for a survey on Toeplitz sequences, and that the 
complexity of a class of Toeplitz sequences has been studied in [20].

\medskip

- For the Rudin-Shapiro sequence one has $p(n) = 8n-8, \ \forall n \geq 8$.
This result can be extended to two generalizations of the Rudin-Shapiro 
sequence, see [2] and [6], see also [4]: let us just say here that one
obtains again ultimately affine complexities.

\medskip

- We quote finally in this paragraph the paper [26], where it is shown that,
if $u$ is an automatic sequence which satisfies some mild conditions, then
its complexity function $p_u$ has the property that $(p_u(n+1) - p_u(n))_n$
is also an automatic sequence. The simple fact that this last
sequence takes only finitely many values is by no means trivial. 

\bigskip

\noindent
{\bf 2.3. The case of non-constant length morphisms}

\medskip

After the Fibonacci sequence which is Sturmian as we saw above, one can ask
for the complexity function of the fixed points of non-constant length 
morphisms. The problem is more difficult, reflecting the fact that these
sequences are more ``complicated" than the automatic sequences.

It has been shown that the complexity of a sequence $u$ fixed point of a
non-constant length morphism satisfies $p(n) = O(n^2)$. In the case where
the morphism is ``primitive" (i.e. if there exists a power of the morphism
such that the image of every single letter contains all the letters), this
bound can be lowered to $p(n) = O(n)$. The most precise result can be found 
in [27] and [28] where it is proven that the complexity of any fixed point of a 
morphism has one of the following orders of magnitude:
$$\exists A, B > 0, \ \exists f \in \{1, n, n \log\log n, n\log n, n^2\}, \
\forall n \geq 1, \ Af(n) \leq p(n) \leq Bf(n).$$

\medskip

Kolakoski's sequence, which has already been discussed in Part I, is probably
not generated by a non-constant length morphism. Its complexity is not known.
It has been conjectured to be of the order of $n^q$, where 
$q = \frac{\log 3}{\log \frac{3}{2}}$, see [17].

\bigskip
\bigskip

\references

\refis{1}
Alessandri P., Codage des rotations, M\'emoire de D. E. A.
(Universit\'e Claude Bernard, Lyon I, 1993).

\refis{2}
Allouche J.-P., The number of factors in a paperfolding sequence,
\review Bull. Austr. Math. Soc., 46, 1992, 23--32.

\refis{3}
Allouche J.-P., Bacher R., Toeplitz sequences, paperfolding, Hanoi towers 
and progression-free sequences of integers, \review Enseign. Math., 38, 1992,
315--327.

\refis{4} 
Allouche J.-P., Bousquet-M\'elou M.,
Facteurs des suites de Rudin-Shapiro g\'en\'eralis\'ees, \review 
Bull. Belg. Math. Soc., 1, 1994, 145--164. 

\refis{5}
Allouche J.-P., Bousquet-M\'elou M., Canonical positions for the factors 
in the paperfolding sequences, \review Theoret. Comput. Sci., 129, 1994, 
263--278.

\refis{6} 
Allouche J.-P., Shallit J., Complexit\'e des suites de Rudin-Shapiro 
g\'en\'era\-li\-s\'ees, \review Journal de Th\'eorie des Nombres de Bordeaux,
5, 1993, 283--302. 

\refis{7} 
Arnoux P., Mauduit C., Meigniez G.,
Sur la complexit\'e de certaines suites arith\-m\'e\-tiques, 
{\it in preparation}.

\refis{8}
Arnoux P., Mauduit C., Shiokawa I., Tamura J. I.,
Complexity of sequences defined by billiards in the cube,
\review Bull. Soc. math. France, 122, 1994, 1--12. 

\refis{9}
Arnoux P., Mauduit C., Shiokawa I., Tamura J. I.,
Rauzy's conjecture on the cubic billiards, \review Tokyo J. Math.,
17, 1994, 211--218.

\refis{10}
Arnoux P., Rauzy G., Repr\'esentation g\'eom\'etrique de suites de 
complexit\'e $2n+1$, \review Bull. Soc. math. France, 119, 1991, 
199--215.

\refis{11}
Berstel J., S\'e\'ebold P., Mots de Sturm - un survol, {\it Preprint}
(1994).

\refis{12}
Berth\'e V., {\it this volume}.

\refis{13}
Brlek S., Enumeration of factors in the Thue-Morse word,
\review Discrete Appl. Math., 24, 1989, 83--96.

\refis{14}
Cobham A., Uniform tag sequences, \review Math. Systems Theory, 6, 1972, 
164--192.

\refis{15}
Coven E. M., Hedlund G. A., Sequences with minimal block growth,
\review Math. Systems Theory, 7, 1973, 138--153.

\refis{16}
Crisp D., Moran W., Pollington A., Shiue P., Substitution invariant cutting 
sequences, \review Journal de Th\'eorie des Nombres de Bordeaux, 5,
1993, 123--137.
 
\refis{17}
Dekking F. M.,  On the structure of self-generating sequences,
\review {S\'eminaire de Th\'eorie des Nombres de Bordeaux, Expos\'e 31},
\! , 1980--1981, 31-01--31-06.

\refis{18}
Ferenczi S., Les transformations de Chacon : combinatoire,
structure g\'eo\-m\'etrique, lien avec les syst\` emes de complexit\'e
$2n+1$, \review Bull. Soc. math. France, , 1995, to \ appear.

\refis{19}
Hubert P., Complexit\'e des suites d\'efinies par des billards rationnels,
\review Bull. Soc. math. France, , 1995, to \ appear.

\refis{20}
Koskas M., Complexity functions for some Toeplitz sequences,
\review Theoret. Comput. Sci., , 1995, to \ appear.

\refis{21}
Li M., Vit\'anyi P., Kolmogorov complexity and its applications, in: 
Handbook of Theoretical Computer Science, Volume A: Algorithms and 
Complexity, J. van Leeuwen Ed (MIT Press, 1990) p. 187--254.

\refis{22}
de Luca A., Varricchio S., On the factors of the Thue-Morse word on three 
symbols, \review Information Processing Letters, 27, 1988, 281--285.
 
\refis{23}
de Luca A., Varricchio S., Some combinatorial properties of the Thue-Morse 
sequence, \review Theoret. Comput. Sci., 63, 1989, 333--348.
 
\refis{24}
Morse M., Hedlund G. A., Symbolic Dynamics, \review Amer. J. Math., 60,
1938, 815--866.

\refis{25}
Morse M.,  Hedlund G. A., Symbolic Dynamics II. Sturmian trajectories, 
\review  Amer. J. Math., 62, 1940, 1--42.

\refis{26}
Moss\'e M., Notions de reconnaissabilit\'e pour les substitutions et 
complexit\'e des suites automatiques, {\it Pr\'epublication du LMD} 
{\bf 93-21} (1993).

\refis{27}
Pansiot J.-J., Bornes inf\'erieures sur la complexit\'e des facteurs des 
mots infinis engendr\'es par morphismes it\'er\'es, \review Lecture Notes 
in Computer Science, 166, 1984, 230--240.

\refis{28}
Pansiot J.-J., Complexit\'e des facteurs des mots infinis engendr\'es par 
morphismes it\'er\'es, \review Lecture Notes in Computer Science,
172, 1984, 380--389.

\refis{29}
Rauzy G., Suites \`a termes dans un alphabet fini, \review
{S\'eminaire de Th\'eorie des Nombres de Bordeaux, Expos\'e 25},
\! , 1982--1983, 25-01--25-16.

\refis{30}
Rote G., Sequences with subword complexity $2n$, \review J. Number Theory,
46, 1994, 196--213.

\refis{31}
Santini M.-L., \'Echanges de trois intervalles (Th\`ese, Universit\'e 
Aix-Mar\-seille II, 1994).

\endreferences

\newpage

\centerline{\big PART VI}

\bigskip

\centerline{\big Opacity of an automaton}

\bigskip

\noindent
{\bf 1. AUTOMATA REVISITED}

\bigskip

In this part all the automata we consider are $2$-automata. The arrows are
named $(+)$ and $(-)$.

\medskip

An automaton $\Cal A$ is determined by a finite set $S$ (set of states), an
initial state $A\in S$ and a map $\tau : \{-, +\}\times S\rightarrow S$
represented by $(\pm)$ arrows which join two states:
$$\Cal A=(S,\ A,\ \tau).$$
Given an infinite sequence $\varepsilon=(\varepsilon_n),\
\varepsilon_n=\pm\,1$, it acts on $\Cal A$ in the following way.
$\varepsilon_0$ puts the automaton in state $\varepsilon_0(A)$. Then
$\varepsilon_1$ sends $\varepsilon_0(A)$ onto
$\varepsilon_1(\varepsilon_0(A))$, etc... The sequence of states
$$A,\ \varepsilon_0(A),\ \varepsilon_1(\varepsilon_0(A)),\ \cdots$$
is an infinite sequence on the alphabet $S$ denoted $\Cal A\varepsilon$.

\medskip

We now define the output function
$$\varphi: S\rightarrow\Bbb R$$
which maps $\Cal A\varepsilon$ on $\varphi(\Cal A\varepsilon)$
coordinate-wise.

\medskip

We propose to measure the discrepancy between the
input sequence $\varepsilon$ and the output sequence
$\varphi(\Cal A\varepsilon)$.

\medskip

Given two real bounded sequences
$\delta'=(\delta'_n)$ and $\delta''=(\delta''_n)$ we define their distance by
$$\Vert\delta'-\delta''\Vert=\limsup\limits_{N \to \infty}
\left(\frac{1}{N} \sum\limits_{n<N} |\delta'_n-\delta''_n|^2\right)^{1/2}.$$
This norm was already discussed in Part IV.

\medskip

$\Vert\varphi(\Cal A\varepsilon)-\varepsilon\Vert$ is to be thought as the
distorsion of the sequence $\varepsilon$ when mapped through $(\Cal A,
\varphi)$. We minimize the distorsion by choosing $\varphi$ as ``best as we
can'':
$$\inf\limits_{\varphi}\ \Vert\varphi(\Cal
A\varepsilon)-\varepsilon\Vert.$$
The $\inf$ is to be taken over all maps
$\varphi:\ S\rightarrow\Bbb R$. The {\it opacity} of $\Cal A$ is by
definition
$$\omega(\Cal A)=\sup\limits_{\varepsilon\in UP}\
\inf\limits_{\varphi} \Vert\varphi(\Cal A\varepsilon)-\varepsilon\Vert
\tag{1}$$
where the sup is over the set of all $(\pm)$ sequences $\varepsilon$ which
are periodic from some point on ($UP$ stands for Ultimately Periodic).

\medskip

The $UP$ restriction is a technical requirement which may perhaps not be
necessary. At the time of writing, Nathalie Loraud [1] is studying the
pertinence of the restriction. She proposes to modify the definition of the
opacity:
$$\widetilde\omega(\Cal A)=\sup\limits_{\varepsilon}
\limsup\limits_{N \to \infty} \left(\frac{1}{N} \inf\limits_{\varphi}
\sum\limits_{n<N} |\varphi(\Cal A\varepsilon)-\varepsilon|^2\right)^{1/2}$$
(notice the permutation of the two operations $\limsup\limits_{N}$
and $\inf\limits_{\varphi}$). With this definition she can show that
$\sup\limits_{\varepsilon\in UP}$ and $\sup\limits_{\varepsilon}$ coincide.
She can also prove that $\omega(\Cal A)=\widetilde\omega(\Cal A)$. 

\smallskip

The question remains: is it true that
$$\omega(\Cal A)=\sup\limits_{\varepsilon} \inf\limits_{\varphi} 
\Vert\varphi(\Cal A\varepsilon)-\varepsilon\Vert ?$$
We shall not discuss the matter and we maintain our original definition (1).

\medskip

Let us describe some simple properties of the opacity. Trivially
$\omega(\Cal A)\ge0$. We also remark
$$\inf\limits_{\varphi} \Vert\varphi(\Cal A\varepsilon)-\varepsilon\Vert \le
\Vert-\varepsilon\Vert=1$$
since one can choose $\varphi$ to be the constant $0$ function. Therefore
$$0\le \omega(\Cal A)\le 1.$$
An automaton $\Cal A$ for which $\omega(\Cal A)=0$ is said to be transparent.
If $\omega(\Cal A)=1$ then $\Cal A$ is completely opaque.

\bigskip

\noindent
{\bf Example 1.} The identity automaton

\bigskip

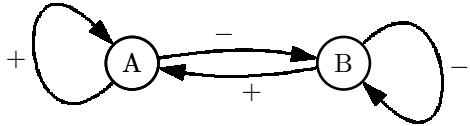
\Draw
\NewCIRCNode(\StateNode,103,)
\Define\StateAt(3){ \MoveTo(#2,#3) \StateNode(#1)(--#1--)}
\CycleEdgeSpec(45,50)
\DiagramSpec(\StateAt&\TransEdge)
\ArrowHeads(1) 
\SaveAll
\Diagram
(A,0,0 & B,80,0) (A,B,$-$,$+$ & B,B,0,$-$) 
\RecallAll
\CycleEdgeSpec(225,50)
\Diagram  
(A,0,0) (A,A,0,$+$)
\EndDraw

\bigskip

\noindent
Consider for example the input sequence
$$\varepsilon :\ \ +\ -\ -\ +\ +\ -\ \cdots$$
The automaton sends it on
$$\Cal A\varepsilon :\ A\ B\ B\ A\ A\ B\ \cdots$$
Apart from the symbols, both sequences are identical. Define
$$\varphi(A)=+1,\ \varphi(B)=-1.$$
Then for all $\varepsilon$,
$\varphi(\Cal A\varepsilon)=\varepsilon$ hence $\omega(\Cal A)=0$.

\bigskip

\noindent
{\bf Example 2.} The constant automaton

\bigskip

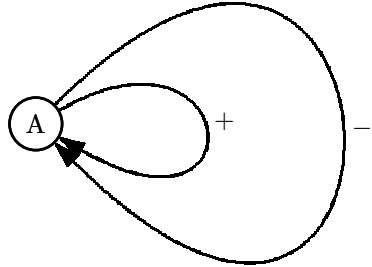
\Draw
\NewCIRCNode(\StateNode,103,)
\Define\StateAt(3){ \MoveTo(#2,#3) \StateNode(#1)(--#1--)}
\CycleEdgeSpec(45,200)
\DiagramSpec(\StateAt&\TransEdge)
\ArrowHeads(1) 
\SaveAll
\Diagram
(A,0,0) (A,A,0,$-$)
\RecallAll
\CycleEdgeSpec(30,80)
\Diagram
(A,0,0) (A,A,0,$+$)
\EndDraw

\bigskip

\noindent
Put $\varphi(A)=a$. Then
$$\omega(\Cal A)^2=\sup\limits_{\varepsilon\in UP} \inf\limits_{a}
\limsup\limits_{N} \frac{1}{N} \sum\limits_{n<N} |a-\varepsilon_n|^2.$$
In the $\limsup$ choose $\varepsilon_n=(-1)^n$:
$$\aligned
\limsup\limits_{N} \frac{1}{N} \sum\limits_{n<N} |a-(-1)^n|^2
&= \limsup\limits_{N} \frac{1}{N} \sum\limits_{n<N}
(a^2+1-2a(-1)^n) \\
&=a^2+1.
\endaligned
$$
Trivially
$$\inf\limits_{a} (a^2+1)=1,$$
hence
$$\omega(\Cal A)\ge1.$$
But $\omega(\Cal A)\le1$ therefore $\omega(\Cal A)=1$.
\medskip
The output sequence gives no information on the input sequence, and it is
then not surprising to discover that this automaton is completely opaque.

\bigskip

\noindent
{\bf Example 3.} The Thue-Morse automaton

\bigskip

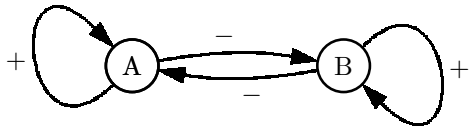
\Draw
\NewCIRCNode(\StateNode,103,)
\Define\StateAt(3){ \MoveTo(#2,#3) \StateNode(#1)(--#1--)}
\CycleEdgeSpec(45,50)
\DiagramSpec(\StateAt&\TransEdge)
\ArrowHeads(1) 
\SaveAll
\Diagram
(A,0,0 & B,80,0) (A,B,$-$,$-$ & B,B,0,$+$) 
\RecallAll
\CycleEdgeSpec(225,50)
\Diagram  
(A,0,0) (A,A,0,$+$)
\EndDraw

\bigskip

\noindent
Put $\varphi(A)=a$, $\varphi(B)=b$. If the input sequence is
$\varepsilon=(\varepsilon_n)$ then it can be easily seen that the output
sequence is
$$\delta_n=\frac{1}{2} (a+b) + \frac{1}{2} (a-b) \pi_n,$$
where $$\pi_n=\varepsilon_0 \varepsilon_1\cdots \varepsilon_n.$$
Let us choose as above $\varepsilon_n=(-1)^n$. Then
$$\Vert\delta-\varepsilon\Vert^2=\limsup\limits_{N} \frac{1}{N}
\sum\limits_{n<N} \left(\frac{1}{2} (a+b) + \frac{1}{2} (a-b)
\pi_n-\varepsilon_n\right)^2.$$
Observing that
$$\lim \frac{1}{N} \sum \varepsilon_n=\lim \frac{1}{N} \sum
\pi_n=\lim \frac{1}{N} \sum \varepsilon_n \pi_n=0,$$
we easily see that
$$\Vert\delta-\varepsilon\Vert^2=\frac{1}{2}(a^2+b^2)+1.$$
Then
$$\omega(\Cal A)\ge \inf\limits_{a,b} \left[\frac{1}{2}(a^2+b^2)+1\right]=1,$$
hence
$$\omega(\Cal A)=1.$$

This automaton is also completely opaque though it is possible to reconstruct
$\varepsilon$ from $\Cal A_\varepsilon$.

\medskip

We shall see below how to compute $\omega(\Cal A)$ in many cases; we leave it to
the reader to show that $\omega(\Cal A)$ can take all values of the form 
$\sqrt{\rho}$ where $\rho$ is any rational number in the interval $(0,1)$.

\bigskip

\noindent
{\bf 2. COMPUTING THE OPACITY}

\bigskip

An automaton is said to be strongly connected if from any two states, there
exists at least one path of arrows joining the first state to the second.

\bigskip

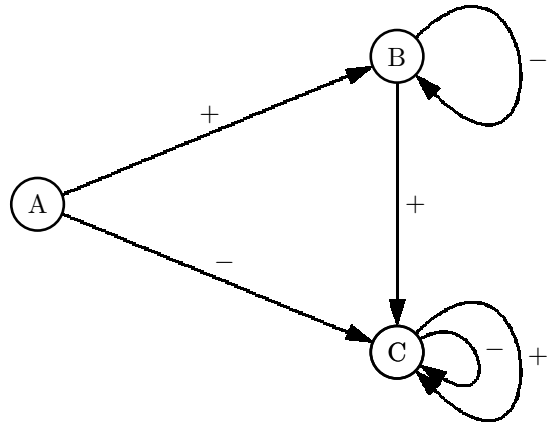
\Draw
\NewCIRCNode(\StateNode,103,)
\Define\StateAt(3){ \MoveTo(#2,#3) \StateNode(#1)(--#1--)}
\CycleEdgeSpec(45,68)
\DiagramSpec(\StateAt&\TransEdge)
\ArrowHeads(1) 
\SaveAll
\Diagram
(A,0,0 & B,136,56 & C,136,-56) 
(A,B,$+$, & B,C,$+$, & A,C,$-$, & B,B,0,$-$ 
& C,C,0,$+$)
\RecallAll
\CycleEdgeSpec(30,27)
\Diagram
(C,136,-56) (C,C,0,$-$)
\EndDraw

\bigskip

\noindent
The above automaton is not strongly connected since no arrows go to
$A$.

\medskip

An automaton is homogeneous if each state has exactly two incident arrows.
This is not the case in the preceding automaton. Automata in examples 1, 2,
3 are both strongly connected and homogeneous.

\medskip

We now define strong states on a path. Let $P$ be a closed path on an
automaton. Its length $\ell(P)$ is the number of states it contains, counted
with multiplicity.

\medskip

A state $B$ on the path $P$ is a {\it strong state} if it is attained
twice at least, once through a $(+)$ arrow and once through a $(-)$ arrows on
$P$. $\nu(P)$ counts the number of strong states on $P$.

\proclaim{Theorem} The opacity of a strongly connected homogeneous automaton
$\Cal A$ is given by the formula
$$\omega(\Cal A)=\sup\limits_{P} \sqrt{\frac{2\nu(P)}{\ell(P)}}.$$
Given any rational number $\rho$ in the interval $(0,1)$ there exist a
strongly connected homogeneous automaton $\Cal A$ such that
$\omega(\Cal A)=\sqrt{\rho}$. 
\endproclaim

We shall not give the proof of this result (see [2]). It can
be shown that the result extends to a larger class of automata, namely those 
for which the states are of three kinds:

\smallskip

\roster
\item "a)" exactly two incident arrows, one $(+)$ and one $(-)$, \par
\item "b)" any number of $(+)$ arrows and no $(-)$ arrows, \par
\item "c)" any number of $(-)$ arrows and no $(+)$ arrows.
\endroster

\smallskip

Let us illustrate this theorem by computing the opacity of the following
automaton

\bigskip

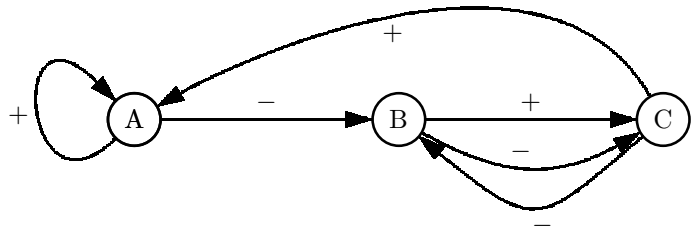
\Draw
\NewCIRCNode(\StateNode,103,)
\Define\StateAt(3){ \MoveTo(#2,#3) \StateNode(#1)(--#1--)}
\CycleEdgeSpec(45,50)
\CurvedEdgeSpec(20,.3,-20,.3)
\DiagramSpec(\StateAt&\TransEdge)
\ArrowHeads(1) 
\SaveAll
\Diagram
(A,0,0 & B,100,0 & C,200,0) (A,B,$-$,  & B,C,$+$,) 

\CurvedEdgeSpec(-60,.4,30,.1)
\CurvedEdge(C,A) 
\EdgeLabel(--$+$--)

\CurvedEdgeSpec(-30,.3,30,.3)
\CurvedEdge(B,C) 
\EdgeLabel(--$-$--)

\CurvedEdgeSpec(40,.5,-40,.5)
\CurvedEdge(C,B)
\EdgeLabel(--$-$--)

\RecallAll
\CycleEdgeSpec(225,50)
\Diagram  
(A,0,0) (A,A,0,$+$)

\EndDraw

\bigskip

\noindent
The shortest closed path $P$ which contains both the $+$ and $-$ arrows incident 
to $C$ is
$$B\overset +\to\longrightarrow C\overset -\to\longrightarrow B\overset
-\to\longrightarrow C\overset -\to\longrightarrow B$$
for which $\ell(P)=4$ and $\nu(P)=1$ thus
$$\omega(\Cal A)=\sqrt{\frac{2\nu}{\ell}}=\frac{1}{\sqrt 2}.$$

We shall apply these results to the Ising chain in Part VII.

\bigskip
\bigskip

\references

\refis{1}
Loraud N., Contribution \`a l'\'etude de l'opacit\'e d'un automate,
{\it Work in progress}.

\refis{2} 
Mend\`es France M., Opacity of an automaton. Application to the
inhomogeneous Ising chain, \review Comm. Math. Phys., 139, 1991,
341--352.

\endreferences

\newpage

\centerline{\big PART VII}

\bigskip

\centerline{\big The Ising automaton}

\bigskip

A large part of this chapter (Paragraphs 1 and 2) is due to B. Derrida who 
introduced us to the idea of induced field and who showed us how to compute it. 
See [6] and the forthcoming paper [8].

\bigskip

\noindent {\bf 1. THE INHOMOGENEOUS ISING CHAIN}

\bigskip

An Ising chain consists of $N+1$ sites in a row occupied by $N+1$ particles 
with spin $\pm 1$. At site $q\in\{0,1\cdots, N\}$ the spin is denoted
$$\sigma_q\in \{-1, +1\}.$$
A configuration is a finite sequence of spins
$$\sigma=(\sigma_0, \sigma_1, \cdots, \sigma_N).$$

Let
$$\varepsilon=(\varepsilon_0, \varepsilon_1, \cdots, \varepsilon_{N-1})\in \{-1,
+1\}^N.$$
be a given sequence which could be thought of as the distribution of two 
substances in an alloy.
The Hamiltonian or energy of a configuration $\sigma$ is defined by
$$\Cal H_\varepsilon(\sigma)=-J \sum\limits_{q=0}^{N-1}
\varepsilon_q \sigma_q \sigma_{q+1} - H \sum\limits_{q=0}^{N} \sigma_q,$$
where $J>0$ is a given parameter known as the coupling constant or binding 
energy and where $H$  is a real parameter called the external field.

\medskip

One of the main problems is to find the equilibrium state (or fundamental state)
of the chain i.e. to discover the configuration $\sigma$ which minimizes 
$\Cal H_\varepsilon(\sigma)$. 

\medskip

Let $T>0$ be the temperature; put $\beta=1/T$. The fundamental axioms of Statistical Mechanics
show that the probability of configuration $\sigma$ at temperature $T$ is
$$p_T(\sigma)=\frac{1}{Z(\beta, N)} \exp(-\beta \Cal H_\varepsilon(\sigma)),$$
where $Z(\beta, N)$, the partition function, is defined by
$$Z(\beta, N)=\sum\limits_{\sigma'\in\{-1, +1\}^{N+1}} \exp(-\beta \Cal
H_\varepsilon(\sigma')).$$
Suppose there exist $L$ fundamental states which minimize $\Cal H_\varepsilon(\sigma)$. Then
it is easy to see that as $T$ vanishes,
$$\lim\limits_{T\searrow 0} p_T(\sigma)=\left\{
\alignedat2
&1/L \  &&\text{ if $\sigma$ is fundamental,} \\
&0 \  &&\text{ if not.}
\endalignedat
\right.
$$

If $L=1$, then, at vanishing temperature, the chain is at equilibrium with 
probability 1. Order prevails.

\medskip

At high temperature, ($T\nearrow+\infty$, $\beta\searrow 0$),
$$\lim\limits_{T\nearrow+\infty} p_T(\sigma)=\frac{1}{2^{N+1}}.$$
All configurations have equal probability: the system is in a state of total disorder.

\bigskip

\noindent
{\bf 2. THE INDUCED FIELD}

\bigskip

We shall now study the sign of the spin $\sigma_N$ of the particle at site $N$ 
(the very last one) at temperature $T=0$. 
Our strategy is based on the simple fact
$$\lim\limits_{\beta\rightarrow+\infty} [Z(\beta, N)]^{1/\beta}=\exp(-\Cal
H_\varepsilon(\widehat\sigma)),$$
where $\widehat\sigma$ is one of the equilibrium configurations.

\medskip

The discussion which we reproduce here in this section is well known to 
physicists. See for example [6].

\medskip

Define the two conditional partition functions
$$\aligned
Z^+(\beta, N)&=\sum\limits_{\Sb \sigma\in\{\pm\}^N \\ \sigma_N=+1\endSb} 
\exp(-\beta \Cal H_\varepsilon(\sigma)) \\
& \\
Z^-(\beta, N)&=\sum\limits_{\Sb \sigma\in\{\pm\}^N \\ \sigma_N=-1\endSb} 
\exp(-\beta \Cal H_\varepsilon(\sigma)).
\endaligned
$$
The probability that $\sigma_N=+1$ at temperature $T$ is
$$Z^+(\beta, N)/Z(\beta, N).$$
Similarly
$$Z^-(\beta, N)/ Z(\beta, N)$$
is the probability that $\sigma_N=-1$ at $T$.

\medskip
If the ratio

$$Z^+(\beta, N)/ Z^-(\beta, N)$$
is larger that 1 then the event $\sigma_N=+1$ occurs with probability larger 
that $1/2$; if the ratio is less than 1 then $\sigma_N=-1$ prevails. 
We are thus led to compute the ratio $Z^+/Z^-$.

\medskip

Define the vector
$$V(N)=\left(
\aligned
&Z^+(\beta, N) \\
&Z^-(\beta, N)
\endaligned
\right).
$$
Then
$$V(N)=M(\varepsilon_{N-1}) V(N-1),$$
where the ``transfer matrix'' is defined by
$$M(\varepsilon_q)=\left(
\matrix
z^{\varepsilon_q+\frac{1}{2}\alpha}\ & z^{-\varepsilon_q+\frac{1}{2} \alpha} \\
z^{-\varepsilon_q-\frac{1}{2}\alpha}\ & z^{\varepsilon_q-\frac{1}{2}\alpha}
\endmatrix
\right).
$$
To simplify notations we have put $\alpha=\frac{2H}{J}$ and $z=e^{\beta J}$.

\noindent Notice that $T\searrow 0 \Longleftrightarrow z\nearrow + \infty$.

\smallskip

By induction
$$V(N)=M(\varepsilon_{N-1}) M(\varepsilon_{N-2})\ \cdots\ M(\varepsilon_0) 
V(0),$$
where
$$V(0)=\left(
\aligned
&\exp(-H) \\
&\exp(+H)
\endaligned
\right).
$$
The product of the transfer matrices is a $2\times 2$ matrix the entries of 
which are ``polynomials'' in $z$. As $T$ decreases 
to 0, $z$ increases to infinity. Only highest degrees are relevant:
$$V(q)\sim \left(\aligned
&z^{a_q} \\
&z^{b_q}
\endaligned
\right).
$$
The symbol $\sim$ signifies that the first (resp. second) coordinate has
order $z^{a_q}$ (resp. $z^{b_q}$).

\medskip

Similarly
$$\aligned
V(q+1)&\sim
\left(\aligned
&z^{a_{q+1}} \\
&z^{b_{q+1}}
\endaligned
\right) \\
&\sim
\left(\aligned
&z^{\varepsilon_q+\frac{1}{2}\alpha}\ \ \,z^{-\varepsilon_q+\frac{1}{2}\alpha} \\
&z^{-\varepsilon_q-\frac{1}{2}\alpha}\ \ z^{\varepsilon_q-\frac{1}{2}\alpha}
\endaligned
\right)
\left(\aligned
&z^{a_q} \\
&z^{b_q}
\endaligned
\right),
\endaligned
$$
therefore
$$\left\{\aligned
&a_{q+1}=\frac{1}{2}\alpha + \max\{a_q+\varepsilon_q,\ b_q-\varepsilon_q\} \\
&b_{q+1}=-\frac{1}{2}\alpha + \max\{a_q-\varepsilon_q,\ b_q+\varepsilon_q\}.
\endaligned
\right.
\tag{1}
$$
On the other hand
$$\dfrac{Z^+(\beta, q)}{Z^-(\beta, q)}\sim z^{a_q-b_q}$$
tends to $+\infty$ or $0$ as $z$ tends to $+\infty$ according to the sign of
$$\delta_q=a_q-b_q.$$
Hence the sign of $\delta_N$ (notice the subscript!) imposes the value of the spin $\sigma_N$
at equilibrium:
$$\sigma_N=\text{sgn} (\delta_N).$$
If $\delta_N=0$ we agree to say that the spin $\sigma_N$ is not determined; it 
can be chosen at will $+1$ or $-1$. (This is not entirely correct. If $a_N=b_N$ 
one should then look at the coefficients of $z^{a_N}$ in both the components of 
$V(N)$. The larger one should then impose the value of $\sigma_N$. This case is 
so singular that we shall adopt the above ``convention''.)

\medskip

The quantity $\delta_N$ is called the induced field at site $N$.\medskip
The two equalities (1) imply
$$\delta_{q+1}=\alpha+\varepsilon_q\text{sgn}(\delta_q) \min\{2, |\delta_q|\}.$$

Changing the sign of $\alpha=2H/J$ only changes the sign of the sequence 
$(\delta_q)$ so without loss of generality we shall assume from now on that 
$\alpha\ge0$, $(H\ge0)$.

\medskip

The above induction formula has a special feature namely that if $\alpha>0$, and
if $q>4/\alpha$, then for almost all $\varepsilon$, $\delta_q$ is independent of
$\delta_0$. If on the other hand $\alpha=0$ we loose no information on 
$\text{sgn }\delta_q$ by fixing $\delta_0$ arbitrarily.

\medskip

It is convenient to choose $\delta_0=\alpha+2$. We are now left to study the 
sequence $(\delta_n)$ defined by
$$\left\{
\aligned
&\delta_0=\alpha+2 \\
&\delta_{n+1}=\alpha+\varepsilon_n\text{sgn}(\delta_n) \min \{2, |\delta_n|\}.
\endaligned
\right.
\tag{2}
$$
As an easy exercise, the reader may verify that for $\alpha=0$, ($H=0$), these 
equations reduce to
$$\delta_{n+1}=2 \varepsilon_0 \varepsilon_1 \cdots \varepsilon_{n-1} 
\varepsilon_n$$
and for $\alpha \ge 4$, $(H\ge 2J)$,
$$\delta_{n+1}=\alpha+2 \varepsilon_n.$$

How are we to interpret the sequence
$\delta_0,\,\delta_1,\,\delta_2,\,\cdots,\,\delta_n,\,\cdots$? The sign of 
$\delta_n$ is the value of the spin $\sigma_n$ at equilibrium under the 
influence of the sites $n-1,\ n-2,\ \cdots$ ignoring sites $n+1,\ n+2,\ \cdots$ 
In particular $\sigma_N = \text{sgn} (\delta_N)$. But it is not true that
 $\sigma_{N-1} = \text{sgn} (\delta_{N-1})$ since the site $N-1$ is influenced
by the sites $N-2$, $N-3$, $\cdots$ and {\it also} by the site $N$. T. Kamae
pointed out that finding the values of $\sigma_n$ for all $n \leq N$ is now
actually possible. The details will appear in a forthcoming paper [8].

\bigskip

\newpage

\noindent
{\bf 3. THE ISING AUTOMATON}

\bigskip

The rest of this chapter is devoted to the study of the sequence 
$\delta=(\delta_n)$ which we denote $\delta^\alpha=(\delta^\alpha_n)$ or
$\delta^\alpha(\varepsilon)=(\delta_n^\alpha(\varepsilon))$ if we wish to 
emphasize the dependence of $\delta$ on $\alpha$ or 
$\varepsilon=(\varepsilon_n)$.

\medskip

We first observe that for all $n$
$$\alpha-2\le \delta_n\le \alpha+2.$$
More precisely, if $\alpha=0$ then
$$\delta_n\in\{-2, +2\}$$
and if $\alpha>0$
$$
\delta_n \in\{(i+1)\alpha-2\ |\ i=0,1,2, \cdots, [4/\alpha]\} 
\cup \{2-(i-1)\alpha\ |\ i=0,1,2, \cdots, [4/\alpha]\}.
$$
More is true: $\alpha$ being fixed, the map 
$$\varepsilon\longmapsto \delta^\alpha(\varepsilon)$$
is described by an automaton $\Cal A_\alpha$ and an output function 
$\varphi_\alpha$ which we discuss now. We agree to call the couple 
$(\Cal A_\alpha, \varphi_\alpha)$ the Ising automaton, so we have a 
family of Ising automata depending on the external field $\alpha=2H/J$.

\bigskip

{\bf Case 1.} $\alpha=0$. The Ising automaton is the Thue-Morse automaton with 
output function $\varphi_0(A)=2,\ \varphi_0(B)=-2$.

\bigskip

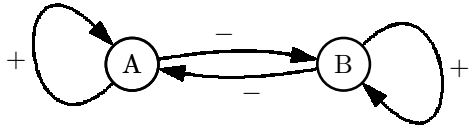
\Draw
\NewCIRCNode(\StateNode,103,)
\Define\StateAt(3){ \MoveTo(#2,#3) \StateNode(#1)(--#1--)}
\CycleEdgeSpec(45,50)
\DiagramSpec(\StateAt&\TransEdge)
\ArrowHeads(1) 
\SaveAll
\Diagram
(A,0,0 & B,80,0) (A,B,$-$,$-$ & B,B,0,$+$) 
\RecallAll
\CycleEdgeSpec(225,50)
\Diagram  
(A,0,0) (A,A,0,$+$)
\EndDraw

\bigskip

{\bf Case 2.} $\alpha>0$. Put $m=[4/\alpha]$. The Ising automaton has 
$2(m+1)$ states.

\bigskip

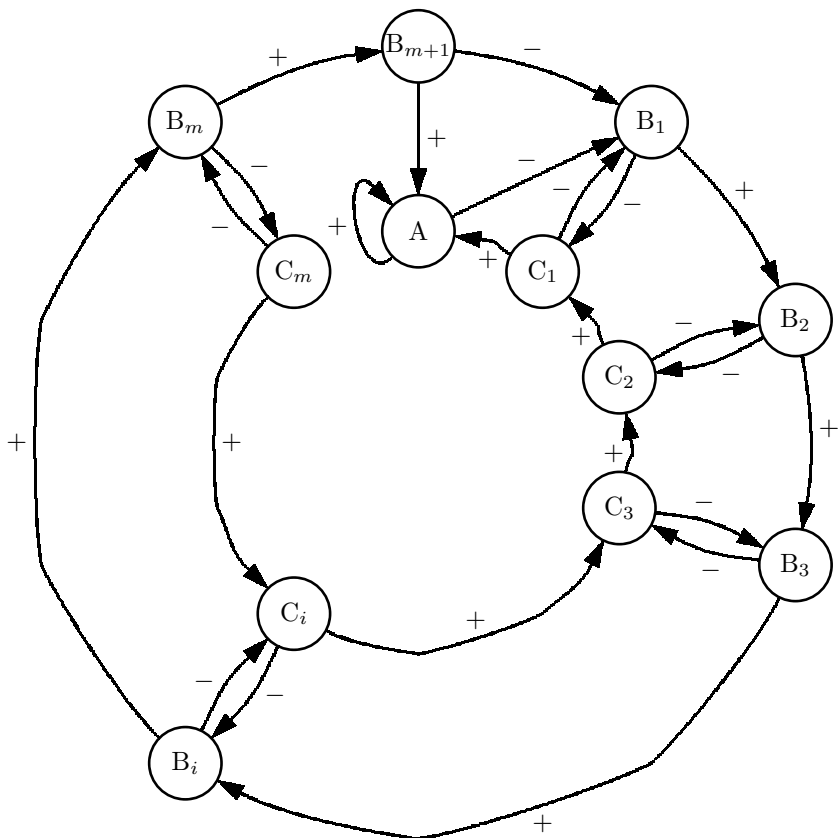
\Draw
\NewCIRCNode(\StateNode,106,)
\Define\StateAt(3){ \MoveTo(0,0) \RotateTo(#2) \MoveF(#3)
\StateNode(#1)(--#1--)}
\Define\FcState(3){ \MoveTo(0,0) \RotateTo(#2)  \MoveF(#3)
\FcNode(#1)}
\CycleEdgeSpec(225,20)
\DiagramSpec(\StateAt  & \FcState &\TransEdge)
\ArrowHeads(1) 
\Diagram
(B$_{m+1}$,90,150 & A,90,80 & B$_1$,54,150 & C$_1$,54,80 
& B$_2$,18,150 & C$_2$,18,80 & B$_3$,-18,150 & C$_3$,-18,80
& B$_i$,-126,150 & C$_i$,-126,80 & B$_m$,126,150 & C$_m$,126,80)
(X$_1$,-54,150 & Y$_1$,-54,80 & X$_2$,-90,150 & Y$_2$,-90,80
& X$_3$,-162,150 & Y$_3$,-162,80 & X$_4$,-198,150 & Y$_4$,-198,80)
(B$_{m+1}$,A,$+$, & B$_1$,C$_1$,$-$,$-$ & B$_2$,C$_2$,$-$,$-$
& B$_3$,C$_3$,$-$,$-$ & B$_i$,C$_i$,$-$,$-$ 
& B$_m$,C$_m$,$-$,$-$
& A,B$_1$,$-$, 
& A,A,0,$+$ )

\CurvedEdge(B$_{m+1}$,B$_1$) \EdgeLabel(--$-$--)
\CurvedEdge(B$_1$,B$_2$) \EdgeLabel(--$+$--)
\CurvedEdge(B$_2$,B$_3$) \EdgeLabel(--$+$--) 
\CurvedEdge(X$_2$,B$_i$) 
\CurvedEdge(X$_4$,B$_m$)  
\CurvedEdge(B$_m$,B$_{m+1}$) \EdgeLabel(--$+$--)

\CurvedEdgeSpec(-10,.2,10,.2)
\CurvedEdge(C$_1$,A) \EdgeLabel(--$+$--)
\CurvedEdge(C$_2$,C$_1$) \EdgeLabel(--$+$--)
\CurvedEdge(C$_3$,C$_2$) \EdgeLabel(--$+$--)
\CurvedEdge(Y$_1$,C$_3$) 
\CurvedEdge(Y$_3$,C$_i$) 

\ArrowHeads(0) 
\CurvedEdgeSpec(10,.2,-10,.2)
\CurvedEdge(B$_3$,X$_1$) 
\CurvedEdge(X$_1$,X$_2$) \EdgeLabel(--$+$--)
\CurvedEdge(B$_i$,X$_3$)
\CurvedEdge(X$_3$,X$_4$) \EdgeLabel(--$+$--)
\CurvedEdgeSpec(-10,.2,10,.2)
\CurvedEdge(C$_i$,Y$_2$)
\CurvedEdge(Y$_2$,Y$_1$) \EdgeLabel(--$+$--)
\CurvedEdge(C$_m$,Y$_4$)
\CurvedEdge(Y$_4$,Y$_3$) \EdgeLabel(--$+$--)

\EndDraw

\smallskip

$\varphi_\alpha(A)=\alpha+2$, $\varphi_\alpha(B_i)
=i\alpha-2$, $i=1,2, \cdots, m+1$, 

$\varphi_\alpha(C_j) = 2-(j-1)\alpha$, $j=1,2, \cdots, m$.

\bigskip

\noindent
{\bf Remark} 

As $\alpha$ tends to $0$ the number of states of $\Cal A_\alpha$ 
goes to infinity. On the other hand, $\Cal A_0$ has only two states so there 
seems to be some discontinuity as $\alpha$ vanishes. The ``paradox'' can be 
solved in the following way. Let $n$ be somewhere between 0 and $[4/\alpha]$,
$0\ll n\ll [4/\alpha]$.
In the vicinity of ``$n$'' the automaton $\Cal A_\alpha$ looks like

\bigskip

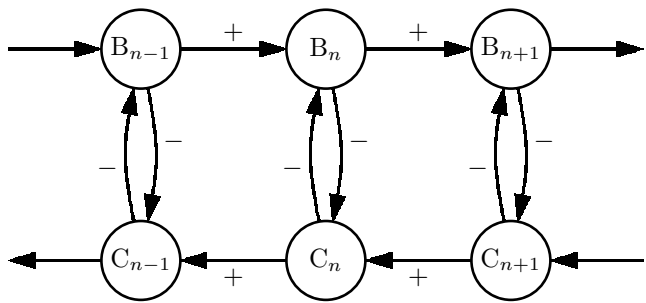
\Draw
\NewCIRCNode(\StateNode,107,)
\Define\StateAt(3){ \MoveTo(#2,#3) \StateNode(#1)(--#1--)}
\Define\FcState(3){ \MoveTo(#2,#3) \FcNode(#1)}
\CycleEdgeSpec(45,50)
\DiagramSpec(\StateAt & \FcState &\TransEdge)
\ArrowHeads(1) 
\Diagram
(B$_{n-1}$,0,0 & B$_n$,70,0 & B$_{n+1}$,140,0
& C$_{n-1}$,0,-80 & C$_n$,70,-80 & C$_{n+1}$,140,-80) 
(X,-50,0 & Y,-50,-80 & Z,190,0 & T,190,-80) 
(X,B$_{n-1}$, , & B$_{n-1}$,B$_n$,$+$, & B$_n$,B$_{n+1}$,$+$,
& B$_{n+1}$,Z, , 
& C$_{n-1}$,Y, , & C$_{n-1}$,C$_n$, ,$+$ & C$_n$,C$_{n+1}$, ,$+$
& T,C$_{n+1}$, ,
& C$_{n-1}$,B$_{n-1}$,$-$,$-$ & C$_n$,B$_n$,$-$,$-$
& C$_{n+1}$,B$_{n+1}$,$-$,$-$)
\EndDraw

\bigskip

\noindent
and
$$\left\{
\aligned
\varphi_\alpha(B_n)&= -2+n \alpha \\
\varphi_\alpha(C_n)&=2-(n-1)\alpha.
\endaligned
\right.
$$
When $\alpha$ vanishes, $\varphi_\alpha$ tends to
$$\left\{
\aligned
\varphi_0(B_n)&= -2\  \text{ for all }\ n \\
\varphi_0(C_n)&=2.
\endaligned
\right.
$$
This automaton which contains infinitely many states collapses and becomes 
indiscernable from $\Cal A_0$! 

\bigskip

\noindent
{\bf 4. AN ERGODIC PROPERTY}

\bigskip

\proclaim{Theorem 1} For all $\alpha$ and for almost all sequences $\varepsilon$
$$\lim\limits_{N\rightarrow\infty} \frac{1}{N} \sum\limits_{n=0}^{N-1} 
(\Cal A_\alpha \varepsilon)_n=\alpha.$$
\endproclaim

The proof we propose here is shorter than in [9]. 
Let $T$ be the shift operator on $\{-1, +1\}^{\Bbb N}$: 
$T(\varepsilon_n) = (\varepsilon_{n+1})$. Let $A$ be the initial state of the
automaton $\Cal A_{\alpha}$. Define the operator $\tilde T$ on the space
$\{-1, +1\}^{\Bbb N} \times S$:
$$\tilde T ((\varepsilon), B) = (T(\varepsilon), \varepsilon_0 B).$$
Here $S$ is the set of states of $\Cal A_{\alpha}$. Then
$$\tilde T^n(\varepsilon, A) = (T^n\varepsilon, \varepsilon_{n-1} 
\varepsilon_{n-2} \cdots \varepsilon_0 A).$$
The operator $\tilde T$ is ergodic with respect to the canonical measure on 
the space $\{-1, +1\}^{\Bbb N} \times S$. Define
$$\aligned
F :  \{-1, +1\}^{\Bbb N} \times S &\to \Bbb R \\
(\delta, B) &\mapsto \varphi_{\alpha}(B).
\endaligned
$$
The ergodic theorem asserts that, for almost all $\varepsilon \in 
\{-1, +1\}^{\Bbb N}$,
$$\lim_{N \to \infty} \frac{1}{N} \sum_{n < N} F(\tilde T^n(\varepsilon, A))
= \int \varphi_{\alpha}(B) \ dB = 
\frac{1}{\sharp S} \sum_{B \in S} \varphi_{\alpha}(B).$$
Therefore
$$\lim_{N \to \infty} \frac{1}{N} \sum_{n < N} 
\varphi_{\alpha}(\varepsilon_{n-1} \varepsilon_{n-2} \cdots  \varepsilon_0 A)
= \alpha, \qquad \qquad \qquad \text{QED.}$$

\bigskip

\noindent
{\bf Remark} 

At the end of the proof we observed that, for almost all $\varepsilon$, 
$$\lim\limits_{N\rightarrow\infty} \frac{1}{N} \sum\limits_{n=0}^{N-1} 
(\Cal A_\alpha \varepsilon)_n = 
\frac{1}{\sharp S} \sum_{B \in S} \varphi_{\alpha}(B).$$
The equality shows that the dynamical average coincides with the spatial average.

\proclaim{Theorem 2} For all $\alpha>0$ and all sequences $\varepsilon$
$$\liminf\limits_{N \to \infty} \frac{1}{N} \sum\limits_{n=0}^{N-1} (\Cal
A_\alpha\varepsilon)_n\ge \frac{1}{2} \alpha.$$
\endproclaim

\smallskip
\noindent
{\it Proof.} Recall
$$(\Cal A_\alpha\varepsilon)_n=\delta_n$$
and
$$\delta_{n+1}=\alpha+\varepsilon_n\text{sgn}(\delta_n)\min \{2, |\delta_n|\}.$$
Hence
$$
\aligned
(\delta_{n+1}-\alpha)^2
&=\min \{4, \delta^2_n\}, \\
2\alpha \delta_{n+1}
&= \alpha^2+\delta^2_{n+1}-\min \{4, \delta^2_n\} \\
&\ge \alpha^2+ \delta^2_{n+1} - \delta^2_n, \\
\delta_{n+1}
&\ge \frac{\alpha}{2} + \frac{1}{2\alpha} (\delta^2_{n+1} - \delta^2_n), \\
\frac{1}{N} \sum\limits_{n=0}^{N-1} \delta_{n+1}
&\ge \frac{\alpha}{2} + \frac{1}{2 \alpha N} (\delta^2_N-\delta^2_0),
\qquad\qquad\qquad \text{QED.}
\endaligned
$$

The following theorem strengthens Theorem 2.

\proclaim{Theorem 3} Let $\alpha>0$ be given and let $\beta, \beta'$ be two real
numbers such that
$$\max\left\{\frac{\alpha}{2}, \alpha-2\right\}\le \beta \le 
\beta'\le \alpha+2.$$
Then there exists a sequence $\varepsilon\in\{-1, +1\}^{\Bbb N}$ such that
$$\aligned
&\liminf\limits_{N \to \infty} \frac{1}{N} \sum\limits_{n=0}^{N-1} (\Cal
A_\alpha \varepsilon)_n= \beta, \\
&\limsup\limits_{N \to \infty} \frac{1}{N} \sum\limits_{n=0}^{N-1} (\Cal
A_\alpha \varepsilon)_n= \beta'.
\endaligned
$$
\endproclaim

\smallskip

\noindent
{\it Proof.} If $\alpha\ge4$ then
$$\delta_{n+1}=\alpha+2\varepsilon_n$$
Then the result is easily established.

\medskip

Let us suppose $0<\alpha<4$. Consider a closed path $P(p,q)$ on the automaton 
$\Cal A_\alpha$ which loops $p$ times around $A$, then joins $B_1$, loops $q$ 
times around $B_1$ and $B_2$ and finally goes back to $A$. The average of 
$\delta_n$ on the closed path is
$$\frac{p(\alpha+2)+ q \alpha+ O(1)}{p+  2q + O(1)}.$$
Let $p_n$ and $q_n$ be two infinite sequences of integers and define
$$\aligned
\underline{\lambda}&= \liminf\limits_{n \to \infty} p_n/q_n \\
\overline{\lambda}&= \limsup\limits_{n \to \infty} p_n/q_n.
\endaligned
$$
Clearly $0\le\underline{\lambda} \le\overline{\lambda}\le +\infty$.

\medskip

\noindent
Consider an infinite path $P$ which consists in the union of 
$P(p_1, q_1)$, $P(p_2,q_2)$, $\cdots$, $P(p_n, q_n)$, $\cdots$. $P$ determines 
a sequences $\varepsilon$ such that
$$\aligned
&\limsup\limits_{N \to \infty} \frac{1}{N} \sum\limits_{n=0}^{N-1}
\delta_n(\varepsilon)= \frac{(\alpha+2)\overline{\lambda}+\alpha}
{\overline\lambda+2} \\
&\liminf\limits_{N \to \infty} \frac{1}{N} \sum\limits_{n=0}^{N-1}
\delta_n(\varepsilon)= \frac{(\alpha+2)\underline{\lambda}+\alpha}
{\underline\lambda+2}
\endaligned
$$
Choose $\underline\lambda$ and $\overline\lambda$ such that
$$\aligned
&\frac{(\alpha+2)\overline{\lambda}+\alpha}{\overline\lambda+2} = \beta' \\
&\frac{(\alpha+2)\underline{\lambda}+\alpha}{\underline\lambda+2} = \beta,
\qquad\qquad\qquad \text{QED.}
\endaligned
$$

\medskip

\noindent
{\bf Remark} 

By choosing $\underline\lambda=0$ one obtains a second proof of Theorem 2.

\bigskip

\noindent
{\bf 5. OPACITY OF THE ISING AUTOMATON} [9]

\bigskip

The Ising automaton is not homogeneous but the theorem of Part VI can be applied,
(see comment following the theorem):
it is a simple matter to maximize $\sqrt{2\nu(P)/\ell(P)}$.

\proclaim{Theorem 4} The opacity of $\Cal A_0$ is 1. If $\alpha>0$, the opacity 
of $\Cal A_\alpha$ is
$$\sqrt{\frac{m'-1}{m'}}, \
\text{ where } m'=\max \{1, [4/\alpha]\}.$$
\endproclaim

\medskip

Many of the results in this part have already been published in different 
journals by several authors. We should point out in particular the papers
of Allouche, Mend\`es France [1], Derrida [6], Mend\`es France [9]. 
In [1] we show that if $\varepsilon$ is an automatic sequence
then the induced field is also automatic.
This result has been generalized by Dekking [5] to sequences which are literal
images of fixed points of any morphism, (not necessarily of constant length).
Theorems 2 and 3 have not appeared in print.

\medskip

The general theory of the Ising model originates with Ising [7]. The literature
concerning the Ising model is so vaste that it would be too long to list all the
books and articles. We single out the work of Baxter [2], Biggs [3], Cipra [4], 
Thompson [10]. 

\bigskip
\bigskip

\references

\refis{1} 
Allouche J.-P., Mend\`es France M., Quasicrystal Ising chain and automata 
theory, \review J. Stat. Phys., 42, 1986, 809--821.

\refis{2} 
Baxter R. J., Exactly solved models in statistical mechanics
(Academic Press, 1982).

\refis{3} 
Biggs N. L., Interaction models, London Mathematical Society
(Lecture Note Series, {\bf 30}, Cambridge University Press, 1977).

\refis{4}
Cipra B. A., An introduction to the Ising model, \review Amer. Math. Monthly,
94, 1987, 937--959.

\refis{5}
Dekking F. M., Iteration of a map by an automaton, \review Discrete Math.,
126, 1994, 81--86. 

\refis{6}
Derrida B., Products of random matrices and one dimensional disordered systems 
in: Nonlinear Equations in Field Theory (Lectures Notes in Physics,
{\bf 226}, Springer-Verlag, 1985).

\refis{7} 
Ising E., Beitrag zur Theorie der Ferromagnetismus, \review Z. Physik, 31, 
1925, 253--258.

\refis{8}
Kamae T., Mend\`es France M., A continuous family of automata:
the Ising automata, {\it Submitted} (1994).

\refis{9}
Mend\`es France M., Opacity of an automaton. Application to the 
inhomogeneous Ising chain, \review Comm. Math. Phys., 139, 1991, 341--352.

\refis{10}
Thompson C. J., Mathematical statistical mechanics
(Princeton University Press, 1972).

\endreferences

\newpage

\centerline{\big PART VIII}

\bigskip

\centerline{\big Extra references}

\bigskip

It may be useful to add to the previous references the following articles
which are more concerned with physical problems and which of course involve
automatic sequences or generalizations. This list is obviously not complete.
We beg pardon to all those whom we inadvertently forgot to mention. 
Needless to say their number is much greater than the number of authors
quoted here.

\bigskip
\bigskip

\references

\refis{--} Ali M. K., Gumbs G., Quasiperiodic dynamics for a generalized 
third order Fibonacci series, \review Phys. Rev. B (Brief Reports), 39,
1988, 7091--7093.

\refis{--}
Allouche J.-P., Finite automata in $1$-D and $2$-D Physics, in:
Number Theory and Physics, Proceedings of the Winter School,
Les Houches, France, March 7-16 1989, Luck J.-M., Moussa P.,
Waldschmidt M. Eds, Springer Proceedings in Physics {\bf 47} (Springer
Verlag, 1990) p. 177--184.

\refis{--}
Allouche J.-P., Complexity of infinite sequences and the Ising transducer,
in: Cellular Automata and Cooperative Systems, Boccara N., Goles E., Martinez
S., Picco P. Eds, NATO ASI series, Series C: Mathematical and Physical 
Sciences, vol. 396 (Kluwer Academic Publishers, Dordrecht, 1993) p. 1--9.

\refis{--} 
Allouche J.-P., Mend\`es France M., Suite de Rudin-Shapiro et mod\`ele d'Ising,
\review Bull. Soc. math. France, 113, 1985, 273--283.

\refis{--}
Allouche J.-P., Mend\`es France M., Quasi-crystal Ising chain and automata 
theory, \review J. Stat. Phys., 42, 1986, 809--821.

\refis{--}
Allouche J.-P., Peyri\`ere J., Sur une formule de r\'ecurrence sur les traces 
de produits de matrices associ\'es \`a certaines substitutions, 
\review {C. R. Acad. Sci. Paris, S\'erie II}, 302, 1986, 1135--1136.

\refis{--}
Allouche J.-P., Salon O., Robinson tilings and $2$-dimensional automata, in:
Quasicrystals, networks and molecules of fivefold symmetry, Hargittai I. Ed 
(VCH Publishers Inc., 1990) p. 97--105.

\refis{--}
Aubry S., Godr\`eche C., Incommensurate structure with no average lattice:
an example of a one-dimensional quasicrystal, (Gratias D., Michel L. Eds),
\review J. de Phys. Colloq., 47 C3, 1986, 187--196.

\refis{--}
Aubry S., Godr\`eche C., Luck J.-M., A structure intermediate between 
quasiperiodic and random, \review Europhys. Lett., 4, 1987, 639--.

\refis{--}
Aubry S., Godr\`eche C., Luck J.-M., Scaling properties of a structure
intermediate between quasiperiodic and random, \review J. Stat. Phys., 51,
1988, 1033--1075.

\refis{--} 
Aubry S., Godr\`eche C., Vallet F.,  Incommensurate structure with no average 
lattice: an example of a one-dimensional quasicrystal, \review J. Phys., 48,
1987, 327--334.

\refis{--}
Avishai Y., Berend D., Transmission through Fibonacci chain, \review Phys. 
Rev. B, 43, 1991, 6873--6880.

\refis{--}
Avishai Y., Berend D., Transmission through a Thue-Morse chain, \review
Phys. Rev. B, 45, 1992, 2717--2724. 

\refis{--}
Avishai Y., Berend D., Trace maps for arbitrary substitution sequences,
\review J. Phys. A: Math. Gen., 26, 1993, 2437--2443.

\refis{--}
Avishai Y., Berend D., Glaubman D., Minimum-dimension trace maps for 
substitution sequences, \review Phys. Rev. Lett., 72, 1994, 1842--1845. 

\refis{--}
Axel F., Allouche J.-P., Kl\'eman M., Mend\`es France M., Peyri\`ere, J., 
Vibrational modes in a one-dimensional ``quasi alloy'', the Morse case, 
(Gratias D., Michel L. Eds), \review J. de Phys. Colloq., 47 C3, 1986, 
181--187. 
 
\refis{--}
Axel F., Allouche J.-P., Wen Z.-Y., On certain properties of 
high-reso\-lu\-tion X-ray diffraction spectra of finite-size generalized 
Rudin-Shapiro multilayer heterostructures, \review J. Phys.: Condens. Matter, 
4, 1992, 8713--8728.

\refis{--}
Axel F., Peyri\`ere J., Spectrum and extended states in a harmonic chain with 
controlled disorder: effects of the Thue-Morse symmetry, \review
J. Stat. Phys., 57, 1989, 1013--1047. 

\refis{--} 
Axel F., Terauchi H., High-resolution X-ray diffraction spectra of Thue-Morse 
GaAs-AlAs heterostructures: towards a novel description of disorder, \review
Phys. Rev. Lett., 66, 1991, 2223--2226. 

\refis{--} 
Baake M., Grimm U., Joseph D., Trace maps, invariants, and some of their 
applications, \review Int. J. Mod. Phys. B, 7, 1993, 1527--1550.

\refis{--}
Baake M., Grimm U., Pisani C., Partition function zeros for aperiodic systems, 
\review J. Stat. Phys., , 1995, to \ appear.

\refis{--}
Baake M., Joseph D., Kramer P.,  Periodic clustering in the spectrum of 
quasiperiodic Kronig-Penney models, \review Phys. Lett. A, 168, 1992, 199--208.

\refis{--}
Baake M., Roberts J. A. G., Symmetries and reversing symmetries of trace maps,
in: Proceedings 3rd International Wigner Symposium, Boyle L. L., Solomon A. I.
Eds (Oxford, 1993).

\refis{--}
Bajema K., Merlin R., Raman scattering by acoustic phonons in Fibonacci 
GaAs-AlAs superlattices, \review Phys. Rev. B, 36, 1987, 4555--4557.

\refis{--}
Barache D., Luck J.-M., Electronic spectra of strongly modulated aperiodic 
sequences, \review Phys. Rev. B, 49, 1994, 15004--15016. 

\refis{--}
Bellissard J., Spectral properties of Schr\"odinger's operator with a 
Thue-Morse potential, in: Number theory and physics, Luck J.-M., Moussa P., 
Waldschmidt M. Eds, Springer Proceedings in Physics {\bf 47} 
(Springer, Berlin, Heidelberg, New York, 1990) p. 140--150.

\refis{--}
Bellissard J., Gap labelling theorems for Schr\"odinger's operators, in: From 
number theory to physics, Waldschmidt M., Moussa P., Luck J.-M., Itzykson C.
Eds (Springer, Berlin, Heidelberg, New York, 1992) p. 538--630.

\refis{--}
Bellissard J., Bovier A., Ghez J.-M., Spectral properties of a tight binding 
hamiltonian with period doubling potential, \review Comm. Math. Phys., 135, 
1991, 379--399. 

\refis{--}
Bellissard J., Bovier A., Ghez J.-M., Gap labelling theorems for 
one-dimensional discrete Schr\"o\-dinger operators, \review Rev. Math. Phys., 
4, 1992, 1--37.

\refis{--}
Bellissard J., Bovier A., Ghez J.-M., Discrete Schr\"odinger operators with 
potentials generated by substitutions, in: Differential equations with 
applications to mathematical physics, Ames W. F., Harrell E. M., Herod J. V. 
Eds (Academic Press, Boston, 1993) p. 13--23.

\refis{--}
Bellissard J., Iochum B., Scoppola E., Testard D., Spectral properties of 
one-dimensional quasi-crystals, \review Comm. Math. Phys., 125, 1989, 527--543. 

\refis{--}
Benza V. G., Kol\'ar M., Ali M. K., Phase transition in the generalized
Fibonacci quantum Ising models, \review Phys. Rev. B (Rapid Comm.), 41,
1990, 9578--.

\refis{--}
Bovier A., Ghez J.-M., Schr\"odinger operators with substitution potentials, 
in: Cellular Automata and Cooperative Systems, Boccara N., Goles E., Martinez
S., Picco P. Eds, NATO ASI Series, Series C: Mathematical and Physical 
Sciences, vol. 396 (Kluwer Academic Publishers, Dordrecht, 1993) p. 67--83.

\refis{--}
Bovier A., Ghez J.-M., Spectral properties of one-dimensional Schr\"odin\-ger 
operators with potentials generated by substitutions, {\it Comm. Math. Phys.}
{\bf 158} (1993) 45--66 and erratum, {\it Comm. Math. Phys.} to appear.

\refis{--}
Bovier A., Ghez J.-M., Remarks on the spectral properties of tight binding
and Kronig-Penney models with substitution sequences, \review
J. Phys. A: Math. Gen., , 1995, to \ appear.

\refis{--}
Casdagli M., Symbolic dynamics for the renormalization map of a quasi\-periodic 
Schr\"o\-din\-ger equation, \review Comm. Math. Phys., 107, 1986, 295--318. 

\refis{--}
Cheng Z., Savit R., Merlin R., Structure and electronic properties of 
Thue-Morse lattices, \review Phys. Rev. B, 37, 1988, 4375--4382. 

\refis{--} 
Combescure M., Recurrent versus diffusive dynamics for a kicked quantum sytem,
\review J. Stat. Phys., 62, 1991, 779--791. 
 
\refis{--}
Combescure M., Recurrent versus diffusive dynamics for a kicked quantum 
oscillator, \review Ann. Inst. Henri Poincar\'e, 57, 1992, 67--87.  

\refis{--} Doria M. M., Nori F., Satija I. I., Thue-Morse quantum Ising
model, \review Phys. Rev. B, 39, 1989, 6802--6806.

\refis{--}
Dulea M., Physical properties of one-dimensional deterministic aperiodic 
systems (Thesis, Link\"oping, 1992).

\refis{--}
Dulea M., Johansson M., Riklund R., Unusual scaling of the spectrum in a 
deterministic aperiodic tight-binding model, \review Phys. Rev. B, 47, 1993,
8547--8551. 
 
\refis{--}
Ghosh P. K., The Kronig-Penney model on a generalized Fibonacci lattice, 
\review Phys. Lett. A, 161, 1991, 153--157.
 
\refis{--}
Godr\`eche C., Beyond quasiperiodicity: scaling properties of a Fourier
spectrum, in: Universalities in condensed matter, Jullien R. et al. Eds,
Springer Proceedings in Physics {\bf 32} (Springer Verlag, 1988).

\refis{--}
Godr\`eche C., The sphinx: a limit-periodic tiling of the plane, \review
J. Phys. A: Math. Gen., 22, 1989, 1163--1166.

\refis{--}
Godr\`eche C., Types of order and diffraction spectra for tilings of the line,
in: Number theory and physics, Luck J.-M., Moussa P.,
Waldschmidt M. Eds, Springer Proceedings in Physics {\bf 47}
(Springer, Berlin, Heidelberg, New York, 1990) p. 86--99.

\refis{--}
Godr\`eche C., Non-Pisot tilings and singular scattering, \review Phase
Transitions, 32, 1991, 45--.

\refis{--}
Godr\`eche C., Lan\c{c}on F., A simple example of a non-Pisot tiling with
fivefold symmetry, \review J. Physique, 2, 1992, 207--220.

\refis{--}
Godr\`eche C., Luck J.-M.,  Tilings of the plane and their diffraction spectra,
Proceedings of the Anniversary Adriatico Research Conference on Quasicrystals,
Trieste, Italy, July 4-7 1989, Jaric M., Lundqvist S. Eds (World Scientific,
1989).

\refis{--}
Godr\`eche C., Luck J.-M., Quasiperiodicity and randomness in tilings of the
plane, \review J. Stat. Phys., 55, 1989, 1--28.

\refis{--}
Godr\`eche C., Luck J.-M., Multifractal analysis in reciprocal space and the 
nature of the Fourier transform of self-similar structures, \review J. Phys. A: 
Math. Gen., 23, 1990, 3769--3797. 

\refis{--}
Godr\`eche C., Luck J.-M., Indexing the diffraction spectrum of a non-Pisot
self-similar structure, \review Phys. Rev. B, 45, 1992, 176--.

\refis{--}
Godr\`eche C., Luck J.-M., Janner A., Janssen T., Fractal atomic surfaces of
self-similar quasiperiodic tilings of the plane, \review J. Physique, 13, 1993,
1921--.

\refis{--}
Godr\`eche C., Luck J.-M., Orland H., Magnetic phase structure on the
Penrose lattice, \review J. Stat. Phys., 45, 1986, 777--800. 

\refis{--}
Godr\`eche C., Luck J.-M., Vallet F., Quasiperiodicity and types of order:
a study in one dimension, \review  J. Phys. A: Math. Gen., 20, 1987, 
4483--4499.

\refis{--}
Godr\`eche C., Oguey C., Construction of average lattices for quasiperiodic
structure by the section method, \review J. Physique, 51, 1990, 21--37.

\refis{--}
Godr\`eche C., Orland H., Renormalisation on the Penrose lattice,
(Gratias D., Michel L. Eds), \review J. de Phys. Colloq., 47 C3, 1986, 
197--203. 

\refis{--}
Grimm U., Baake M.,  Non-periodic Ising quantum chains and conformal 
invariance, \review J. Stat. Phys., 74, 1994, 1233--1245. 

\refis{--}
Gumbs G., Ali M. K., Dynamical maps, Cantor spectra and localization
for Fibonacci and related quasiperiodic lattices, \review Phys. Rev. Lett.,
60, 1988, 1081--1084.

\refis{--}
Gumbs G., Ali M. K., Scaling and eigenstates for a class of one-dimensio\-nal
quasiperiodic lattices, \review J. Phys. A: Math. Gen., 21, 1988, 517--521.

\refis{--}
Gumbs G., Ali M. K., Electronic properties of the tight-binding Fibonacci
hamiltonian, \review J. Phys. A: Math. Gen., 22, 1989, 951--970.

\refis{--}
Hof A., Quasi-crystals, aperiodicity and lattice systems (Thesis,
Groningen, 1992).

\refis{--}
Hof A., Knill O., Simon B., Singular continuous spectrum for palindromic 
Schr\"odinger operators, {\it Preprint Caltech} (1994).

\refis{--}
Holzer M., Three classes of one-dimensional, two-tile Penrose tilings and the
Fibonacci Kronig-Penney model as a generic case, \review Phys. Rev. B, 38,
1988, 1709--1720.

\refis{--}
H\"ornquist M., Johansson M.,  Singular continuous electron spectrum for a 
class of circle sequences, {\it Preprint Link\"oping University} (1994) 
and {\it J. Phys. A: Math. Gen.} to appear.
 
\refis{--}
Hu P., Ting C. S., Electron transmission in a one-dimensional quasicrystal, 
\review Phys. Rev. B, 34, 1986, 8331--8334. 

\refis{--}
Iochum B., Raymond L., Testard D., Resistance of one-dimensional
quasicrystals, \review Physica A, 187, 1992, 353--368.

\refis{--}
Iochum B., Testard D., Power law growth for the resistance in the Fibonacci 
model, \review J. Stat. Phys., 65, 1991, 715--723. 

\refis{--}
Kalugin P. A., Kitaev A. Y., Levitov L. S., Electron spectrum of a 
one-dimensional quasicrystal, \review Sov. Phys. JETP, 64, 1986, 410--416.

\refis{--}
Kim Y.-J., Oh G. Y., Lee M. H., Electronic structure of a one-dimensio\-nal 
quasicrystal, \review J. Korean Phys. Soc., 22, 1989, 136--140.  
 
\refis{--}
Kohmoto M., Kadanoff L. P., Tang C., Localization problem in one dimension: 
mapping and escape, \review Phys. Rev. Lett., 50, 1983, 1870--1872. 

\refis{--}
Kohmoto M., Oono Y., Cantor spectrum for an almost periodic Schr\"o\-dinger 
equation and a dynamical map, {\it Phys. Lett.} {\bf 102} A (1984) 145--148.

\refis{--}
Kol\'ar M., Ali M. K., Generalized Fibonacci superlattices dynamical trace
maps and magnetic excitations, \review Phys. Rev. B, 39, 1989, 426--432.

\refis{--}
Kol\'ar M., Ali M. K., Attractors of some volume non-preserving trace maps,
\review Phys. Rev. A, 39, 1989, 6538--6543.

\refis{--} 
Kol\'ar M., Ali M. K., Attractors in quantum Ising models, \review Phys. Rev.
B, 40, 1989, 11083--11089.

\refis{--}
Kol\'ar M., Ali M. K., One-dimensional generalized Fibonacci tilings,
\review Phys. Rev. B, 41, 1990, 7108--.

\refis{--}
Kol\'ar M., Ali M. K., Trace maps associated with general two-letter 
substitution rules, \review Phys. Rev. A, 42, 1990, 7112--7124.

\refis{--} 
Kol\'ar M., Ali M. K., Nori F., Generalized Thue-Morse chains and their
physical properties, \review Phys. Rev. B, 43, 1991, 1034--.

\refis{--}
Kol\'ar M., Iochum B., Raymond L., Structure factor of 1D systems 
(superlattices) based on two-letter substitution rules. I. $\delta$ (Bragg)
peaks, \review J. Phys. A: Math. Gen., 26, 1993, 7343--7366.

\refis{--}
Kol\'ar M., Nori F., Trace maps of general substitutional sequences, 
\review Phys. Rev. B, 42, 1990, 1062--1065. 

\refis{--}
Kol\'ar M., S\"ut\H o A., The Kronig-Penney model on a Fibonacci lattice, 
\review Phys. Lett. A, 117, 1986, 203--209. 

\refis{--}
Luck J.-M., Cantor spectra and scaling of gap widths in deterministic
aperiodic systems, \review Phys. Rev. B, 39, 1989, 5834--5849.
  
\refis{--}
Luck J.-M., Critical behavior of the aperiodic quantum Ising chain in a 
transverse magnetic field, \review J. Stat. Phys., 72, 1993, 417--458.

\refis{--}
Luck J.-M., Godr\`eche C., Janner A., Janssen T., The nature of the atomic 
surfaces of quasiperiodic self-similar structures, \review J. Phys. A:
Math. Gen., 26, 1993, 1951--1999.

\refis{--}
Luck J.-M., Orland H., Smilanski U., On the response of a two-level quantum
system to a class of time dependent quasiperiodic perturbations, \review
J. Stat. Phys., 53, 1988, 551--564.

\refis{--}
MacDonald A. H., Aers G. C.,  Continuum-model acoustic and electronic 
properties for a Fibonacci superlattice, \review Phys. Rev. B, 36, 1987,
9142--9145.

\refis{--}
Macia E., Dominguez-Adame F., Exciton optical absorption in self-simi\-lar 
aperiodic lattices, {\it Preprint Madrid University} (1994).

\refis{--}
Merlin R., Bajema K., Raman scattering by acoustic phonons and structural 
properties of Fibonacci, Thue-Morse and random superlattices, \review 
J. de Phys. Colloq., 48 C5, 1987, 503--506. 

\refis{--}
Merlin R., Bajema K., Clarke R., Juang F.-Y., Bhattacharya P. K., 
Quasi-periodic GaAs-AlAs heterostructures, \review Phys. Rev. Lett., 55, 
1985, 1768--1770. 

\refis{--}
Ostlund S., Pandit R., Rand D., Schellnhuber H. J., Siggia E. D.,
Schr\"o\-dinger equation with an almost periodic potential, \review Phys. Rev.
Lett., 50, 1983, 1873--1876. 

\refis{--}
Peyri\`ere J., On the trace map for products of matrices associated with 
substitutive sequences, \review J. Stat. Phys., 62, 1991, 411--414. 

\refis{--}
Peyri\`ere J., Wen Z.-Y., Wen Z.-X., Polyn\^omes associ\'es aux endomorphismes 
de groupes libres, \review Enseign. Math., 39, 1993, 153--175.

\refis{--}
Rivier N., A botanical quasicrystal,  (Gratias D., Michel L. Eds),
\review J. de Phys. Colloq., 47 C3, 1986, 299--309.

\refis{--}
Roberts J. A. G., Baake M., The dynamics of trace maps, in: Hamiltonian
mechanics: integrability and chaotic behaviour, Seimenis J. Ed, NATO ASI
Series, Series B: Physics (Plenum Press, New York, 1994) p. 275--285.
 
\refis{--}
Roberts J. A. G., Baake M., Trace maps as 3D reversible dynamical systems
with an invariant, \review J. Stat. Phys., 74, 1994, 829--888.

\refis{--}
Shechtman D., Blech I., Gratias D., Cahn J. V., Metallic phase with long-range 
orientational order and no translational symmetry, \review Phys. Rev. Lett.,
53, 1984, 1951--1953.

\refis{--}
Steinhardt P. J., Ostlund S., The physics of quasicrystals (World Scientific,
Singapore, Philadelphia, 1987).

\refis{--}
Sutherland B., Kohmoto M., Resistance of a one-dimensional quasicrystal, 
\review Phys. Rev. B, 36, 1987, 5877--5886.

\refis{--}
S\"ut\H o A., The spectrum of a quasiperiodic Schr\"odinger operator, 
\review Comm. Math. Phys., 111, 1987, 409--415. 

\refis{--}
S\"ut\H o A., Singular continuous spectrum on a Cantor set of zero Lebesgue 
measure for the Fibonacci hamiltonian, \review J. Stat. Phys., 56, 1989,
525--531.

\refis{--}
Thomas U., Kramer P., The Fibonacci quasicrystal reconsidered: variety of 
energy spectra for continuous Schr\"odinger equations with simple potentials,
\review Int. J. Mod. Phys. B, 3, 1989, 1205--1235. 

\refis{--}
Todd J., Merlin R., Clarke R., Mohanty K. M., Axe J. D., Synchrotron X-ray 
study of a Fibonacci superlattice, \review Phys. Rev. Lett., 57,
1986, 1157--1160. 

\refis{--}
Tracy C. A., Universality classes of some aperiodic Ising models, \review
J. Phys. A: Math. Gen., 21, 1988, 603--605.

\refis{--}
Wen Z.-X., Wen Z.-Y., On the leading term and the degree of the polynomial 
trace mapping associated with a substitution, \review J. Stat. Phys., 75, 
1994, 627--641.

\refis{--}
W\"urtz D., Soerensen M. P., Schneider T.,  Quasiperiodic Kronig-Penney model 
on a Fibonacci superlattice, \review Helv. Phys. Acta, 61, 1988, 345--362.

\endreferences

\enddocument